\documentclass[a4paper,11pt]{article}
\pdfoutput=1 
\usepackage{jheppub} 
\usepackage[T1]{fontenc} 
\usepackage[numbers,sort&compress]{natbib}
\usepackage{graphicx}  
\usepackage{float}
\usepackage{dcolumn}   
\usepackage{bm}        
\usepackage{amssymb}   
\usepackage{slashed}   
\usepackage{amsmath}   
\usepackage{verbatim}  
\usepackage{hyperref}
\usepackage{multicol}
\usepackage{multirow}
\usepackage{booktabs}
\usepackage{subfigure}
\usepackage{color} 
\usepackage{float}

\newcounter{YJC}

\begin{document}

\title{A quantum machine learning classifier to search for new physics}



\author[a,b]{Ji-Chong Yang}
\author[a,b]{Shuai Zhang}
\author[a,b,1]{Chong-Xing Yue,\note{Corresponding author.}}
\affiliation[a]{Department of Physics, Liaoning Normal University, Dalian 116029, China}
\affiliation[b]{Center for Theoretical and Experimental High Energy Physics, Liaoning Normal University, Dalian 116029, China}
\emailAdd{yangjichong@lnnu.edu.cn}
\emailAdd{2802368240@qq.com}
\emailAdd{cxyue@lnnu.edu.cn}


\abstract{
Due to the success of the Standard Model~(SM), it is reasonable to anticipate that the signal of new physics~(NP) beyond the SM is small.
Consequently, future searches for NP and precision tests of the SM will require high luminosity collider experiments.
Moreover, as precision tests advance, rare processes with many final-state particles require consideration which demands the analysis of a vast number of observables.
The high luminosity produces a large amount of experimental data spanning a large observable space, posing a significant data-processing challenge.
In recent years, quantum machine learning has emerged as a promising approach for processing large amounts of complex data on a quantum computer.
In this study, we propose quantum searching neighbor~(QSN) and variational QSN~(VQSN) algorithms to search for NP.
The QSN is a classification algorithm. 
The VQSN introduces variation to the QSN to process classical data.
As applications, we apply the (V)QSN in the phenomenological study of the NP at the Large Hadron Collider and muon colliders.
Examples are implemented on a real quantum hardware, which confirms reliable performance under noisy conditions.
The results indicate that the VQSN demonstrates superior efficiency in the sense of computational complexity to a classical counterpart k-nearest neighbor algorithm, even when dealing with classical data.
}

\maketitle


\section{\label{sec:1}Introduction}

The standard model~(SM) is supported by a substantial body of experimental evidence and can be concluded to describe and explain the majority of phenomena in particle physics, with a few rare exceptions.
These exceptions include experimental results such as the neutrino mass~\cite{Farzan:2017xzy,Proceedings:2019qno,Arguelles:2022tki}, the $W$ boson mass~\cite{CDF:2022hxs,deBlas:2022hdk}, the muon anomalous magnetic moment~($g-2$)~\cite{Muong-2:2021ojo,Muong-2:2023cdq,Aoyama:2020ynm}, and other issues remain unresolved~\cite{Crivellin:2023zui}.
Furthermore, the SM is unable to account for the existence of dark matter and the force of gravity. 
Consequently, there is a widely held belief in the existence of new physics~(NP) beyond the SM, and the search for NP and precision tests of the SM have become a prominent area of interest within the high-energy physics~(HEP) community~\cite{Ellis:2012zz}.
Given the success of the SM, it is reasonable to posit that any NP signals that have yet to be discovered will be exceedingly weak. 
Therefore, the search for NP and the precision tests of the SM motivate higher luminosity collider experiments. 
The processing of large amounts of data is inherent to such experiments, and thus places demands on our ability to process large amounts of data efficiently.

The application of machine learning~(ML) algorithms represents a promising avenue for the efficient processing of data.
Its applications have already been demonstrated in HEP~\cite{Innocente:1992gq,Albertsson:2018maf,Guest:2018yhq,Radovic:2018dip,Baldi:2014kfa,Ren:2017ymm,Abdughani:2018wrw,Ren:2019xhp,Letizia:2022xbe,DAgnolo:2019vbw,DAgnolo:2018cun,DeSimone:2018efk,Lee:2018xtt}.
Of particular interest in the study of NP phenomenology is the employment of the anomaly detection~(AD) algorithms, which have emerged as a prominent area of research in recent years~\cite{MdAli:2020yzb,Fol:2020tva,Kasieczka:2021xcg,CrispimRomao:2020ucc,vanBeekveld:2020txa,Kuusela:2011aa}.
Since ML methods often eliminate the need for detailed kinematic analysis, their benefits become increasingly evident as kinematics grow more complex. 
Consequently, the advantages will become increasingly apparent as further advances in the precision tests of the SM will inevitably involve the study of rare processes with larger numbers of particles in the final states.

In the meantime, quantum computing represents another effective approach for processing large volumes of data.
Despite quantum computing currently being in the noisy intermediate-scale quantum~(NISQ)~\cite{Chen:2021num,Arute:2019zxq,Preskill:2018jim} era, there has been a notable increase in research activity within the HEP community related to this field~\cite{Feynman:1981tf,Roggero:2018hrn,Roggero:2019myu,Bauer:2022hpo,Carena:2022kpg,Gustafson:2022xdt,Lamm:2024jnl,Carena:2024dzu,Atas:2021ext,Li:2023vwx,Cui:2019sfz,Zou:2021pvl,Georgescu:2013oza,Lamm:2019uyc,Li:2021kcs,Echevarria:2020wct,Perez-Salinas:2020nem,Jordan:2011ci,Mueller:2019qqj,Zhu:2024own}.
Many ML algorithms can be implemented by quantum computing~\cite{Biamonte:2016ugo,qml2,Garcia:2022cqq}.
In conjunction with the advancement of quantum ML algorithms, there has been a growing interest in the deployment of quantum ML in the HEP phenomenological studies, encompassing techniques such as variational quantum classifier~(VQC)~\cite{Guan:2020bdl,Wu:2020cye,Grant:2018oml,Terashi:2020wfi}, quantum support vector machine~(QSVM)~\cite{Guan:2020bdl,Wu:2021xsj,Zhang:2023ykh,Fadol:2022umw}, and quantum kernel k-means~(QKKM)~\cite{Zhang:2024ebl}.
In this study, we propose a variational quantum searching neighbor~(VQSN) algorithm for identifying NP signals.

To illustrate our algorithm, the SM effective field theory~(SMEFT)~\cite{Weinberg:1979sa,Grzadkowski:2010es,Willenbrock:2014bja,Masso:2014xra} are used as the test cases.
The SMEFT has recently been employed extensively in the search for NP.
When the collision energy is insufficient to excite NP particles, the NP particles can be integrated out, and the NP effects become new interactions of known particles. 
These interactions are formally represented as higher-dimensional operators with Wilson coefficients suppressed by powers of a NP scale, denoted as $\Lambda$. 
The signals of the operators that are most likely to be found correspond to those with Wilson coefficients least suppressed by $\Lambda$. 
Instead of dealing with various NP models, the number of operators to be considered at a specific order of $\Lambda$ is finite. 
Moreover, if an operator is not found, all NP models contributing to this operator can be constrained. 
Therefore, the use of the SMEFT is in line with the theme of high efficiency in this work.
The present study will investigate the utility of VQSN for the search for high operators at the large hadron collider~(LHC) and future muon colliders. 
VQSN demonstrates superior performance compared to a classical k-nearest neighbor~(KNN) algorithm, even when applied to classical datasets.

\section{\label{sec:2}Searching for neighbors}

\subsection{\label{sec:2.1}The quantum feature space}

\begin{figure}[htbp]
\begin{center}
\includegraphics[width=0.7\hsize]{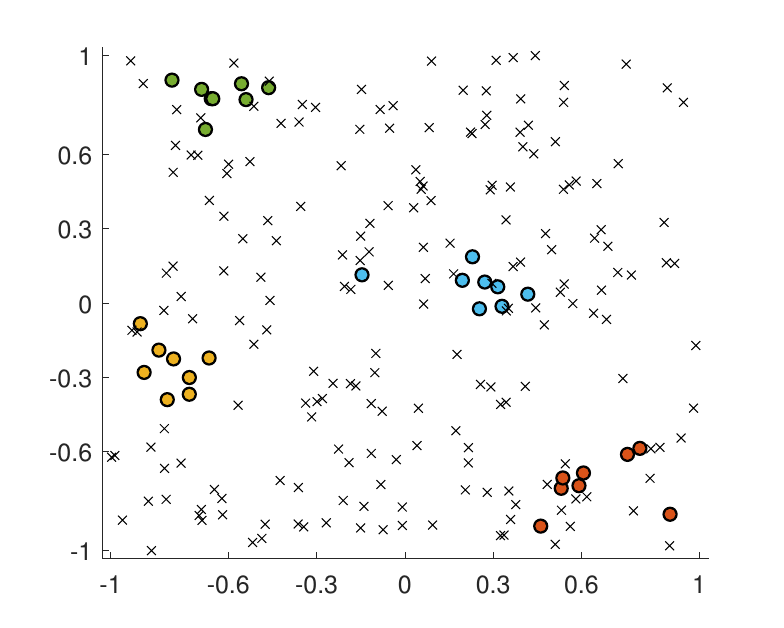}    
\caption{\label{fig:knn2dproblem}The problem addressed by (V)QSN. 
Given known classifications of some points in a feature space (e.g., $32$ colored points in a $2$-dimensional space as shown in the figure), how to classify an unknown point (marked as `$\times$' in the figure). 
(V)QSN classifies `$\times$' based on its distance to the known points, assigning it to the class belonging to the known points closer to it~(its neighbors).}
\end{center}
\end{figure}
(V)QSN is a classification method. 
In the next section, we will explain why a classification method is used to search for NP, but in this section, we will focus on the classification method itself. 
The problem (V)QSN addresses is, if we know some points and their classifications, how should unknown points be classified? 
Fig.~\ref{fig:knn2dproblem} provides an example. 
In Fig.~\ref{fig:knn2dproblem}, $32$ known points are classified into four categories (indicated by different colors). 
These $32$ points with known classification assignment constitute the training set and are generated using \verb"scikit-learn"~\cite{Pedregosa:2011ork}, and rescaled to range $-1<x_i<1$ and $-1<y_i<1$ where $x_i$ and $y_i$ are the coordinates of the points.
How should the unknown point (marked with an '$\times$') be classified?
V(QSN) classifies points based on the distance between known points and unknown points. 
If an unknown point is closer to the known points of a certain class, it will be classified into that category.

To use quantum computing for finding neighbors, the first step is to map the data points to quantum states. 
There are many schemes capable of establishing this one-to-one mapping. 
While these different schemes will impact the algorithm's performance, they do not affect its implementation. 
Taking Fig.~\ref{fig:knn2dproblem} as an example, a straightforward way is to map these two-dimensional points to the points on the Bloch sphere. 
For simplicity, $x$, $y$ coordinates are translated as zenith and azimuth angles of the Bloch sphere, i.e., for the point $\left(x_i,y_i\right)$ the corresponding quantum state is $|\phi _i\rangle = e^{i(y_i+1)\pi/2}\cos \left((x_i+1)\pi/4\right)|0\rangle + e^{-i(y_i+1)\pi/2}\sin\left((x_i+1)\pi/4\right)|1\rangle$.

\subsection{\label{sec:2.2}The circuit to search for neighbors}

\begin{figure}[htbp]
\begin{center}
\includegraphics[width=0.7\hsize]{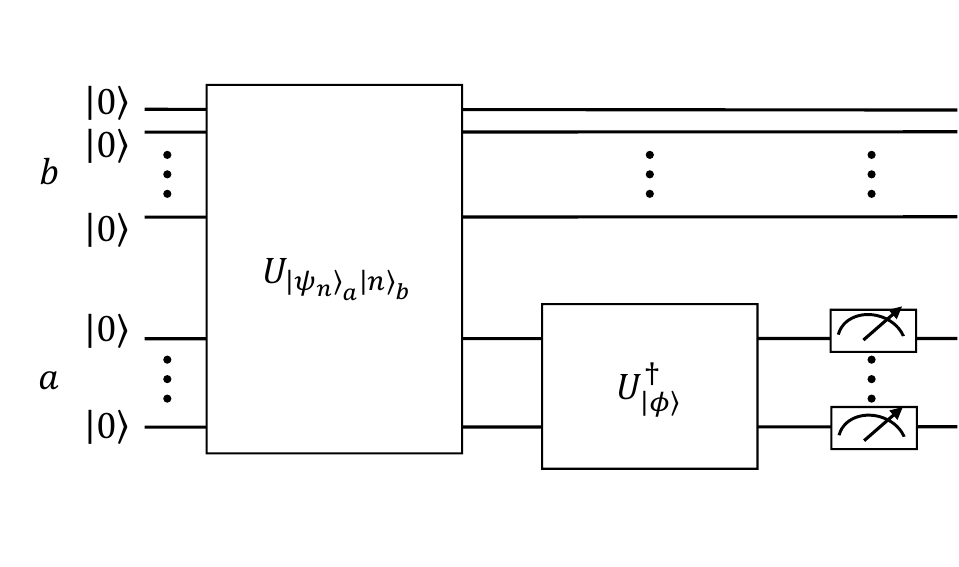}    
\caption{\label{fig:circuits-a}The circuit to construct the state $\langle\phi | \psi _n\rangle|n\rangle$.
$U_{|\psi _j\rangle  _{a}|j\rangle _{b}}$ and $U_{|\phi\rangle}$ are the circuits to encode the training set points $U_{|\psi _j\rangle  _{a}|j\rangle _{b}}|0\rangle_a|0\rangle_b=|\psi _j\rangle  _{a}|j\rangle _{b}$ and test point $U_{|\phi\rangle}|0\rangle=|\phi\rangle$, respectively.
If the measurement on $a$ register yields the $|0\ldots 0\rangle _a$ state, the result is $|0\ldots 0\rangle _a \langle \phi |\psi _j\rangle |j\rangle _b$.
}
\end{center}
\end{figure}
Since the concept of distance is used for classification, a crucial step is defining distance which is convenient to calculate in the context of quantum computing. 
After establishing a one-to-one correspondence between data points and quantum states, a distance measure suitable for quantum computing is the overlap between two states. 
For example, one can define $d_{ij}=\sqrt{1-\left|\langle \phi_i|\psi _j\rangle\right|^2}$, where $i,j$ denote two data points, $\phi _i$ and $\psi _j$ represent their corresponding quantum states. 
In this way, the task of finding neighbors for data point $i$ translates into finding those $\psi$ states which have a larger overlap with the state $\phi _i$.

The overlap between two states can be calculated using the control-swap test. 
However, as will be shown, one can leverage the superposition property of quantum states to directly compute the overlaps of multiple $\psi_j$ states with a state $\phi$, storing the results as amplitude coefficients. 
In this way, when measuring the quantum state where the probability of obtaining a specific outcome is proportional to the square of the absolute value of its amplitude coefficient, states with larger overlaps are naturally measured with higher probability. 
This allows us to identify the neighbors of $\phi$.

As mentioned above, the problem becomes: given one $\phi$ and multiple $\psi _j$, how does one obtain the state $\langle \phi |\psi _j\rangle |j\rangle$.
Assume $U_{|\phi\rangle}$ is the circuit that creates the quantum state $\phi$, i.e., $U_{|\phi\rangle}|0\rangle=|\phi\rangle$. 
For ease of description, we divide the registers into $a$ and $b$, where $a$ corresponds to the qubits storing $\psi_j$ and $b$ corresponds to the indices. 
For the state $|\psi _j\rangle  _{a}|j\rangle _{b}$, if we apply the inverse of $U_{|\phi\rangle}$ to register $a$ and measure the $a$ register yields the $|0\ldots 0\rangle _a$ state, then the state becomes $|0\ldots 0\rangle _a \langle \phi |\psi _j\rangle |j\rangle _b$. 
Note that the state corresponding to register $b$ at this point is precisely what we need.
Denoting $U_{|\psi _j\rangle  _{a}|j\rangle _{b}}$ as the circuit to create the state $|\psi _j\rangle  _{a}|j\rangle _{b}$, i.e., $U_{|\psi _j\rangle  _{a}|j\rangle _{b}}|0\rangle_a|0\rangle_b=|\psi _j\rangle  _{a}|j\rangle _{b}$, the circuit is shown in Fig.~\ref{fig:circuits-a}.
In this work, we use the uniform rotation gates~\cite{Mottonen:2004vly} to create the state $|\psi _j\rangle  _{a}|j\rangle _{b}$

The success rate~(denoted as $r$) to create the state $\langle \phi |\psi _j\rangle |j\rangle$~(the probabilities that $|0\ldots 0\rangle _a$ is measured) is $r=\sum \left|\langle\phi | \psi _n\rangle\right|^2/N$, where $N$ is the number of $\psi _n$ vectors.
Without assumptions on the distribution of the vectors, there is no reason to prefer any result of the measurement on $a$ register.
Therefore, if the vectors are $d$-dimensional complex vectors, $r\approx 1/d$, which is verified by using complex uniform random numbers.
The success rate can be improved to $r\approx 1/\sqrt{d}$ by amplitude amplification, or the `QSearch' algorithm ~\cite{Brassard:2000xvp}.

\subsection{\label{sec:2.3}Fitting the training set}

\begin{figure}[htbp]
\begin{center}
\includegraphics[width=0.7\hsize]{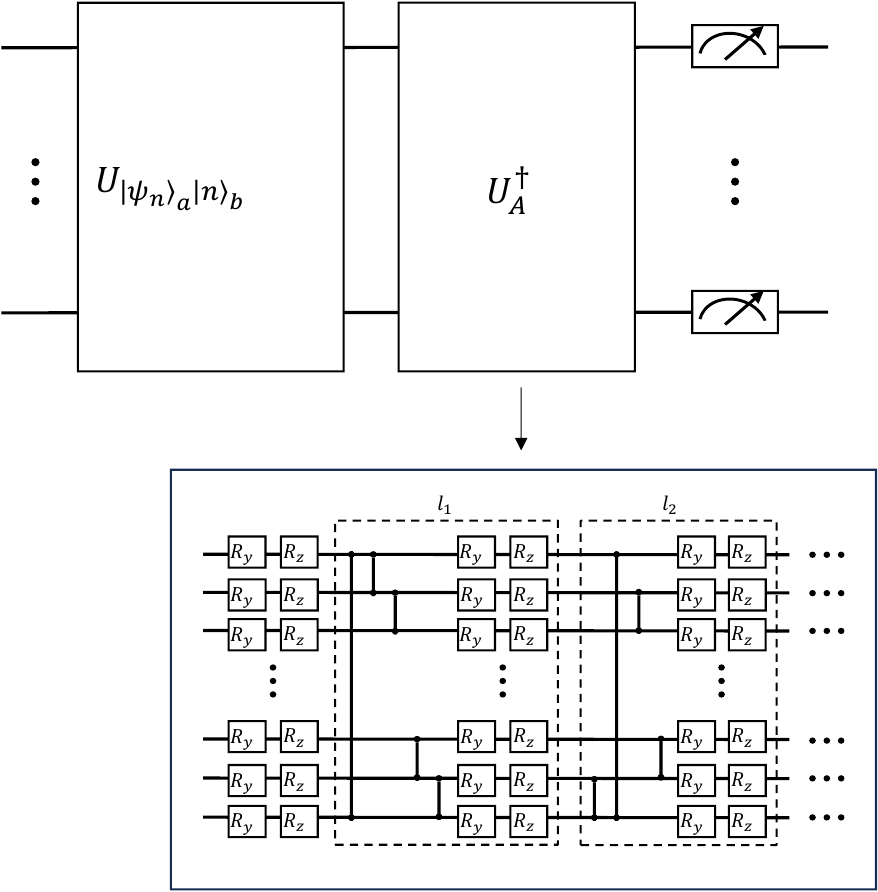}    
\caption{\label{fig:circuits-b}The circuit to fit the ansatz.
$U_A$ denotes the ansatz circuit with trainable parameters $\alpha _i$.
The goal is to adjust $\alpha _i$ to maximize the probability where $|0\ldots 0\rangle$ is measured.
An example of $U_A$ with the `SCA' layer is shown.}
\end{center}
\end{figure}
The complexity of finding neighbor algorithm is mainly the query complexity.
In this work, $|\psi _n\rangle |n\rangle$ represent events that would be generated by a Monte Carlo~(MC) simulation.
Since at present, there is no quantum MC algorithm to generate events, one has to deal with a classical dataset.
An improvement is to approximate the state $|\psi _n\rangle |n\rangle$ with a parameterized ansatz circuit.

Denoting the circuit of the ansatz as $U_A(\alpha_i)$ where $\alpha_i$ are trainable parameters, the circuit of fitting an ansatz is shown in Fig.~\ref{fig:circuits-b}, where the probability of the outcome $|0\ldots 0\rangle$ in the measurements is $\left|\langle \Psi(\alpha_i)|\psi _j\rangle  _{a}|j\rangle _{b}\right|^2$ with $U_A(\alpha_i)|0\rangle = |\Psi(\alpha_i)\rangle$.
The training process adjusts the parameters $\alpha _i$ to maximize the probability of obtaining $|0\ldots 0\rangle$ in the measurements.

In this work, we use the `hardware efficient' ansatz~\cite{Kandala:2017vok,Bravo-Prieto:2019kld} with entanglement layers, 
\begin{equation}
\begin{split}
U_A(\alpha _i) &= \prod _{i}^l\left(\prod _q e^{{\rm i}\rho^{(q)} _i \sigma^{(q)} _z}\prod _q e^{{\rm i}\beta^{(q)} _i \sigma^{(q)} _y}\prod _{\langle q_1,q_2\rangle}c_z^{q_1,q_2}\right)\\
&\times \prod _q e^{{\rm i}\rho^{(q)} _0 \sigma^{(q)} _z}\prod _q e^{{\rm i}\beta^{(q)} _0 \sigma^{(q)} _y},
\end{split}
\end{equation}
where $\sigma ^{q}_{y,z}$ are Pauli matrices acting on $q$ qubit, the product over $q$ is the product over all qubits, $c_z^{q_1,q_2}$ is the controlled-Z gate acting on $q_{1,2}$ qubits with $c_z=diag\left(1,1,1,-1\right)$, and $\langle q_1,q_2\rangle$ is the set of pairs of qubits depending on the type of entanglement layer, $l$ is the number of layers of the ansatz, $\{\alpha _i\} = \{\beta^{(q)} _0, \rho^{(q)} _0, \beta ^{(q)}_1, \rho^{(q)} _1, \ldots \beta^{(q)} _l, \rho^{(q)} _l\}$ is a set of trainable parameters.
The number of parameters is $2(l+1)n_q$ where $n_q$ is the number of qubits.
Two different entanglement layers are used in this work, which are the `shifted-circular-alternating'~(SCA) layer~\cite{Sim:2019yyv} and the `pairwise' layer. 
An example of $U_A$ with the `SCA' layer is shown in Fig.~\ref{fig:circuits-b}.
The `pairwise' layer is shown in Sec.~\ref{sec:2.5}.

The relationship between training cost and training set is not straightforward to determine, as it depends on multiple factors: the expressive power of the ansatz, the gradient descent algorithm used during training, the distribution of the training dataset, and the method of encoding classical data.
In the paper, the ansatz, gradient descent algorithm, and classical data encoding scheme were chosen naively, each of which admits further optimization, a direction we defer to future work. 
A key point is that, as with most machine learning algorithms, the training dataset plays a critical role. 
However, the distribution of the training data depends on the case to be studied, making it difficult to predict a priori whether or to what extent additional training cost would ensure convergence.
One can, however, roughly estimate its scalability. 
Since the information content of the classical data is proportional to $Nd$, the cost of preparing the training dataset is at least $\mathcal{O}(Nd)$, regardless of the encoding method. 
Accordingly, the training cost is roughly on the order of the number of convergence iterations multiplied by $Nd$.  

\subsection{\label{sec:2.4}Variation quantum searching neighbor algorithm}

\begin{figure}[htbp]
\begin{center}
\includegraphics[width=0.7\hsize]{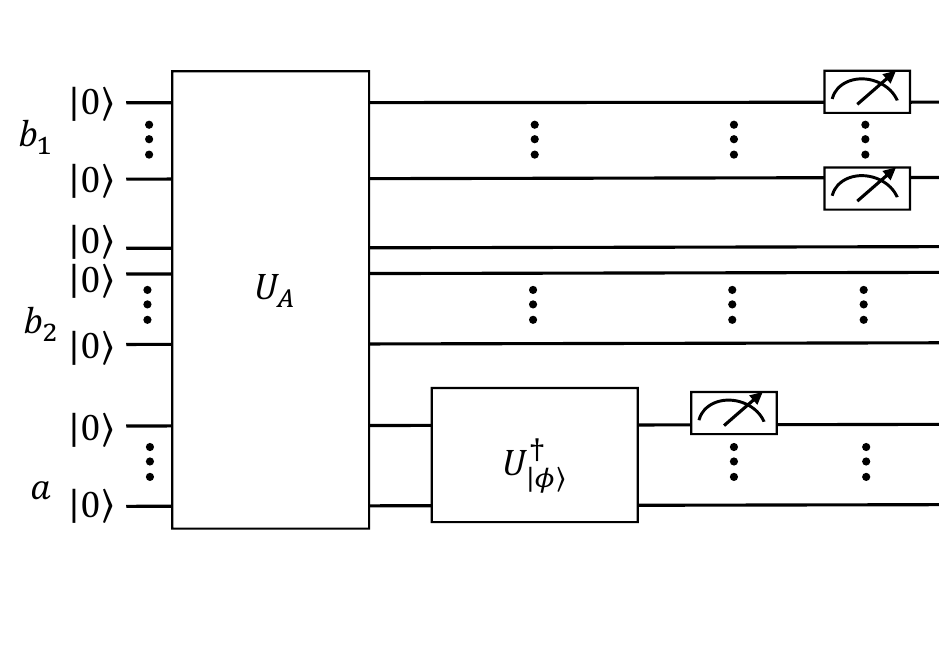}    
\caption{\label{fig:circuits-c}The circuit to find the classification assignment of $|\phi\rangle$.
If the measurement on the $a$ register yields $|0\ldots 0\rangle$, then $b_1$ is measured.
If the quantum states corresponding to the points classified as $k$ in the training set have larger overlaps with $|\phi\rangle$, the measurement on $b_1$ yields outcome $|k\rangle _{b_1}$ with higher probability.}
\end{center}
\end{figure}

Since our goal is to classify, in the following, we assume that each $\psi$ vector carries a classification label and that the states are stored accordingly.
Thus, the state to be created is $|\psi _{k,n}\rangle _a|k\rangle _{b_1}|n\rangle _{b_2}$, where $k$ indicates the class, the $b$ register is split to $b_1$ and $b_2$ for clarity.
With this ordering, it becomes unnecessary to measure the entire $b$ register to find neighbors of $\phi$. 
Instead, only measuring $b_1$ reveals which class $\phi$'s neighbors belong to.

To summarize, the steps of the VQSN are listed as follows,
\begin{enumerate}
\item Arrange the training dataset as $|\psi _{k,n}\rangle _a|k\rangle _{b_1}|n\rangle _{b_2}$, where $k$ is the class of $\psi _{k,n}$, $n$ is the index of $\psi _{k,n}$ in class $k$.
\item Use an ansatz circuit $|\Psi(\alpha _i)\rangle$ to fit $|\psi _{k,n}\rangle _a|k\rangle _{b_1}|n\rangle _{b_2}$. 
\item For a test vector $\phi$, apply the inverse of amplitude encoding for $|\phi\rangle _a$ after the ansatz circuit, then measure register $a$. If $|0\rangle _a$ is measured, the state is successfully created. Apply amplitude amplification if necessary.
\item Measure the $b_1$ register to find out the classification. Similarly to the KNN, we use a majority voting strategy, whereby we repeat $c$ times and designate the classification of $\phi$ as the majority result.
\end{enumerate}

In the case of QSN the step-2 is omitted and the exact $\langle \phi | \psi _{k,n}\rangle |k\rangle_{b_1} |n\rangle _{b_2}$ is created. 
The circuit of VQSN is shown in Fig.~\ref{fig:circuits-c}.

In essence, the VQSN is a variant of the VQC algorithms. 
As a variant of VQC, given ansatze of comparable complexity their computational complexities are of the same order of magnitude.
However, VQSN possesses two advantages. 
Firstly, conventional VQC can be seen as the quantum analogue of an artificial neural network, which is a `black box' difficult to interpret.
The reason is that VQC trains a parameterized quantum circuit to achieve the mapping $f(\phi_{sig})=1$ and $f(\phi_{bg})=0$, where $\phi_{sig}$ and $\phi_{bg}$ refer to the signal and background, respectively. 
This parameterized quantum circuit is constructed as a layered, interconnected ansatz, structurally resembling a neural network.
The VQSN also trains a parameterized quantum circuit to realize the same discriminative effect.
In contrast, the state to build in VQSN has a discernible geometric interpretation, and the target of the training is guaranteed to exist.
Secondly, QSN is accompanied by a classical algorithm, KNN, which allows for a comparison between quantum and classical algorithms.
This is clearer for a binary classification problem, in which QSN can be viewed as using $\sum w_n t_n/N$ to determine the classification of $\phi$ where $w_n = \left|\langle \phi | \psi _n\rangle \right|^2$, and $t_n=0,1$ is the class of $\psi _n$.
Therefore, it uses a weighted average of the entire training dataset.
KNN can also be viewed as a weighted average except that $\{w _n = 1 | \psi _n \in S\}$ and $\{w _n = 0 | \psi _n \notin S\}$, where $S$ is the set contains the $k$ nearest $\psi _n$ vectors to $\phi$.

Although our approach resembles KNN, it is fundamentally distinct. 
Unlike Refs.~\cite{Lloyd:2013lby,Wiebe:2014imz}, we define distance through quantum overlap~(same as Ref.~\cite{basheer2021quantumknearestneighborsalgorithm}). 
This enables efficient identification of classification assignments via measurement. 
While QSN could extend to quantum KNN using the maxima finding algorithm from Ref.~\cite{basheer2021quantumknearestneighborsalgorithm}, VQSN deliberately eliminates this step to reduce quantum gate count and circuit depth. 
Furthermore, replacing data queries with an ansatz circuit significantly decreases gate requirements and depth, enabling execution of VQSN on a current NISQ hardware.

\subsection{\label{sec:2.5}The numerical result}

Before moving on to the search for NP, a numerical example is implemented to illustrate the algorithm.

\begin{figure*}[htbp]
\begin{center}
\includegraphics[width=0.95\hsize]{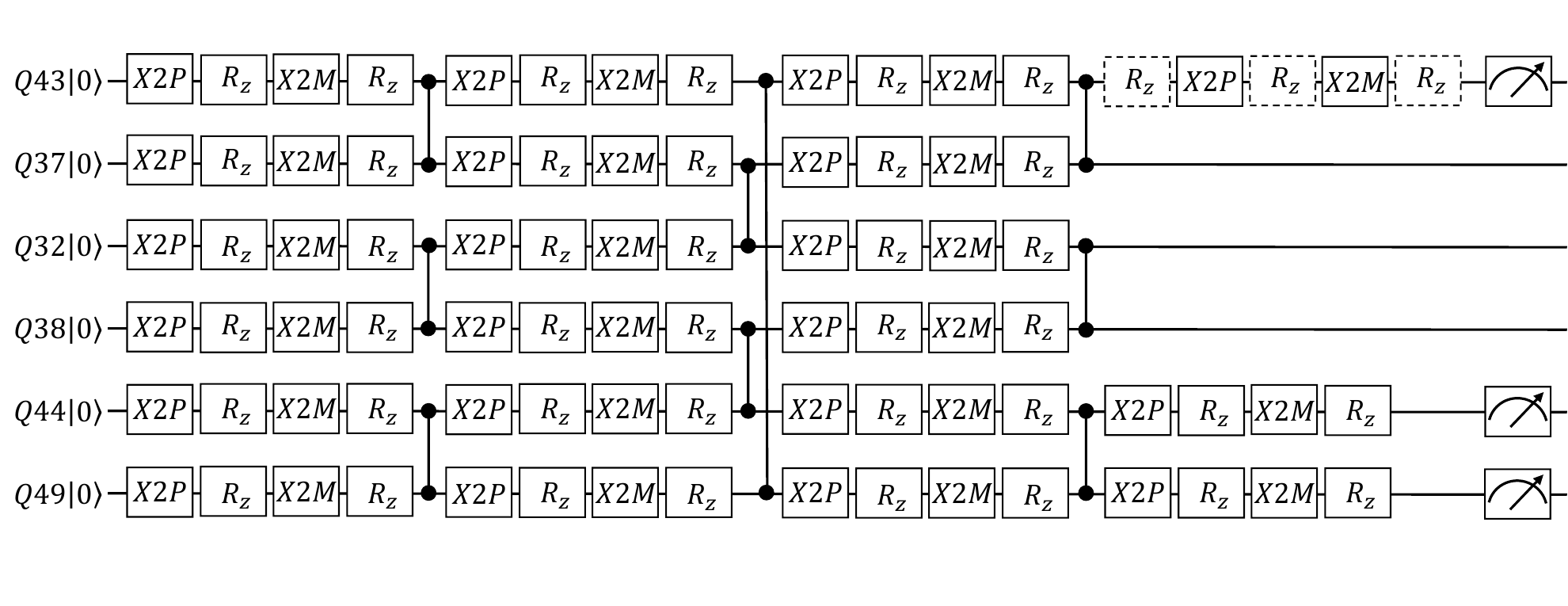}    
\caption{\label{fig:qcis1} The circuit running on \textit{tianyan176-2} to classify a test point. The $X2M$ is $R_x(-\pi/2)$ gate, $X2P$ is $R_x(\pi/2)$ gate. The parameters of the gates depicted using dashed edged boxes are different for different test points. The parameters of the gates depicted using solid edged boxes are the same because they are used to build the ansatz.
The $Q43$ is $a$ register, $Q44$ and $Q49$ make up the $b_1$ register, and the others make up the $b_2$ register.}
\end{center}
\end{figure*}

\begin{figure*}[htbp]
\begin{center}
\includegraphics[width=0.48\hsize]{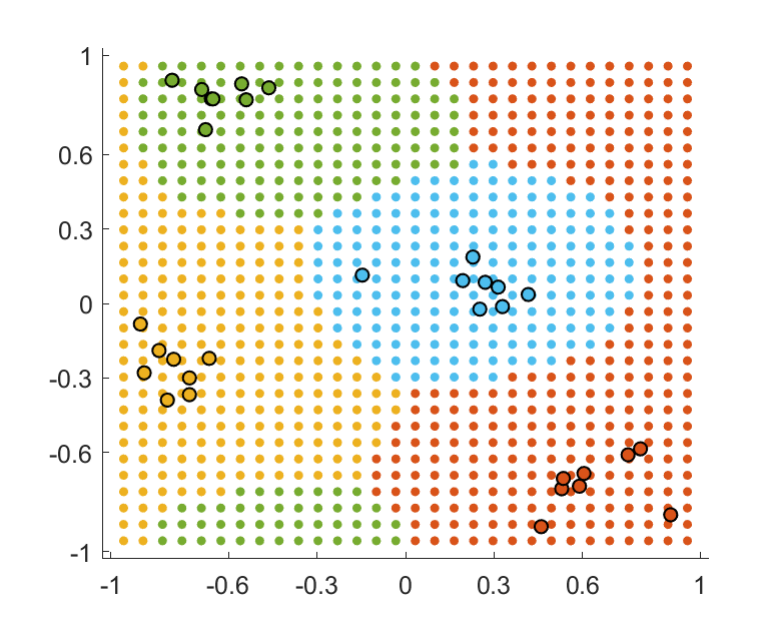}    
\includegraphics[width=0.48\hsize]{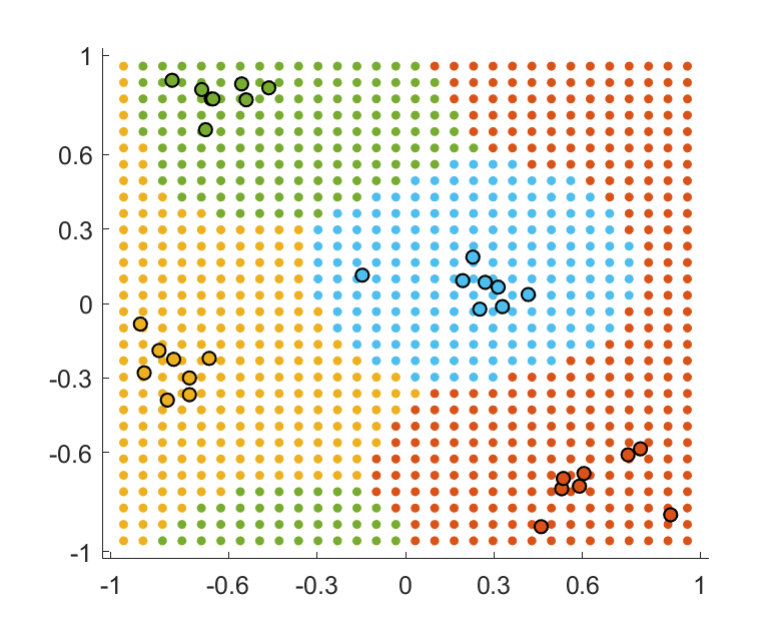}\\
\includegraphics[width=0.48\hsize]{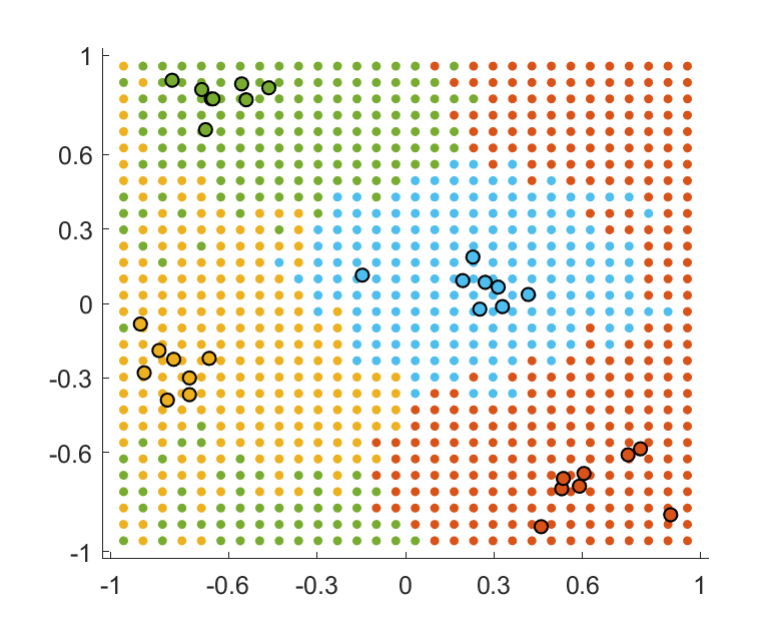}    
\includegraphics[width=0.48\hsize]{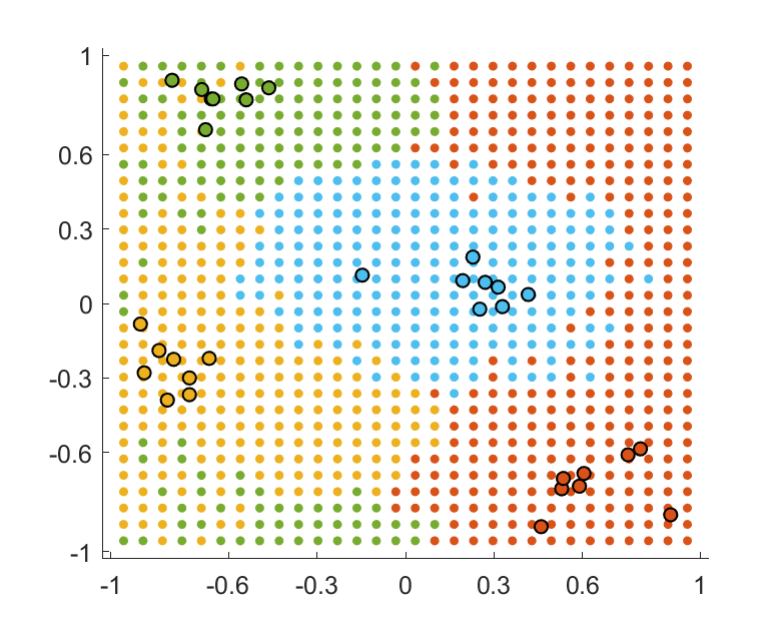}
\caption{\label{fig:knn2d}The example of classifier with two dimensional points. 
The top left panel is the result of QSN with $c\to \infty$~(with direct comparison of probabilities of each amplitudes). 
The top-right panel is the result of VQSN with $l=3$ and $c\to \infty$. 
The bottom-left panel is the result of VQSN with $l=3$ and $c=1000$. 
The bottom-right is the result of VQSN on \textit{tianyan176-2} with $l=3$ and $c=1000$.}
\end{center}
\end{figure*}
In the numerical example, we use $N=32$ two dimensional points which are divided into four classes equally.
The training set is generated using \verb"scikit-learn"~\cite{Pedregosa:2011ork}, and $x$, $y$ coordinates are translated as zenith and azimuth angles of the Bloch sphere, $\theta = (x+1)\pi/2$, $\varphi = (y+1)\pi$.
When training, we use `Adam' optimizer~\cite{Kingma:2014vow}, and the initialization of the ansatz parameters is randomly sampled from the many-body localized distribution~\cite{Park:2024rim}, the training is done by using the \verb"QuEST" toolkit~\cite{Jones:2019knd}.
When $l=3$, the ansatz is trained to a fidelity of $99.4\%$.
In terms of quantum control instruction set~(QCIS), the circuit for classification is depicted in Fig.~\ref{fig:qcis1}.
Then we scan the Bloch sphere and use QSN and VQSN to determine the classification assignment of each point on the Bloch sphere.
The discrimination circuit is also run on \textit{tianyan176-2} of \textit{Tianyan quantum computing cloud platform}, with \textit{Zuchongzhi} superconducting quantum computing system~\cite{Zhu:2021gkn}.

The training set and the results with different methods are shown in Fig.~\ref{fig:knn2d}.
The difference between the QSN and VQSN with $c\to \infty$ indicates the effect of approximation of $|\Phi\rangle$ with an ansatz.
The difference between the VQSN with $c\to \infty$ and $c=1000$ indicates the effect of randomness in measurements.
The difference between the results obtaiend by \verb"QuEST" and \textit{tianyan176-2} indicates the effect of noise on a quantum computer.
It can be shown that (V)QSN works as expected, the VQSN demonstrates resilience to noise, and the primary error originates from the randomness in measurements.

On the \textit{tianyan176-2} quantum processor, the average two-qubit gate error is $1.6\%$, the average single-qubit gate error is $0.28\%$, and the average readout error is $3.4\%$. 
The median decoherence times are $T_1 = 22.89 \mu s$ and $T_2 = 2.94 \mu s$. 
If we consider only quantum state fidelity and neglecting single-qubit gate errors, the probability of obtaining a correct result can be estimated as $(1 - 0.016)^L \times (1 - 0.034)^{n_m}$, where $L = 3$ is the circuit depth in the sense of two-qubit gates, and $n_m$ is the number of qubits measured. 
This yields an estimated accuracy of $85.9\%$.

However, as observed in Fig.~\ref{fig:knn2d}, in the absence of quantum noise and with a finite number of measurement shots $c$, mis-tagged points occur near decision boundaries. 
When real quantum noise is introduced, no significant discrepancies attributable to quantum errors were detected. 
To evaluate the impact of noise, we treat the result with $l = 3$ and $c\to \infty$ as the ground truth. 
The mis-tag rate is approximately $11\%$ for the quantum simulator (due solely to finite measurements) and about $15\%$ on the real quantum device. 
This indicates that the majority of mis-tags are due to finite sampling rather than quantum noises. 

The noise tolerance of the measurement outcomes depends on both the distribution of the training dataset and the specific test point in question. 
For instance, if the training set contains only two points, a signal event $\psi_{sig}$ and a background event $\psi_{bg}$, then a more precise measurement is required when $\langle \phi|\psi_{sig}\rangle$ is close to $\langle \phi|\psi_{bg}\rangle$ (i.e., when the test point $\phi$ lies near the decision boundary). 
In contrast, when the two inner products differ significantly, the measurement outcome is robust to the noise.

Kullback-Leibler~(KL) divergence can be used to uncover errors that are masked by this robustness. 
Denoting $P_i$ as the probability of outcome class $i$, KL divergence is defined as $KL = H(P^{truth}, P^{test}) - H(P^{truth}, P^{truth})$, where $H$ denotes cross entropy, $H(P^1, P^2)=-\sum _i P_i^1 \log(P_i^2)$.
The average KL divergence over all test points is $0.003$ for the simulator and $0.026$ for the real quantum computer. 
Unlike the mis-tag rate, the KL divergence significantly increases for the real quantum computer.
That means, although noise alters the probabilities of outcomes, these changes are generally insufficient to alter the most probable class assignment, which is the reason of the robustness of the method.  

\section{Applications in the search for new physics}

The searching for NP signals can be considered as a binary classification, therefore the (V)QSN is expected to be feasible in the phenomenological study of NP.
In this section, three applications to use (V)QSN to search for NP and set expected constraints on the NP parameters are present.

\subsection{Preparation of the datasets}

The datasets are generated using \verb"MadGraph5" toolkit\cite{Alwall:2014hca,Christensen:2008py,Degrande:2011ua}. 
A parton shower is applied using \verb"Pythia8" with NNPDF~\cite{Sjostrand:2014zea,Ball:2013hta}, the fast detector simulation is applied using \verb"Delphes"~\cite{deFavereau:2013fsa} with either the CMS or muon collider card, depending on the processes under study.
The preparation of the datasets are applied using \verb"MLAnalysis"~\cite{Guo:2023nfu}.

After the collision events are generated, a set of observables are picked to form the feature space, an event can be mapped into this feature space as a vector $\vec{v}$.
Assuming $D$ observables are picked, then $\vec{v}$ is a $D$-dimensional vector.
The z-score standardization~\cite{Donoho_2004} is applied, $x^j_m=\left(v^j_m-\bar{v^j}\right)/\epsilon ^j$, where $v^j_m$ is the $j$-th component~($0\leq j \leq D-1$) of $\vec{v}_m$ for the $m$-th collision event, $\bar{v^j}$ and $\epsilon ^j$ are mean values and standard deviations of the $j$-th component of $\vec{v}$ over the training dataset.
$\vec{x}$ is $\vec{v}$ after z-score standardization.

In the search for NP, the data to be dealt with are the collision events.
The idea of this work is to decide whether some unknown events are neighbors of the SM events, or the NP events.
In the case of HEP, there are different methods to define the distance between two events, which can have an impact on the results of event selection strategies~\cite{Komiske:2019fks}.
In (V)QSN, we use the overlap of two quantum states to calculate the distance, therefore the definition of the distance depends on the method to encode vector $\vec{x}$ into a Hilbert space.
There exist various methods for mapping vectors into a Hilbert space. 
Among these, the most straightforward approach is amplitude encoding.
While different schemes may impact the efficiency of event selection, we employ the most straightforward amplitude encoding method for brevity and to maintain focus on (V)QSN, leaving the optimization of the encoding scheme for future work.
For example, this study has not addressed improvements such as encoding the vectors in the dataset into a quantum state using a nonlinear approach~\cite{Havlicek:2018nqz}. 
It is also possible to utilize the vast Hilbert space of quantum computers to separate the signal from the background more when mapping to the quantum feature space~\cite{Lloyd:2020eeh}.
The vector $\vec{x}$ is mapped to the quantum state,
\begin{equation}
\begin{split}
&|\psi _m\rangle =\frac{1}{\sqrt{1+\sum _{j=0}^{d'-1} (x^j_m)^2}}\left(\sum _{j=0}^{d-1}\left(x^{2j}_m+x^{2j+1}_m{\rm i}\right) |j\rangle\right),
\end{split}
\label{eq.encoding}
\end{equation}
where $d$ is the least power of $2$ larger than $d'/2$, for $x^j_m$ with $j>D-1$, $x^{j=d'}_m=1$ and $x^{j>d'}_m=0$ are used.
The amplitude encoding is also implemented using the uniform rotation gates.

\subsection{Event selection strategy}

Both (V)QSN and (t-)KNN fundamentally remain traditional cut-and-count approaches. 
Whereas conventional event selection strategies apply cuts directly to observables in collision events, (V)QSN and (t-)KNN instead target anomaly scores $\hat{a}$ and $\hat{b}$. 
Essentially, $\hat{a}$ and $\hat{b}$ are themselves observables generated from final-state observable information.
The efficacy of these methods hinges critically on using events with SM kinematic features and events exhibiting NP kinematic features (i.e., the events from NP squared matrix element) during training.

During the event selection stage, interference terms must be taken into account, the search for NP is no longer a classification problem. 
Similar as traditional cut-and-count schemes, the differential cross-section can now be regarded as a function of the anomaly scores $\hat{a}$ or $\hat{b}$. 
Selecting events is equivalent to adjusting the integration limits of the phase space integral. 
Therefore, at their core, (V)QSN and (t-)KNN utilize kinematic information from final-state observables to suppress the cross-section of events `appear SM-like' while retaining cross-section that 'appear non-SM-like', thereby enhancing signal significance.
In other words, we use the difference between the total cross-sections with and without NP after cut to `discover' NP or set constraints on the parameters of NP.

Because only two classes are involved, we change the last step of (V)QSN in Sec.~\ref{sec:2.3}.
Denoting $\psi _{n,{\rm SM}}$ and $\psi _{n,{\rm NP}}$ as states for the SM and NP events, respectively, the state to be trained is arranged as,
\begin{equation}
\begin{split}
&\Psi_{\rm train}=\sum _n^{\frac{N}{2}}\left(|\psi_{n,{\rm NP}}\rangle _a|0\rangle _{b_1} |n\rangle _{b_2} + |\psi_{n,{\rm SM}}\rangle _a|1\rangle _{b_1}|n\rangle _{b_2}\right).
\end{split}
\label{eq.train}
\end{equation}
The number of $|0\rangle _{b_1}$ outcome is denoted as $a$ and $\hat{a}=a/(rc)$ is used as a score to indicate the probability such how likely $\phi$ is from the NP, where $r\times c$ is the number of times $\langle \phi |\Psi\rangle$ is successfully built.
And after $\hat{a}$ is obtained, we select events with $\hat{a}>\hat{a}_{th}$ where $\hat{a}_{th}$ is a tunable threshold that maximizes the signal significance which is defined as~\cite{ParticleDataGroup:2020ssz,Cowan:2010js},
\begin{equation}
\begin{split}
&\mathcal{S}_{stat}=\sqrt{2 \left[(N_{\rm bg}+N_{s}) \ln (1+N_{s}/N_{\rm bg})-N_{s}\right]},
\end{split}
\label{eq.ss}
\end{equation}
where $N_{bg}$ and $N_s$ are numbers of background and signal events, respectively, and $N_{s,bg}=L\sigma_{s,bg}$, where $L$ is the luminosity, $\sigma _{s,bg}$ are the cross-sections of signal and background contributions.
The accuracy of $\hat{a}$ can be estimated as $\mathcal{O}(1/\sqrt{rc})$.

For comparison with a classical algorithm, we also introduce the classical KNN algorithm which is already used in the search for NP~\cite{DeSimone:2018efk}.
We use a modified KNN~(denoted as `t-KNN' where `t' stands for tunable) with the following strategy for event selection.
\begin{enumerate}
\item For a data point in the test dataset, find the $k$ closest data points to this point in the training set. 
\item Among these $k$ points, find out how many of them are from the NP, assuming the number of those points is $b$, use $\hat{b} =b/k$ as the anomaly score for the point being tested.
\item Determine a tunable threshold $\hat{b}_{th}$ according to signal significance, and select the events with $\hat{b}>\hat{b}_{th}$.
\end{enumerate}
When $\hat{b}_{th}=0.5$, this is the conventional KNN algorithm.
When using the (t-)KNN, the training dataset is the same one as the (V)QSN but consists of $D$-dimensional real vectors $\vec{x}$. 
However, when finding the $k$ closest data points, we use the Euclidean distance. 

For the (t-)KNN algorithm, we only incorporate the cost needed to compute the distances, so the computational complexity~(cc) can be estimated as $\mathcal{O}(d'NM)$ for $M$ test events.
In the context of quantum computers, the number of quantum gates is typically employed as a means of assessing the cc.
Even if the dataset is classical, the cc for VQSN is $n_t+\mathcal{O}(c'M(n_{ans}+d)\sqrt{d})$, where $n_t$ is the cost to train the ansatz which is a constant in the view of test dataset, $n_{ans}$ is the cc of the ansatz, the cc of $U_{|\phi\rangle}^{\dagger}$ is $\mathcal{O}(d)$, and $c=c'/\sqrt{d}$ where $1/\sqrt{d}$ comes from the success rate to create the state $\langle \phi|\Psi (\alpha _i)\rangle$ when amplitude amplification is used and the number of times of measurement on $a$ is required to be a fixed number $c'$.
Note that, $n_t$ depends on $N$ and $d$, but not on $M$. 
Unlike $M$ which grows with the luminosity, $N$ and $d$ remain effectively constant in practical applications, as a result, for practical applications it can be expected that $M\gg N$. 
In other words, $n_t$ is independent of the luminosity of future colliders.
The ansatz can be viewed as a compressed representation of the training set. 
Its complexity should be lower than that of the full dataset; otherwise, using an ansatz would offer no advantage. 
Thus, for each selection operation, the ansatz provides computational savings compared to ansatz-free methods, until the initial training cost is amortized. 
As $M$ grows (i.e., as the luminosity of future colliders increases), the advantage of VQSN over (t-)kNN becomes more pronounced.
By omitting $n_t$ because $M\gg N$, and by the fact that $d$ is at the same order of $D$, it can be estimated that the VQSN outperforms (t-)KNN even with a classical dataset when $c(n_{ans}+d)<N\sqrt{d}$.
From the principle of the algorithm, the larger $N$ and $d$, the better the classification, as long as $M\gg N$ is kept, this means that ideally VQSN can outperform (t-)KNN in the sense of cc even dealing with a classical dataset.

\subsection{Search for the gluon quartic gauge couplings at a muon collider}

\subsubsection{The contribution of the gluon quartic gauge coupling at a muon collider}

\begin{figure*}[htbp]
\begin{center}
\includegraphics[width=0.8\hsize]{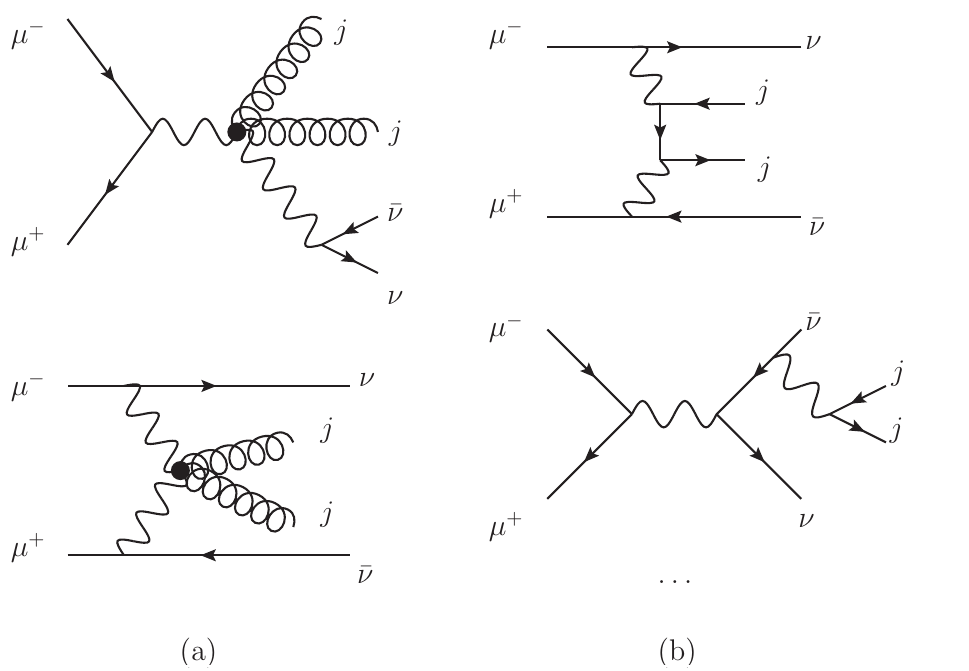}
\caption{\label{fig:feyndiagmuongqgc} Feynman diagrams of the process $\mu^+\mu^-\to jj\nu\bar{\nu}$~(diagrams in (a) represent the contribution of $O_{gT,0}$, while diagrams in (b) represent the SM contribution).}
\end{center}
\end{figure*}
The gluon quartic gauge couplings~(gQGCs) emerge from the Born-Infeld~(BI) extension of the SM, which was originally driven by the idea that there should be an upper limit on the strength of the electromagnetic field~\cite{Born:1934gh}.
It has been demonstrated that the BI model is also associated with M-theory-inspired models~\cite{Ellis:2018cos,Ellis:2021dfa,Fradkin:1985qd,Tseytlin:1999dj,Cheung:2018oki}.
In the SMEFT, the operators contributing to gQGCs appear at dimension-8.
As the importance of the dimension-8 operators in the SMEFT has been realized~\cite{Green:2016trm,Zhang:2020jyn,Murphy:2020rsh,Li:2020gnx,Anders:2018oin,Henning:2015alf}, more and more phenomenological studies have been devoted to the dimension-8 operators in recent years~\cite{Zhang:2018shp,Bi:2019phv,ATLAS:2017vqm,CMS:2017rin,CMS:2020ioi,CMS:2016gct,CMS:2017zmo,CMS:2018ccg,ATLAS:2018mxa,CMS:2019uys,CMS:2016rtz,CMS:2017fhs,Ellis:2023ucy,Spor:2022hhn,Spor:2022zob,Yilmaz:2021ule,Ellis:2020ljj,Senol:2019swu,Yilmaz:2019cue,Ellis:2019zex,Jahedi:2022duc,Jahedi:2023myu}.
The case of gQGC is among the operators of dimension-8 in the SMEFT that have received attention. 
We focus on the $O_{gT,0}$ operator,
\begin{equation}
\begin{aligned}
O_{gT,0}=\frac{1}{16M_0^4}\sum_a G_{\mu\nu}^a G^{a,\mu\nu}\times\sum_i W_{\alpha\beta}^i W^{i,\alpha\beta},\\
\end{aligned}
\end{equation}
where ${G_{\mu\nu}^a}$ is gluon field strengths, ${W_{\mu\nu}^i}$ denote electroweak field strengths, and $M_0$ is the NP scale.
The expected constraint is $M_0\geq 1040\;{\rm MeV}$ at the LHC with the center of mass~(c.m.) energy $\sqrt{s}=13\;{\rm TeV}$, and luminosity $L=36.7\;{\rm fb}^{-1}$ obtained by using the process $gg\to \gamma\gamma$~\cite{Ellis:2018cos}.
The combined sensitivities of the $Z\gamma$ and $\gamma\gamma$ channels at the LHC with $\sqrt{s}=13\;{\rm TeV}$ and $L=137\;{\rm fb}^{-1}$ are about three times of the above~\cite{Ellis:2021dfa}.
For convenience, we define $f_0\equiv 1/(16M_0^4)$.

At muon colliders, the gQGCs contribute to the process $\mu^+\mu^- \to \nu\bar{\nu}jj$~\cite{Yang:2023gos}.
The Feynman diagrams are shown in Fig.~\ref{fig:feyndiagmuongqgc}.
This process is chosen as one of the processes to validate the VQSN since it is convenient to handle due to the absence of interference between the SM and NP.

\subsubsection{Training of the ansatz}

\begin{table}[hbtp]
\centering
\begin{tabular}{c|c|c|c|c|c|c}
\hline
           & $v^0$ & $v^1$ & $v^2$ & $v^3$ & $v^4$ & $v^5$ \\
\hline           
observables & $E_{j_1}$ & $p_{j_1}^T$ & $E_{j_2}$ & $p_{j_2}^T$ & $\slashed{E}$ & $\slashed{p}^T$ \\
\hline
\end{tabular}
\caption{The components of the feature space and the corresponding observables.}
\label{table:featurespacegqgc}
\end{table}
The events are generated for a $3$ TeV muon collider.
The final states are required to contain at least two jets.
At lepton colliders, the missing energy is also accessible, the chosen observables are listed in Table~\ref{table:featurespacegqgc}, where $E_{j_{1,2}}$ and $p_{j_{1,2}}^T$ are the energies and transverse momenta of the hardest and second hardest jets, $\slashed{E}$ is the missing energy, and $\slashed{p}^T$ is the transverse missing momentum, respectively.
Therefore, in this case, $d'=6$ and $d=4$.

$2^{17}+2^{17}$ events are selected from the SM and NP events to form the training set.
Since the $XY2M(\theta)=R_z(\theta) R_x(-\pi/2) R_z(-\theta)$ and $XY2P(\theta)=R_z(\theta) R_x(\pi/2) R_z(-\theta)$ are supported gates on \textit{tianyan176-2}, the single-qubit layers of the ansatz are chosen to be $XY2M(\theta _1) XY2P(\theta _2)$ on each qubit.
The entanglement layer is chosen to be pairwise.
The ansatz is chosen to have $l=3$ entanglement layers, with $N=2^{18}$ and $d=4$, $n_q=20$ qubits are used.
The ansatz is trained using \verb"QuEST", the fidelity reaches $78.5\%$.

\begin{figure*}[htbp]
\begin{center}
\includegraphics[width=0.95\hsize]{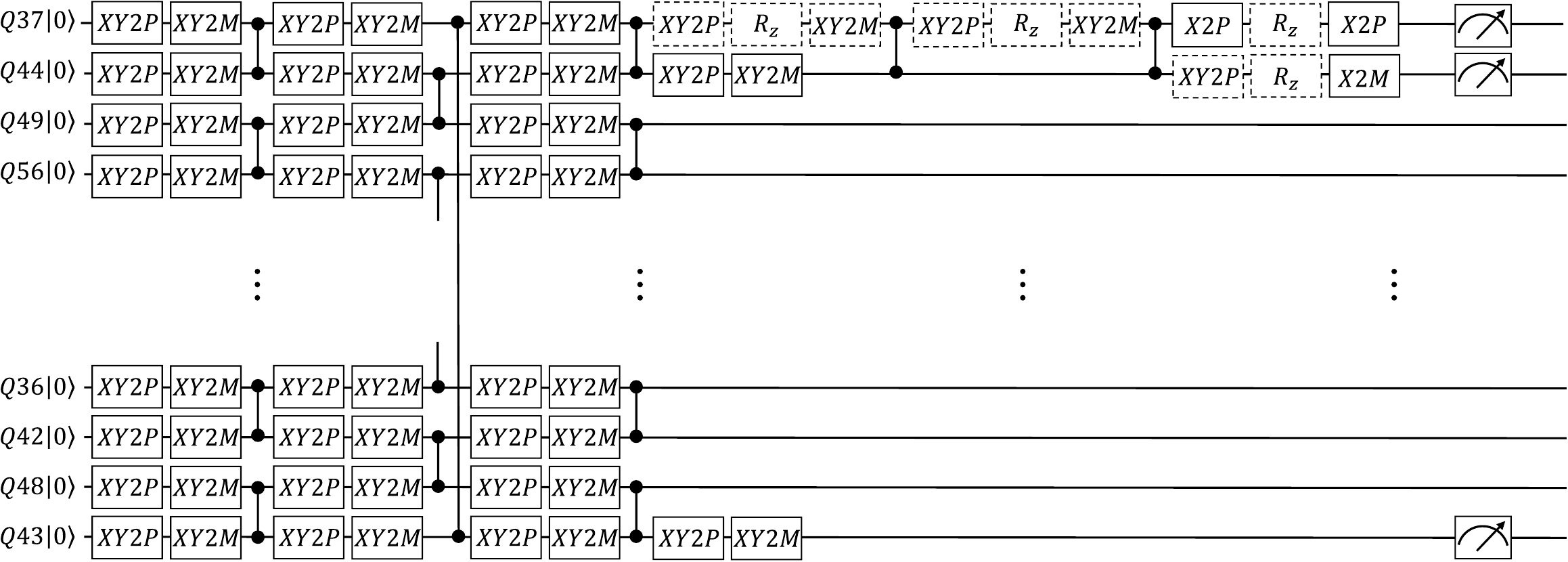}    
\caption{\label{fig:qcis2}Same as Fig.~\ref{fig:qcis1} but for the study of gQGCs in the process $\mu^+\mu^- \to \nu\bar{\nu}jj$.
The $Q37$ and $Q44$ make up the $a$ register, $Q43$ is the $b_1$ register, and the others make up the $b_2$ register.}
\end{center}
\end{figure*}
The test dataset is $50000+50000$ events selected from the SM and NP contributions, which are all different from the training events.
In the case of running on the \textit{tianyan176-2}, a sub set consist of $500+500$ events are selected.
The circuit for discrimination in terms of QCIS on \textit{tianyan176-2} is shown in Fig.~\ref{fig:qcis2}.

\subsubsection{Numerical results}

\begin{figure}[htbp]
\begin{center}
\includegraphics[width=0.48\hsize]{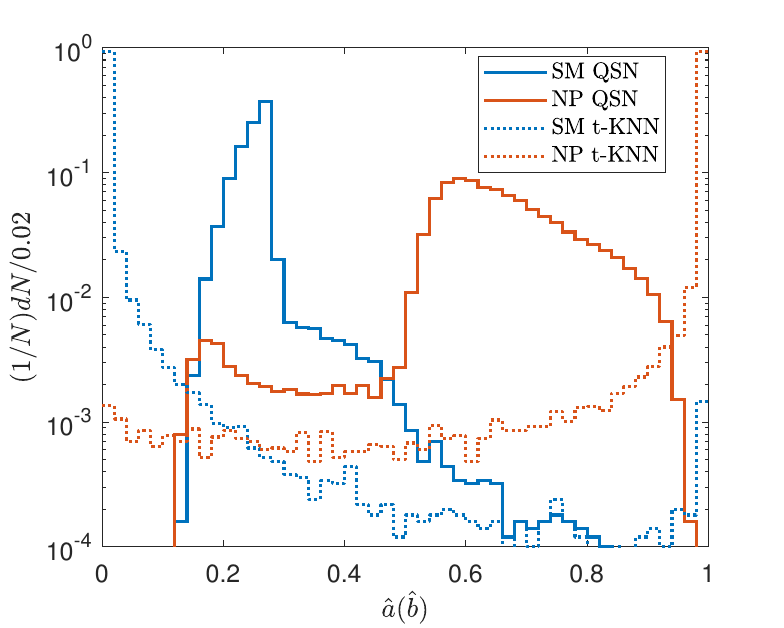}    
\includegraphics[width=0.48\hsize]{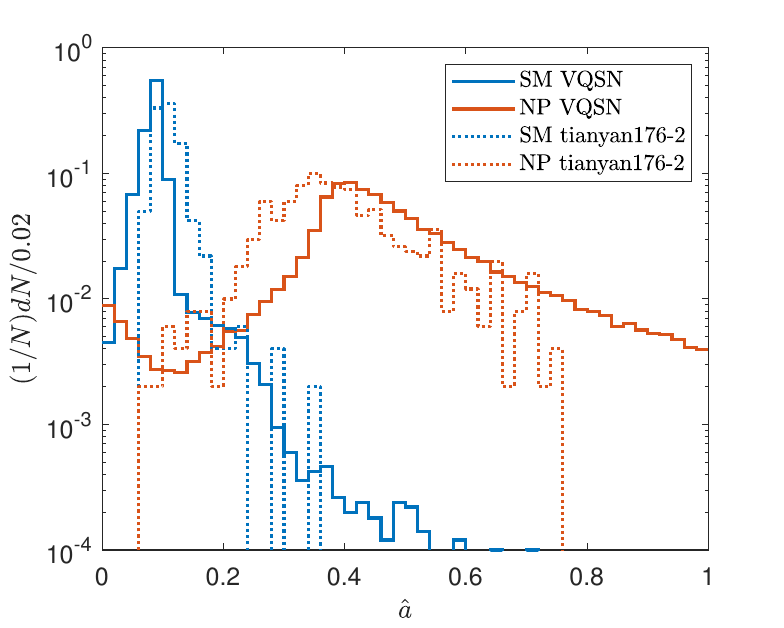}    
\caption{\label{fig:bingqgc} The normalized distributions of $\hat{a}$ and $\hat{b}$ for different methods.}
\end{center}
\end{figure}
To compare the results, we also use QSN and VQSN with $c\to \infty$, and with (t-)KNN with $k=1000$~(the results of $k=10$, $100$ are similar with the case of $k=1000$).
On \textit{tianyan176-2}, we use $c=1000$.
For the test set, the normalized distributions of $\hat{a}$ and $\hat{b}$ are shown in Fig.~\ref{fig:bingqgc}.
It can be seen that, both $\hat{a}$ and $\hat{b}$ can be used for discriminating the NP signals from the SM background.
The difference between the QSN and t-KNN comes from the fact that the QSN algorithm weights all points into account, which may bring the two peaks closer.
However, QSN also has an advantage that the tail of the SM background at the large $\hat{a}$ region is suppressed.
The difference between QSN and VQSN indicates that both peaks are shifted to a lower $\hat{a}$ region by VQSN, but the discriminative power is not significantly affected.

\begin{figure}[htbp]
\begin{center}
\includegraphics[width=0.6\hsize]{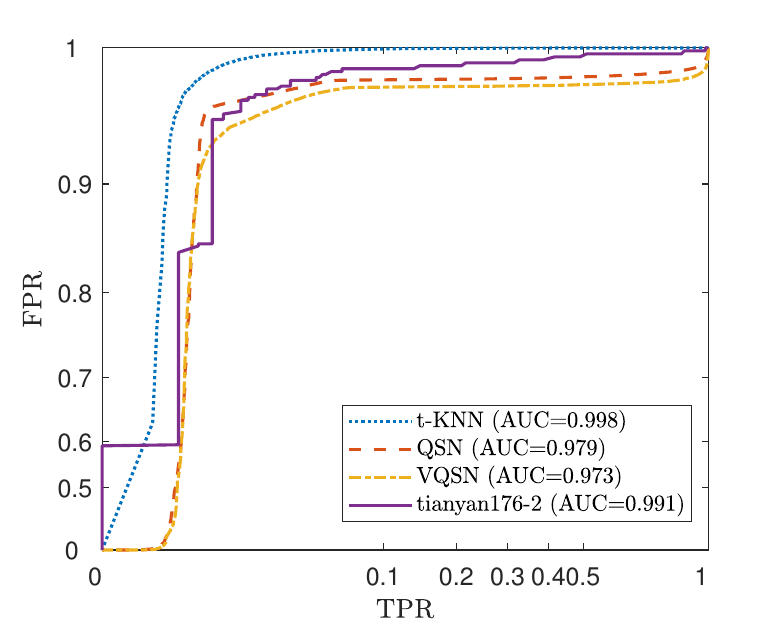}    
\caption{\label{fig:rocgqgc}The ROCs for different event selection strategies and the corresponding AUCs, the $x$-axis~(TPR) is scaled as $x^{1/3}$ and the $y$-axis~(FPR) is scaled as $y^3$ for clarity.}
\end{center}
\end{figure}
The receiver operating characteristic curve~(ROC) represents a crucial metric for evaluating the efficacy of a classification algorithm. 
ROC is obtained by mapping the $\hat{a}_{th}$~($\hat{b}_{th}$) into a two-dimensional plane with false positive rate~(FPR) and true positive rate~(TPR) as axes. 
The area under the ROC~(AUC) can be utilized as a quantitative metric to assess the efficacy of a classification algorithm.
The ROCs for (V)QSN and t-KNN are shown in Fig.~\ref{fig:rocgqgc}.
The large AUC in the case of \textit{tianyan176-2} can be attributed to the fact that the difference between all cases is small, therefore the results are affected by a limited number of events, and the test set running on \textit{tianyan176-2} consists of only $1\%$ events.


\begin{table}[hbtp]
\centering
\begin{tabular}{c|c|c|c|c|c|c}
\hline
 & $\hat{a}_{th}(\hat{b}_{th})$ & $\varepsilon_{\rm SM}$ & $\varepsilon_{\rm NP}$ & $\mathcal{S}_{stat}=2$ & $\mathcal{S}_{stat}=3$ & $\mathcal{S}_{stat}=5$ \\
 & & & & $({\rm TeV}^{-4})$ & $({\rm TeV}^{-4})$ & $({\rm TeV}^{-4})$ \\
\hline           
KNN & & $0.484\%$ & $98.2\%$ & $0.194$ & $0.238$ & $0.308$\\
t-KNN & $0.003$ & $0.074\%$ & $76.43\%$ & $0.138$ & $0.170$ & $0.221$ \\
QSN &  $0.44$ & $0.33\%$ & $85.29\%$ & $0.190$ & $0.233$ & $0.301$ \\
VQSN & $0.64$ & $0.322\%$ & $85.67\%$ & $0.189$ & $0.232$ & $0.301$ \\
tianyan176-2 & $0.71$ & $0.2\%$ & $82.2\%$ & $0.171$ & $0.209$ & $0.271$ \\
\hline
\end{tabular}
\caption{The $\hat{a}_{th}$, $\hat{b}_{th}$, cut efficiencies and the expected coefficient constraints.}
\label{table:gqgcconstraints}
\end{table}
The expected constraints on $|f_0|$ can be obtained by Eq.~(\ref{eq.ss}) and,
\begin{equation}
\begin{split}
&\sigma ^{\mu^+\mu^- \to jj\nu\bar{\nu}}(f_0)=\varepsilon _{\rm SM}\sigma ^{\mu^+\mu^- \to jj\nu\bar{\nu}}_{\rm SM} + \frac{f_0^2}{\tilde{f}_0^2}\varepsilon _{\rm NP}\sigma ^{\mu^+\mu^- \to jj\nu\bar{\nu}}_{\rm NP},
\end{split}
\label{eq.csgqgc}
\end{equation}
where $\sigma ^{\mu^+\mu^- \to jj\nu\bar{\nu}}_{\rm SM}$ and $\sigma ^{\mu^+\mu^- \to jj\nu\bar{\nu}}_{\rm NP}$ are the cross-section of the SM and NP after requiring at least two jets, $\sigma ^{\mu^+\mu^- \to jj\nu\bar{\nu}}_{\rm NP}$ is obtained when $f_0=\tilde{f}_0$.
At $\sqrt{s}=3\;{\rm TeV}$ and $\tilde{f}_0=1\;{\rm TeV}^{-4}$, $\sigma ^{\mu^+\mu^- \to jj\nu\bar{\nu}}_{\rm SM}=722.9\;{\rm fb}$ and $\sigma ^{\mu^+\mu^- \to jj\nu\bar{\nu}}_{\rm NP}=3.21\;{\rm fb}$~\cite{Yang:2023gos}.
By choosing proper $\hat{a}_{th}$ and $\hat{b}_{th}$, the expected constraints on $|f_0|$ at $\sqrt{s}=3\;{\rm TeV}$ and $L=1\;{\rm ab}^{-1}$~(where $L$ is the luminosity chosen according to the design of the muon colliders~\cite{Black:2022cth,Accettura:2023ked}) are listed in Table~\ref{table:gqgcconstraints}.
The expected constraints on $|f_0|$ by using a traditional event selection strategy at $\mathcal{S}_{start}=2$, $3$ and $5$ are $<0.155$, $<0.191$ and $0.248\;{\rm TeV}^{-4}$, respectively.
The dataset in this paper is in fact a subset~(about $28\%$) of the one used in Ref.~\cite{Yang:2023gos}, it can be seen that the VQSN can reach a same order of magnitude, while t-KNN outperforms the traditional method.

\begin{figure}[htbp]
\begin{center}
\includegraphics[width=0.6\hsize]{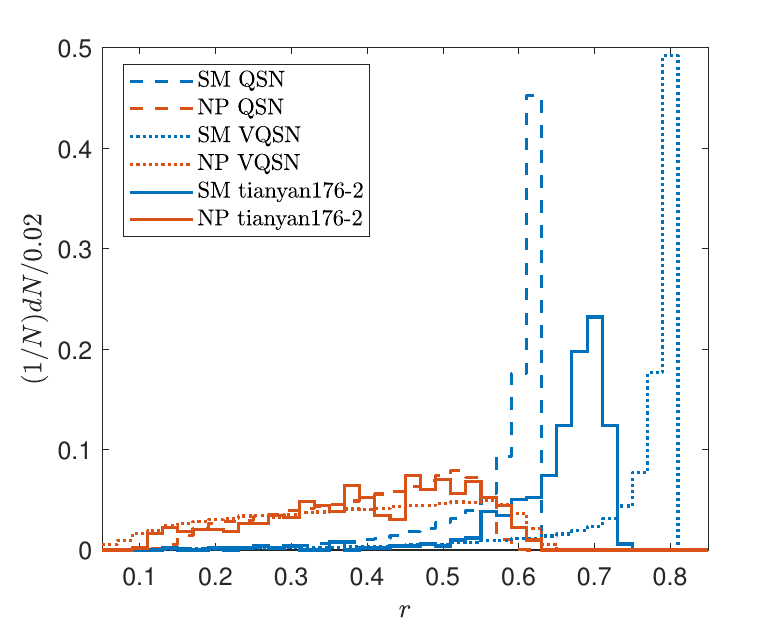}    
\caption{\label{fig:successrategqgc} The normalized distributions of $r$ for different methods.}
\end{center}
\end{figure}

It has been shown that, the randomness in measurement contributes to the primary of the errors.
In this subsection, $c=1000$ is used. 
When $c$ is a constant, the accuracy of $\hat{a}$ is affected by $r$, the success rates of building the states $\langle\phi | \Psi _{\rm train}\rangle$ and $\langle \phi|\Psi(\alpha _i)\rangle$~(where $\phi$ is the state for a test vector). 
Fig.~\ref{fig:successrategqgc} illustrates the normalized distributions of success rates where $\phi$ are the signal and background events in the validation dataset. 
The background and signal events are considered separately because the majority of the dataset from experiments will be the latter. 
It can be observed that the success rate is typically higher for background events than for signal events. 
This is due to the more centralized distribution of background events since the success rate is $\sum _n^N\left|\langle\phi | \psi _n\rangle\right|^2/N$ for QSN. 
Furthermore, it was found that the ansatz not only did not reduce but rather enhanced the success rates for the background events. 
This further reduces the cc of VQSN because for a fixed $c'$, a smaller $c$ is needed.
However, this improvement depends on the training data distribution, the ansatz structure, and the data encoding method, and should not be considered a universal result. 
The ansatz serves as an approximation and simplification of the training set. 
When the ansatz perfectly matches the training dataset, the state preparation success rates should be identical.  

The cc of VQSN and its classical counterpart, i.e. (t-)KNN are compared.
For one test event, $168$ gates need to be applied in terms of QCIS, therefore, when $c=1000$, the cc is estimated as $168000$.
For (t-)KNN, the cc is for one test event is $1.57\times 10^6$ because the training set contains $2^{18}$ $6$-dimensional vectors.
It can be seen that, even with a classical dataset, the cc of VQSN can be one order of magnitude smaller.

\subsection{Search for the gluon quartic gauge couplings at the LHC}

\subsubsection{The gluon quartic gauge couplings and the process to be considered}

One of the reasons muon collider is better suited to the search for NP is that the QCD background is suppressed.
ML algorithms are more suitable for processes with more intricate final states, which necessitates greater investment in kinematic analysis when traditional methods are employed.
There are certain complexities involved in discussing QCD background at the LHC, which offer an opportunity for utilizing ML algorithms in this regard.
Therefore, we also investigate an example of the search for gQGCs at the LHC.

\begin{figure}[htbp]
\begin{center}
\includegraphics[width=0.95\hsize]{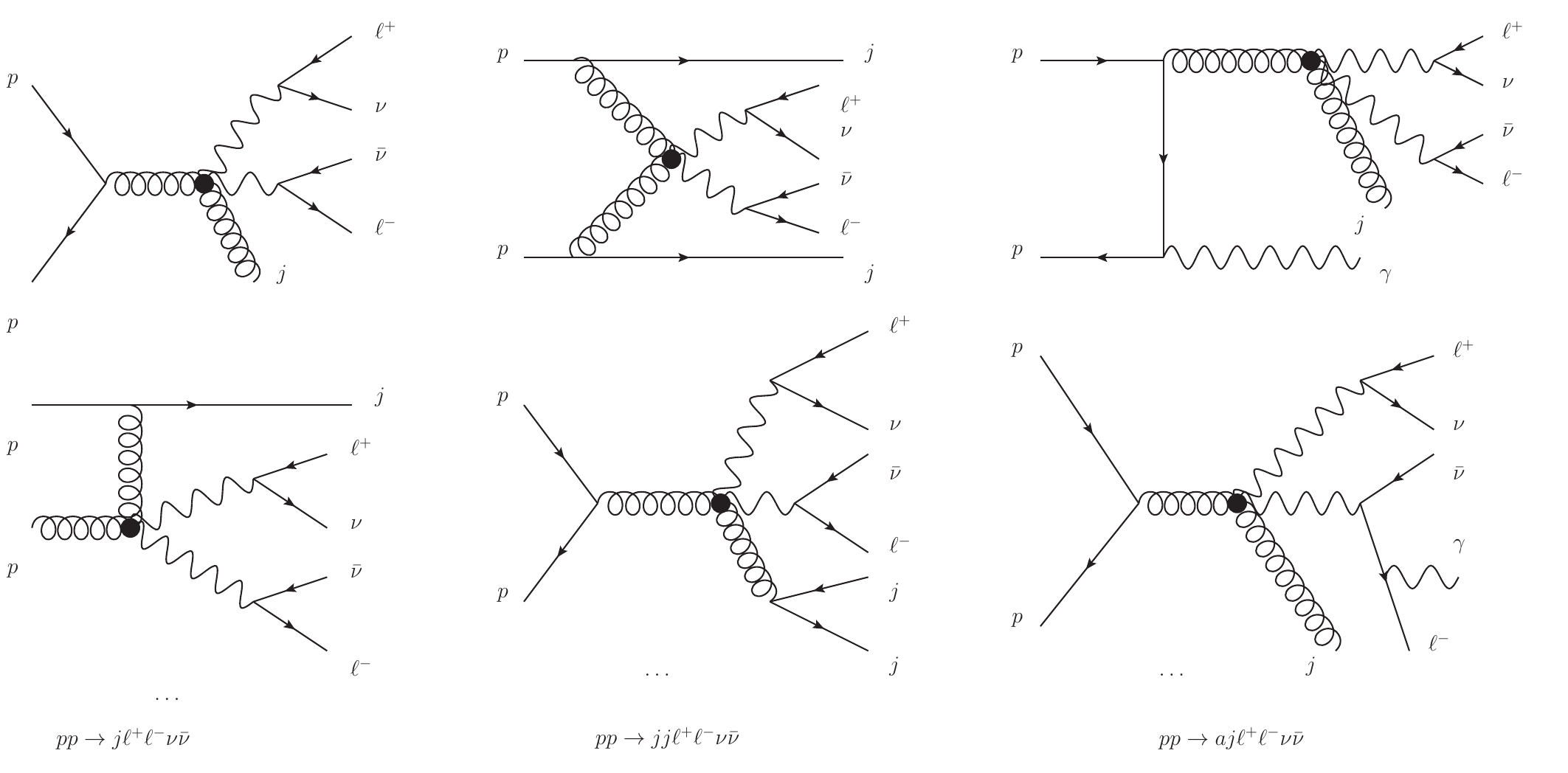}
\caption{\label{fig:feyndiagjllvv} Feynman diagrams the processes $pp\to j\ell^+\ell^-\bar{\nu}\nu$~(left panels), $pp\to jj\ell^+\ell^-\bar{\nu}\nu$~(middle panels) and  $pp\to \gamma j\ell^+\ell^-\bar{\nu}\nu$~(right panels), respectively, induced by the gQGCs.}
\end{center}
\end{figure}
\begin{table}[hbtp]
\centering
\begin{tabular}{c|c|c}
& processes & $\sigma\;({\rm pb})$ \\
 \hline
 $t\bar{t}$ & $pp\to t\bar{t}+jets$ & $832$ \\
 \hline
 \multirow{3}{*}{$\ell\ell+jets$} & $pp\to jj\ell^+\ell^-$ & $208$ \\
 & $pp\to jjj\ell^+\ell^-$ & $84.3$ \\
 & $pp\to jjj\ell^+\ell^-$ & $32.9$ \\
 \hline
 \multirow{2}{*}{$j\ell\ell\ell\nu$} & $pp\to j\ell^+\ell^-\ell^+\nu$ & $0.106$ \\
 & $pp\to j\ell^+\ell^-\ell^-\bar{\nu}$ & $0.0703$ \\
\hline
\multirow{2}{*}{$\gamma\ell\ell + jets$} & $pp\to \gamma j\ell^+\ell^-$ & $6.51$ \\
& $pp\to \gamma jj\ell^+\ell^-$ & $2.85$ \\
\end{tabular}
\caption{The backgrounds due to mis-tagging, or due to particles missed by the detectors, and their cross-sections.}
\label{table:otherbackgrounds}
\end{table}
In this paper, three processes $pp\to j\ell^+\ell^-\bar{\nu}\nu$, $pp\to jj\ell^+\ell^-\bar{\nu}\nu$ and  $pp\to \gamma j\ell^+\ell^-\bar{\nu}\nu$ are considered together.
The Feynman diagrams are shown in Fig.~\ref{fig:feyndiagjllvv}.
The three processes are combined together because they have similar backgrounds.
The presence of two (anti-)neutrinos in the final state inevitably entails a certain degree of information loss, which offers an opportunity to assess the efficacy of our algorithm.

Except for the SM contribution, other processes contributing to background due to detector effects are considered, including $pp\to (\gamma)\ell\ell+jets$ with missing jets, $pp\to t\bar{t}\to jjW^+W^-\to jj\ell^+\ell^-\nu\bar{\nu}$ with b-jets mis-tagged, and $pp\to j\ell\ell\ell\nu$ with a missing charged lepton.
They are listed in Table~\ref{table:otherbackgrounds}, with the cross-sections obtained by using \verb"MadGraph5" toolkit~\cite{Alwall:2014hca}, except for the case of $t\bar{t}+jets$, in which we use inclusive cross-section of $t\bar{t}+jets$ production~\cite{Czakon:2011xx,CMS:2020grm,ATLAS:2024aht}.

\subsubsection{Training of the ansatz}

To suppress the background with many jets, it is required that the number of jets~(denoted as $N_j$) in the final state satisfies $1\leq N_j \leq 6$, and it is required that there is no $b$-jet.
Furthermore, to suppress the $pp\to \ell^+\ell^- + jets$ background, it is required that the transverse missing energy~(denoted as $\slashed{E}_T$) satisfies $\slashed{E}_T>200\;{\rm GeV}$.
Except for that, the final state is required to have exactly two opposite charged leptons.
The above requirement is called particle number cut, and is denoted as $N_{part}$ cut.

\begin{table}[hbtp]
\centering
\begin{tabular}{c|c|c|c|c|c|c|c}
\hline
           & $v^0$ & $v^1$ & $v^2$ & $v^3$ & $v^4$ & $v^5$ & \\
observables & $E_{j_1}$ & $p_{j_1}^T$ & $E_{j_2}$ & $p_{j_2}^T$ & $E_{\ell^+}$ & $p_{\ell^+}^T$ & \\
\hline
           & $v^6$ & $v^7$ & $v^8$ & $v^{9}$ & $v^{10}$ & $v^{11}$ & $v^{12}$ \\
observables & $E_{\ell^-}$ & $p_{\ell^-}^T$ & $E_{\gamma}$ & $p_{\gamma}^T$ & $\slashed{E}_T$ & $f_{\ell^+}$ & $f_{\ell^-}$ \\
\hline
\end{tabular}
\caption{Same as Table~\ref{table:featurespacegqgc} but for the processes at the LHC.}
\label{table:featurespace}
\end{table}

A natural way is to use a feature space whose axes are observables in a collision event.
As a ML algorithm, instead of analyzing the kinematics, we simply choose some typical observables arbitrarily, and the more observables are expected to be better.
In this work we mainly use the momentum space of the final particles, i.e., the energies~($E$) and transverse momenta~($p^T$) of the hardest three jets, the hardest photon, and the charged leptons are picked as the features, as well as the missing momentum~($\slashed{E}_T$).
Since degrees of freedom of a $n$-qubit state is $2^n-2$, to better utilize the Hilbert space, the $f_{\ell^{\pm}}$ are added which are the flavors of charged leptons, and $f_{\ell^{\pm}}=0, 1$ when the charged leptons are electron or muon.
In summary, an event is mapped into a $13$-dimensional vector $\vec{v}$ whose components are listed in Table~\ref{table:featurespace}.
If there are no photons or less than three jets, the corresponding $E_{\gamma,j_i}$ and $p_{\gamma,j_i}^T$ are set to be zero.
Using this feature space, $d=8$ and $d'=13$, respectively.

\begin{table}[hbtp]
\centering
\begin{tabular}{c|c}
\hline
processes & number of events \\
 \hline
 $pp\to jj \ell^+\ell^-\nu\bar{\nu}$ for $O_{gT,0}$ & $16384$ \\
 $pp\to j \ell^+\ell^-\nu\bar{\nu}$ for $O_{gT,0}$ & $12288$ \\
 $pp\to \gamma j \ell^+\ell^-\nu\bar{\nu}$ for $O_{gT,0}$ & $4096$ \\
 \hline
 $pp\to jj \ell^+\ell^-\nu\bar{\nu}$ in the SM & $16384$ \\
 $pp\to j \ell^+\ell^-\nu\bar{\nu}$ in the SM & $8192$ \\
 $pp\to t\bar{t}\to jjW^+W^-\to jj \ell^+\ell^-\nu\bar{\nu}$ & $8192$ \\
\hline
\end{tabular}
\caption{The processes and the number of events included in the training dataset.}
\label{table:training}
\end{table}
The training dateset consists of events after $N_{part}$ cut.
The number of events picked to form the training dataset is listed in Table~\ref{table:training}.
Training dataset has $N=65536$ vectors, with three qubits to encode one vector, with $d=8$, the number of qubits to store $\Psi_{\rm train}$~(or the ansatz $\Psi(\alpha _i)$) is $n_q=19$.
In this subsection, the `shifted-circular-alternating'~(SCA) entanglement layer~\cite{Sim:2019yyv} is used.
The circuit of the ansatz is the one shown in Fig.~\ref{fig:circuits-c}.

Without the amplitude amplification, the cc for an ansatz with SCA entanglement layers is about $\left((2+3l)n_q+2(2d-\log _2(d)-2)\right)c'd$ for a test event. 
In this subsection, the calculation is performed by using \verb"QuEST", taking advantages of the controlled collapse operation of a simulator, we use a fixed $c'=100$ for simplicity.
If the cc of VQSN is to be less than that of its KNN counterpart (which is $d'N=8.5\times 10^5$ with $d'=13$ and $N=65536$), the ansatz should satisfy $l<17.6$.
In this work, we consider three cases, $l=5$, $10$, and $15$.
As a comparison, we consider (t-)KNN with $k=1$ , $10$, $100$, and $1000$.
The improvement brought by a larger $k$ is negligible when $k$ is larger than $1000$.

\subsubsection{Numerical results}

\begin{table}[hbtp]
\centering
\begin{tabular}{c|c|c|c|c|c}
\hline
& \multirow{2}{*}{processes} & $\sigma$ & \multicolumn{2}{c|}{$\sigma_{N_{part}}$} & \multirow{2}{*}{$\varepsilon _{N_{part}}$} \\
& & (fb) & \multicolumn{2}{c|}{(fb)} & \\
 \hline
 \multirow{3}{*}{NP} & $jj \ell^+\ell^-\nu\bar{\nu}$ & $1.485$ & \multicolumn{2}{c|}{$0.7034$} & $47.5\%$ \\
 & $j \ell^+\ell^-\nu\bar{\nu}$ & $1.549$ & \multicolumn{2}{c|}{$0.6923$} & $44.6\%$\\
 & $\gamma j \ell^+\ell^-\nu\bar{\nu}$ & $0.1738$ & \multicolumn{2}{c|}{$0.06879$} & $39.5\%$ \\
 \hline
 \multirow{7}{*}{SM} & $jj \ell^+\ell^-\nu\bar{\nu}$ & $679.3$ & \multicolumn{2}{c|}{$16.76$} & $2.47\%$ \\
 & $j \ell^+\ell^-\nu\bar{\nu}$ & $1.147\times 10^3$ & \multicolumn{2}{c|}{$14.01$} & $1.22\%$\\
 & $\gamma j \ell^+\ell^-\nu\bar{\nu}$ & $18.90$ & \multicolumn{2}{c|}{$0.3530$} & $1.87\%$\\
\cline{2-6} 
 & $t\bar{t}$ & $2.519\times 10^4$ & \multicolumn{2}{c|}{$101.1$} & $0.401\%$\\
\cline{4-5}
 & $\ell\ell + jets$ & $3.251\times 10^5$ & $44.10$ & \multirow{3}{*}{$45.18$} & $0.0136\%$\\
 & $j \ell\ell\ell\nu$ & $1.826\times 10^3$ & $0.7914$ & & $0.0433\%$\\
 & $\gamma \ell\ell + jets$ & $9.367\times 10^3$ & $0.2954$ & & $0.00315\%$\\
\hline
\end{tabular}
\caption{The cross-sections before cuts~($\sigma$), and the $N_{part}$ cut~($\sigma_{N_{part}}$), and the cut efficiency of the $N_{part}$ cut~($\varepsilon_{N_{part}}$) for different processes.}
\label{table:signalandbackground1}
\end{table}
For simplicity, we consider only the cut efficiency of $pp\to t\bar{t}$ and evaluate the $t\bar{t}+jets$ using the inclusive cross-section of the process $t\bar{t}+jets$ with the cut efficiency of $pp\to t\bar{t}$. 
The considered signal and background processes and the cross-section after $N_{part}$ cut are listed in Table~\ref{table:signalandbackground1}.
The requirement for a minimal $\slashed{E}_T$ significantly suppresses the background events without neutrinos.
In Table~\ref{table:signalandbackground1}, the $b$-tagging efficiency is assumed to be $77\%$~\cite{ATLAS:2019bwq,Guo:2019agy}.

\begin{figure*}[htbp]
\begin{center}
\includegraphics[width=0.9\hsize]{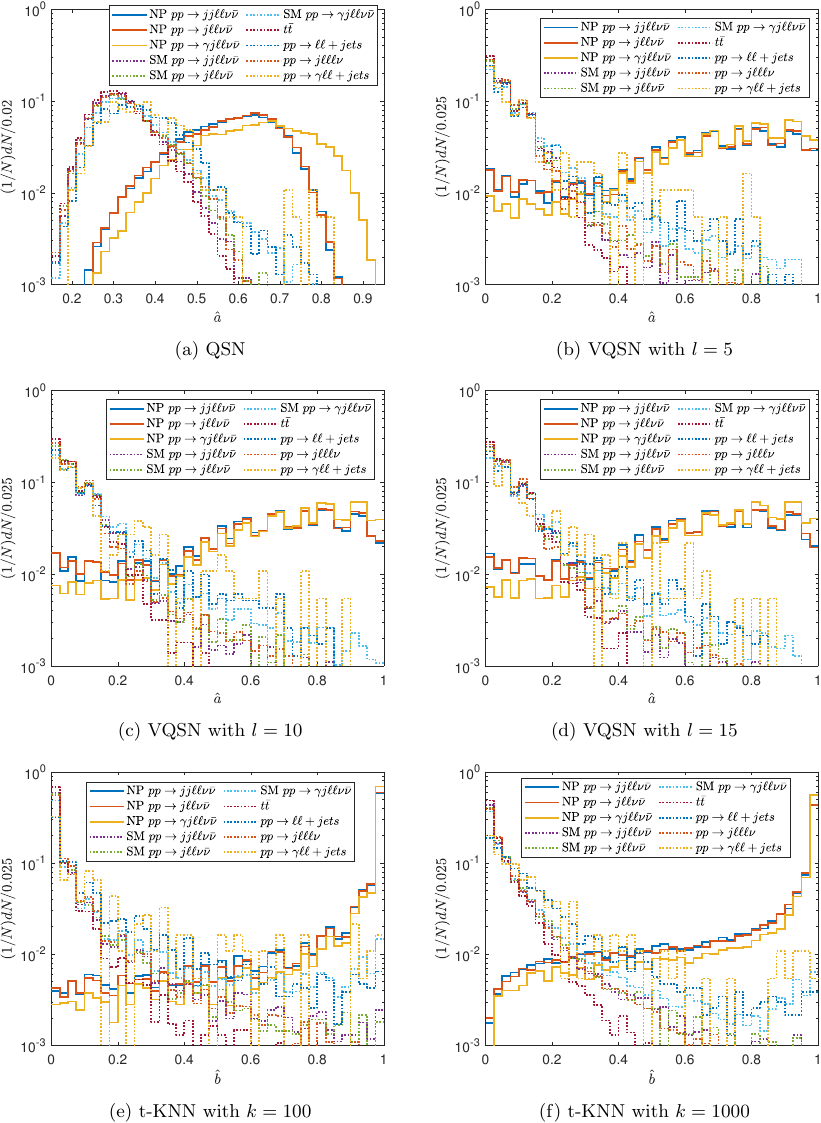}
\caption{\label{fig:histfig}The normalized distributions of $\hat{a}$ and $\hat{b}$ for different processes using different event selection strategies. For t-KNN, only the cases of $k=100$ and $1000$ are shown as examples.}
\end{center}
\end{figure*}
After the $N_{part}$ cut, the ansatz is trained.
The fidelities for $l=5$, $10$ and $15$ ansatz are $47.12\%$, $47.47\%$ and $47.53\%$, respectively.
A validation set is made which consists of events after $N_{part}$ cut with $73400$ and $108000$ NP and SM events~(all different from the training set), respectively.
The normalized distributions of $\hat{a}$ and $\hat{b}$ for the validation set are shown in Fig.~\ref{fig:histfig}.
It can be seen that, for both VQSN and t-KNN event selection strategies, $\hat{a}$ and $\hat{b}$ can be used for discriminating the signal events from background events.
The distributions of $\hat{a}$ for the VQSN show that the difference from different $l$s is small. 
Furthermore, the $\hat{a}$ of a background event is more closely aligned with $0$ when using an ansatz.

When NP signals are not found, the constraints on the operator coefficients are of interest. 
The expected coefficient constraints can be estimated using Eq.~(\ref{eq.ss}) and the cross-section.
When the interference term is considered, the total cross-section after cuts can be written as,
\begin{equation}
\begin{split}
\sigma (f_0)&= \left(\sum _p \sigma^{(p)} _{\rm SM} + \sigma _{t\bar{t}} \varepsilon _b^2 Br_{W\to \ell \nu}^2 \varepsilon ^{t\bar{t}}_{N_{part}} \varepsilon _{t\bar{t}} + \tilde{\sigma} _{bg}\right)\\
& + f_0 \sum _p \sigma^{(p)} _{\rm int} + f_0^2 \sum _p \sigma^{(p)} _{\rm NP},
\end{split}
\label{eq.cs}
\end{equation}
where the summation $\sum _p$ is over the three processes $pp\to jj \ell^+\ell^-\nu\bar{\nu}$, $pp\to j \ell^+\ell^-\nu\bar{\nu}$ and $pp\to \gamma j \ell^+\ell^-\nu\bar{\nu}$, $\sigma _{t\bar{t}}$ is the inclusive cross-section of the production of $t\bar{t}+jets$ which is taken as $832\;({\rm pb})$, $\varepsilon _b=1-77\%$ is from mis-tagging of $b$-jets, $Br_{W\to \ell \nu}=21.7\%$~\cite{ParticleDataGroup:2020ssz}, $\varepsilon ^{t\bar{t}}_{N_{part}}$ is the cut efficiency of $N_{part}$ cut in Table~\ref{table:signalandbackground1}, $\varepsilon _{t\bar{t}}$ is the cut efficiency of (V)QSN and (t-)KNN estimated using the $t\bar{t}$ events, and $\tilde{\sigma} _{bg}$ is the sum of $\ell\ell + jets$, $j \ell\ell\ell\nu$, and $\gamma \ell\ell + jets$ backgrounds after all cuts.
Then, $\sigma _{bg}=\sigma(f_0=0)$ and $\sigma _s = \sigma (f_0) - \sigma _{bg}$,

\begin{table}[hbtp]
\centering
\begin{tabular}{c|c|c|c|c|c}
& & & $\sigma _{t\bar{t}}$ & $\varepsilon _{t\bar{t}}$ & $\tilde{\sigma} _{bg}$ \\
& & & (fb) & & (fb) \\
\hline
& & $\hat{a}_{th}$ & & & \\
QSN                    &        & $0.56$  & $0.0277$  & $2.74\times 10^{-4}$ & $1.58$ \\
\hline
& $l$ & $\hat{a}_{th}$ & & & \\
\multirow{3}{*}{VQSN}  & $5$    & $0.81$  & $0.00498$ & $4.92\times 10^{-5}$ & $0.195$ \\
                       & $10$   & $0.88$  & $0.00213$ & $2.11\times 10^{-5}$ & $0.0135$ \\
                       & $15$   & $0.9$   & $0.00213$ & $2.11\times 10^{-5}$ & $1.97\times 10^{-4}$ \\
\hline 
                       & $k$    & & & & \\                      
\multirow{4}{*}{KNN}   & $1$    &         & $0.322$   & $3.19\times 10^{-3}$ & $7.52$ \\
                       & $10$   &         & $0.133$   & $1.32\times 10^{-3}$ & $4.61$ \\
                       & $100$  &         & $0.119$   & $1.18\times 10^{-3}$ & $4.35$ \\
                       & $1000$ &         & $0.0547$  & $5.42\times 10^{-4}$ & $2.92$ \\
\hline                 
                       & $k$    & $\hat{b}_{th}$ & & & \\      
\multirow{3}{*}{t-KNN} & $10$   & $0.9$   & $0.0163$   & $1.62\times 10^{-4}$ & $0.880$ \\
                       & $100$  & $0.99$  & $0.000711$ & $7.03\times 10^{-6}$ & $0.0583$ \\
                       & $1000$ & $0.998$ & $0.000711$ & $7.03\times 10^{-6}$ & $2.34\times 10^{-5}$ \\                                          
\end{tabular}
\caption{The cross-sections after the (V)QSN and (t-)KNN event selection strategies for the $t\bar{t}$ process~($\sigma _{t\bar{t}}$) as well as the $\ell\ell+jets$, $j\ell\ell\ell\nu$ and $\gamma \ell\ell+jets$ backgrounds~($\tilde{\sigma} _{bg}$). The cut efficiencies of the (V)QSN and (t-)KNN event selection strategies for the $t\bar{t}$ process~($\sigma _{t\bar{t}}$) are listed. The optimal $\hat{a}_{th}$ and $\hat{b}_{th}$ for VQSN and t-KNN are also listed.}
\label{table:signalandbackground2}
\end{table}

\begin{figure*}[htbp]
\begin{center}
\includegraphics[width=0.48\hsize]{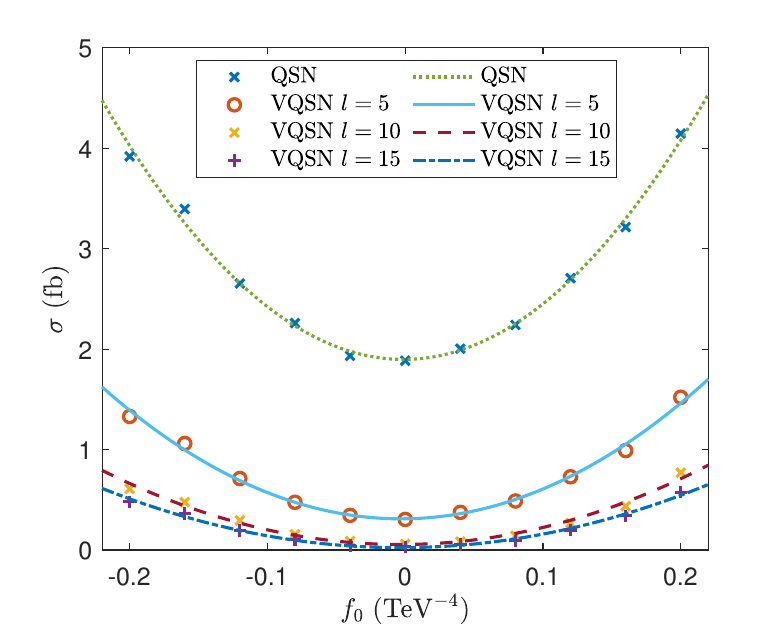}
\includegraphics[width=0.48\hsize]{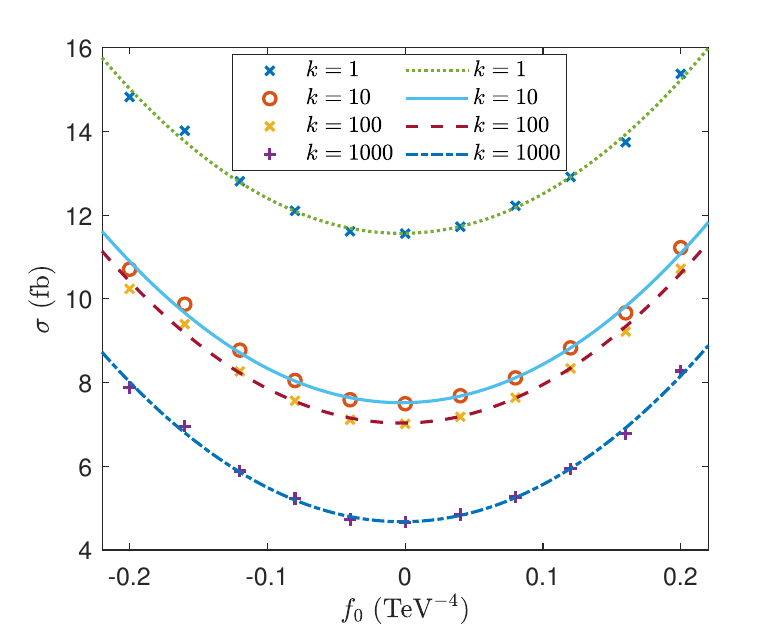}\\
\includegraphics[width=0.48\hsize]{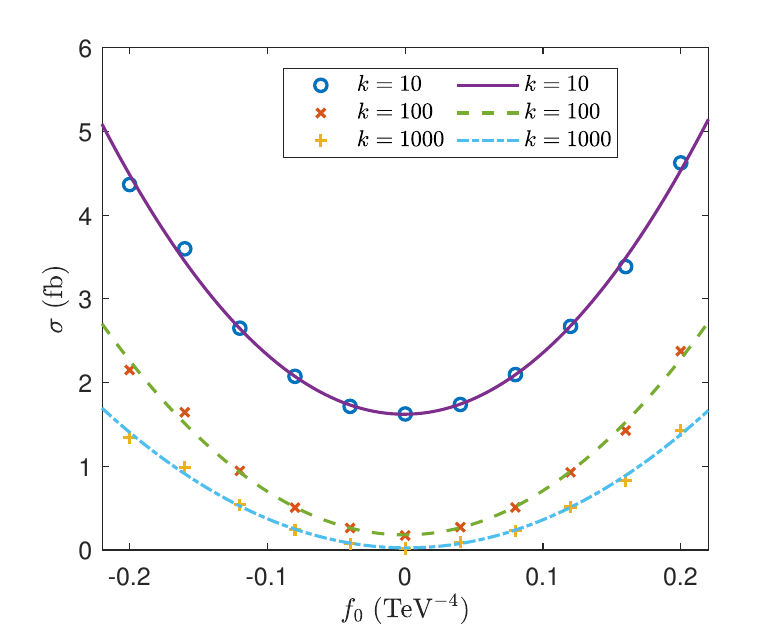}
\caption{\label{fig:crosssection}The total cross-sections $\sigma (f_0)$ after the (V)QSN and (t-)KNN event selection strategies at different $f_0$, as well as the fitted total cross-sections. The cases of (V)QSN are shown in the left panel, the cases of KNN are shown in the middle panel, and the cases of t-KNN are shown in the right panel, respectively.}
\end{center}
\end{figure*}

To consider the interference, we scanned the interval $f_0 \in [-0.2\;{\rm TeV}^{-4},0.2\;{\rm TeV}^{-4}]$. 
For a given $\hat{a}_{th}$~(or $\hat{b}_{th}$) and a given $f_0$, one can obtain the cross-section $\sigma^{(p)}(f_0)$ after cuts as well as $\varepsilon _{t\bar{t}}$ and $\tilde{\sigma}_{bg}$, then $\sigma(f_0)$ and $\mathcal{S}_{stat}$ can be obtained.
The tunable parameters $\hat{a}_{th}$ and $\hat{b}_{th}$ are chosen to maximize $\mathcal{S}_{stat}$ when $f_0=0.2\;{\rm TeV}^{-4}$.
The optimal $\hat{a}_{th}$, $\hat{b}_{th}$, as well as $\varepsilon _{t\bar{t}}$ and $\tilde{\sigma} _{bg}$ at this point are listed in Table~\ref{table:signalandbackground2}.
$\sigma(f_0)$ with different $f_0$ can be fitted according to Eq.~(\ref{eq.cs}).
The fittings of $\sigma (f_0)$ at the optimal $\hat{a}_{th}$ and $\hat{b}_{th}$ are shown in Fig.~\ref{fig:crosssection}.
It can be seen that the bilinear function fits well.
In the region of coefficient in consideration, the interference is relatively small, and is found to be positive.

\begin{table}[hbtp]
\centering
\begin{tabular}{c|c|c|c|c}
& & \multicolumn{3}{c}{$\mathcal{S}_{stat}$} \\
\cline{3-5}
& & $2$ & $3$ & $5$ \\
\hline
QSN                    &        & $[-0.068, 0.065]$ & $[-0.083, 0.081]$ & $[-0.108, 0.106]$ \\
\hline
& $l$ & & & \\
\multirow{3}{*}{VQSN}  & $5$    & $[-0.063, 0.057]$ & $[-0.077, 0.071]$ & $[-0.101, 0.095]$ \\
                       & $10$   & $[-0.057, 0.049]$ & $[-0.070, 0,063]$ & $[-0.094, 0.086]$ \\
                       & $15$   & $[-0.053, 0.046]$ & $[-0.066, 0.059]$ & $[-0.089, 0.082]$ \\
\hline 
                       & $k$    & & & \\                      
\multirow{4}{*}{KNN}   & $1$    & $[-0.084, 0.078]$ & $[-0.102, 0.097]$ & $[-0.132, 0.126]$ \\
                       & $10$   & $[-0.076, 0.071]$ & $[-0.093, 0.088]$ & $[-0.120, 0.115]$ \\
                       & $100$  & $[-0.075, 0.070]$ & $[-0.092, 0.086]$ & $[-0.118, 0.113]$ \\
                       & $1000$ & $[-0.068, 0.064]$ & $[-0.084, 0.079]$ & $[-0.108, 0.104]$ \\
\hline                 
                       & $k$    & & & \\      
\multirow{3}{*}{t-KNN} & $10$   & $[-0.056, 0.055]$ & $[-0.069, 0.067]$ & $[-0.090, 0.088]$ \\
                       & $100$  & $[-0.039, 0.038]$ & $[-0.048, 0.047]$ & $[-0.064, 0.063]$ \\
                       & $1000$ & $[-0.031, 0.031]$ & $[-0.039, 0.039]$ & $[-0.053, 0.053]$ \\                                          
\end{tabular}
\caption{The expected constraints on the $f_0\;({\rm TeV}^{-4})$ at $\sqrt{s}=13\;{\rm TeV}$, $L=137\;{\rm fb}^{-1}$ and $S_{stat}=2$, $3$, and $5$ using different event selection strategies.}
\label{table:constraints}
\end{table}

When the fittings of $\sigma (f_0)$ are obtained, the expected constraints at $\mathcal{S}_{stat}=2,3$ and $5$ can be solved.
The results at the $13\;{\rm TeV}$ LHC with $L=137\;{\rm fb}^{-1}$ are listed in Table~\ref{table:constraints}.
The results indicate that the expected coefficient constraints of the various algorithms are generally at a same order of magnitude. 
It is evident that the constraints obtained by VQSN are more off-centered, indicating that the contribution of the interference terms is more preserved. 
This may confer an advantage to VQSN when the coefficient is in a more stringent region as the luminosity increases.

\begin{figure}[htbp]
\begin{center}
\includegraphics[width=0.6\hsize]{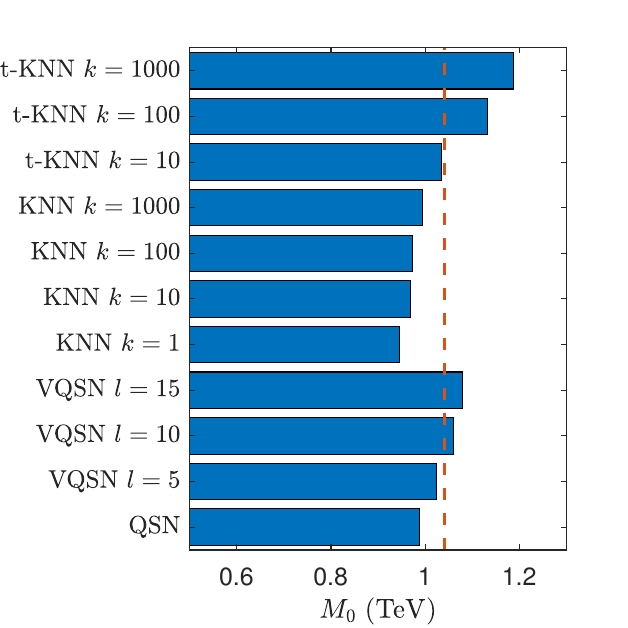}
\caption{\label{fig:compare}Comparison of the constraints on $M_0$ when $f_0$ is positive at $\mathcal{S}_{stat}=2$ obtained from different event selection strategies. The result of Ref.~\cite{Ellis:2018cos} is shown as the dashed line.}
\end{center}
\end{figure}
Using the upper bound of coefficient constraint, one can set a constraint on $M_0$ when $f_0$ is positive. 
The results at $\mathcal{S}_{stat}=2$ are shown in Fig.~\ref{fig:compare}, which provides a more clear comparison of different algorithms. 
It can be seen that t-KNN~($k=1000$) is the most effective, followed by t-KNN~($k=100$), VQSN~($l=15$), and VQSN~($l=10$). 
Furthermore, the result presented in Ref.~\cite{Ellis:2018cos}~($M_0\geq 1040\;{\rm MeV}$) is shown as a dashed line, and the four aforementioned cases demonstrate tighter constraints. 
This suggests that the processes under consideration are sensitive to the gQGCs and are worthy of consideration in the global fit of the constraints on gQGCs. 
Furthermore, the results demonstrate the efficacy of the t-KNN and VQSN algorithms employed. 
The validity is a crucial consideration when using an EFT~\cite{Dong:2023nir,Yang:2021pcf,Guo:2020lim,Yue:2021snv,Fu:2021mub,Yang:2022ilt,Layssac:1993vfp,Corbett:2017qgl,Perez:2018kav,Almeida:2020ylr,Kilian:2018bhs,Kilian:2021whd}, and can usually be tested through unitarity. 
According to the Ref.~\cite{Ellis:2018cos}, the region of coefficients in this study is free from the violation of unitarity.

\begin{table}[hbtp]
\centering
\begin{tabular}{c|c|c|c|c|c}
     & QSN & \multicolumn{3}{c|}{VQSN} & (t-)KNN \\
\hline     
$l$  & & $5$ & $10$ & $15$ & \\
\hline
$R_y$   & $458759$  & $121$ & $216$ & $311$ & \\
$R_z$   & $458759$  & $121$ & $216$ & $311$ & \\
\hline
$U$     & $914865$  & $106$ & $201$ & $296$ & \\
\hline
CZ      & $0$      & $95$ & $190$ & $285$ & \\
CNOT    & $917506$  & $8$ & $8$ & $8$ & \\
\hline
total   & $1832371$ & $209$ & $399$ & $589$ & \\
\hline
cc & $9.9\times 10^8$ & $8.5\times 10^4$ & $1.6\times 10^5$ & $2.4\times 10^5$ & $8.5\times 10^5$ \\
\hline
depth   & $1832354$ & $105$ & $210$ & $315$ & \\
\end{tabular}
\caption{The number of single-qubit gates, two-qubit gates used in (V)QSN event selection strategies.
The `total' represents the number of gates to measure the classification assignment once for one test vector.
The depths of circuits are also listed.}
\label{table:numberofgates}
\end{table}

We use the average success rate of background events to estimate the cc necessary to determine the classification of one vector when $c'=100$. 
The requisite number of quantum gates is presented in Table~\ref{table:numberofgates}. 
When counting the number of quantum gates, the single-qubit gates are combined as the single-qubit unitary gates~(denoted as $U$), and the gates irrelevant of the measurements are removed.
The total number of quantum gates is defined as the sum of the number of $U$ gates and the number of two-qubit gates. 
It can be observed that the cc for VQSN is less than that of (t-)KNN. 
This indicates that even with classical data, VQSN can still be more efficient than (t-)KNN. 
Furthermore, the cc is reduced by an order of magnitude when $l = 5$. 
Note that, the superiority of VQSN can be even further evident for a larger training dataset or a larger $d$. 

Meanwhile, the depth of the quantum circuit is also a significant metric. 
Given the noise in quantum computers at this stage, the depth of a quantum circuit is limited. 
In addition, since quantum gates on different qubits can act in parallel, the depth is a measure of time complexity.
For VQSN, the depth of the circuit when a SCA entanglement layer is employed is less than or equal to $dep_{ans}+dep_{ae}$ with $dep_{ans}=(n_q+1)l+1$ and $dep_{ae}=4d-3-4\log_2(d)$ are the depths of the ansatz and inverse of amplitude encode, respectively. 
It can be seen that the SCA entanglement layer introduces much more depth than the pairwise entanglement layer.
Due to the large amount of test data with interference between the SM and NP, which would consume too much machine time, we only performed the simulation of quantum computation on a simulator, so we choose the better-fitting SCA instead of pairwise in this subsection.

Given the substantially higher error rate of two-qubit gates compared to single-qubit gates, evaluating the circuit depth based on two-qubit gates provides a better estimate of the overall error.
For $d$-dimensional normalized complex vector~($2d-1$ dimensional feature space), $n_q=\log _2(Md)$.
For pairwise entanglement layer, $dep_{ans}=l$ in the sense of two-qubit gates, and $dep_{ans}=n_ql$ for the case of SCA entanglement layer.
$dep_{ae}=2\left(d-1-\log_2(d)\right)$ in the sense of two-qubit gates.
The number of qubits to be measured is $\log_2d + 1$
$l$ depends on the expressive power of the ansatz and the distribution of the training data.
If we assume that the ansatz can express the training dataset with a moderate $l$, then both the error and computational complexity scale logarithmically with $M$.
This is the reason why the ansatz leads to optimization.

\subsection{Search for the anomalous quartic gauge couplings in the tri-photon process at muon colliders}

\subsubsection{The contribution of the anomalous quartic gauge couplings in the tri-photon process}

As machine learning algorithms, (V)QSN and (t-)KNN should not depend on the NP or process to be searched. 
Thus, as an example, we consider the dimension-8 operators contributing to the anomalous quartic gauge couplings~(aQGCs) in the tri-photon process at muon colliders.
Due to the attention paid to vector boson scattering~(VBS) processes at the LHC, the aQGCs which are suitable to be studied in VBS processes have also been widely studied phenomenologically~\cite{Green:2016trm,Chang:2013aya,Anders:2018oin,Zhang:2018shp,Bi:2019phv} and experimentally~\cite{ATLAS:2014jzl,CMS:2020gfh,ATLAS:2017vqm,CMS:2017rin,CMS:2020ioi,CMS:2016gct,CMS:2017zmo,CMS:2018ccg,ATLAS:2018mxa,CMS:2019uys,CMS:2016rtz,CMS:2017fhs,CMS:2019qfk,CMS:2020ypo,CMS:2020fqz}.

\begin{figure}[htbp]
\begin{center}
\includegraphics[width=0.8\hsize]{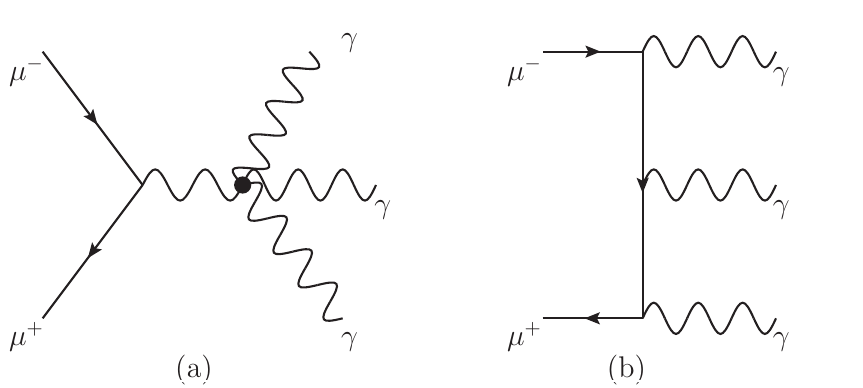}
\caption{\label{fig:feyndiagtriphoton} Feynman diagrams the tri-photon process. There are two diagrams of aQGCs shown in (a) where the s-channel propagator can be either $Z$ or $\gamma$. There are six diagrams in the SM, one of which is shown in (b) and other five diagrams can be obtained by permuting the photons in the final state.}
\end{center}
\end{figure}
The dimension-8 operators contributing to the aQGCs are often classified as scalar/longitudinal operators $O_{S_i}$, mixed transverse and longitudinal operators $O_{M_i}$ and transverse operators $O_{T_i}$.
It has been found that the tri-photon processes at the muon colliders are sensitive to the $O_{T_i}$ operators~\cite{Yang:2020rjt}, they are~\cite{Eboli:2006wa,Eboli:2016kko,Anders:2018oin},
\begin{equation}
\begin{split}
&O_{T,0}={\rm Tr}\left[\widehat{W}_{\mu\nu}\widehat{W}^{\mu\nu}\right]\times {\rm Tr}\left[\widehat{W}_{\alpha\beta}\widehat{W}^{\alpha\beta}\right],\\
&O_{T,1}={\rm Tr}\left[\widehat{W}_{\alpha\nu}\widehat{W}^{\mu\beta}\right]\times {\rm Tr}\left[\widehat{W}_{\mu\beta}\widehat{W}^{\alpha\nu}\right],\\
&O_{T,2}={\rm Tr}\left[\widehat{W}_{\alpha\mu}\widehat{W}^{\mu\beta}\right]\times {\rm Tr}\left[\widehat{W}_{\beta\nu}\widehat{W}^{\nu\alpha}\right],\\
&O_{T,5}={\rm Tr}\left[\widehat{W}_{\mu\nu}\widehat{W}^{\mu\nu}\right]\times B_{\alpha\beta}B^{\alpha\beta},\\
&O_{T,6}={\rm Tr}\left[\widehat{W}_{\alpha\nu}\widehat{W}^{\mu\beta}\right]\times B_{\mu\beta}B^{\alpha\nu},\\
&O_{T,7}={\rm Tr}\left[\widehat{W}_{\alpha\mu}\widehat{W}^{\mu\beta}\right]\times B_{\beta\nu}B^{\nu\alpha},\\
&O_{T,8}=B_{\mu\nu}B^{\mu\nu}\times B_{\alpha\beta}B^{\alpha\beta},\\
&O_{T,9}=B_{\alpha\mu}B^{\mu\beta}\times B_{\beta\nu}B^{\nu\alpha},\\
\end{split}
\label{eq.aqgcoperators}
\end{equation}
where $\widehat{W}\equiv \vec{\sigma}\cdot {\vec W}/2$ with $\sigma$ being the Pauli matrices and ${\vec W}=\{W^1,W^2,W^3\}$, $B_{\mu}$ and $W_{\mu}^i$ are $U(1)_{\rm Y}$ and $SU(2)_{\rm I}$ gauge fields, and $B_{\mu\nu}$ and $W_{\mu\nu}$ correspond to the field strength tensors.
The Feynman diagrams are shown in Fig.~\ref{fig:feyndiagtriphoton}.
The contributions of $O_{T,1}$ and $O_{T,6}$ operator are as same as the ones of $O_{T,0}$ and $O_{T,5}$, respectively, therefore are not studied.

\subsubsection{Training of the ansatz}

\begin{table}[hbtp]
\centering
\begin{tabular}{c|c|c|c|c|c|c}
\hline
           & $v^0$ & $v^1$ & $v^2$ & $v^3$ & $v^4$ & $v^5$ \\
\hline           
observables & $E_{\gamma_1}$ & $p^T_{\gamma_1}$ & $E_{\gamma_2}$ & $p^T_{\gamma_2}$ & $E_{\gamma_3}$ & $p^T_{\gamma_3}$ \\
\hline
\end{tabular}
\caption{Same as Table~\ref{table:featurespacegqgc} but for the tri-photon processes at the muon colliders.}
\label{table:featurespaceaqgc}
\end{table}
Different from the previous subsections, in the generation of the events, the standard cuts are set to be as same as the default except for the $p_{\gamma}^T$ cut which is set to be $p_{\gamma}^T \geq 0.1 E_{beam}$ where $E_{beam}$ is the beam energy which can suppress the SM background and ease the pressure on numerical computation consumption.
The cases of $\sqrt{s}=3\;{\rm TeV}$, $10\;{\rm TeV}$, $14\;{\rm TeV}$ and $30\;{\rm TeV}$ are considered~\cite{Black:2022cth,Accettura:2023ked}.
To generate the training set, $2^{16}+2^{16}$ events are chosen from the SM background and the NP events.
It has been shown that, the kinematic features are similar for different operators, therefore in the training set, we use only the events from $O_{T_0}$.
It is required that the final state to have at least three photons.
The observables chosen to form the feature space are listed in Table~\ref{table:featurespaceaqgc}, where $E_{\gamma _{1,2,3}}$ and $p^T_{\gamma _{1,2,3}}$ are energies and transverse momenta of the hardest, second hardest and the third hardest photons.
Using this feature space, $d=6$ and $d'=4$, and with $2^{17}$ events, the ansatz takes $n_q=19$ qubits.
In this subsection, we also use the SCA entanglement layer and with $l=10$, the circuit is the one shown in Fig.~\ref{fig:circuits-c}.
After training, the fidelities are $46.8\%$, $46.8\%$, $46.9\%$ and $46.9\%$ for the cases of $\sqrt{s}=3\;{\rm TeV}$, $10\;{\rm TeV}$, $14\;{\rm TeV}$ and $30\;{\rm TeV}$, respectively.

\subsubsection{Numerical results}

\begin{figure*}[htpb]
\begin{center}
\includegraphics[width=0.3\hsize]{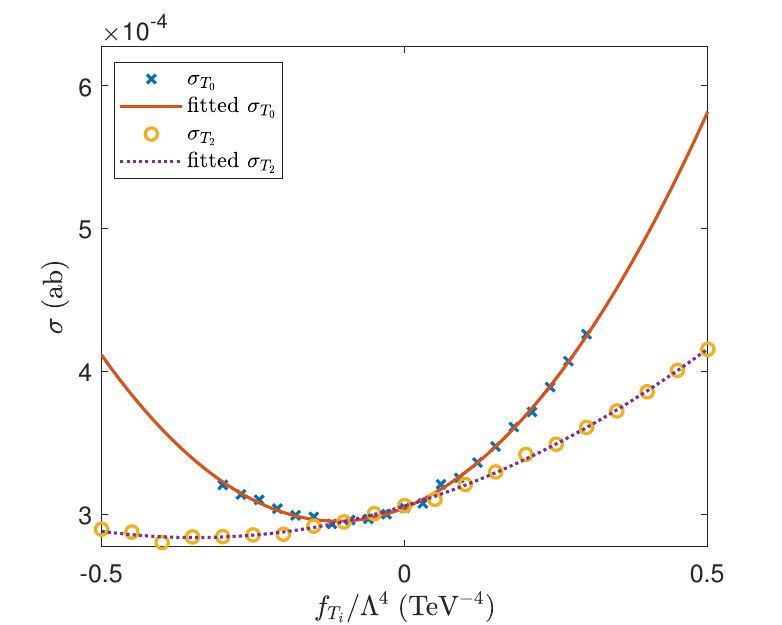}
\includegraphics[width=0.3\hsize]{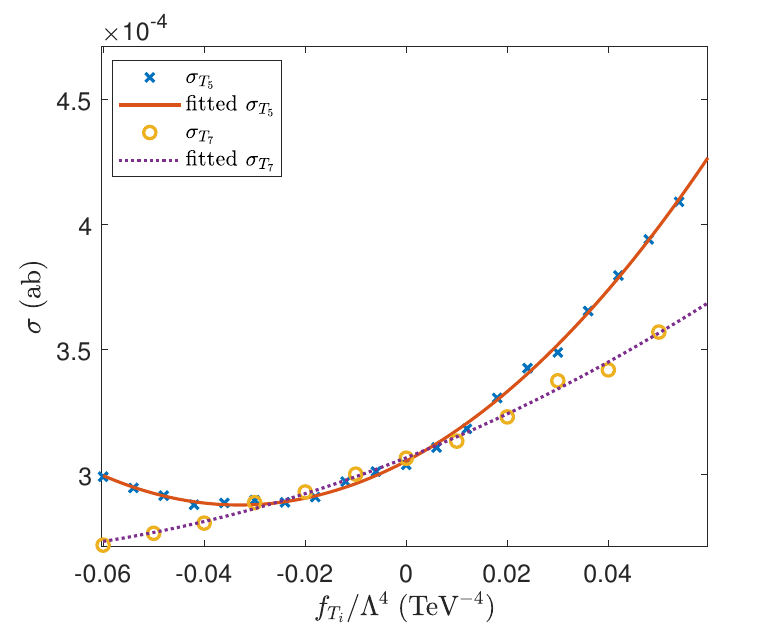}
\includegraphics[width=0.3\hsize]{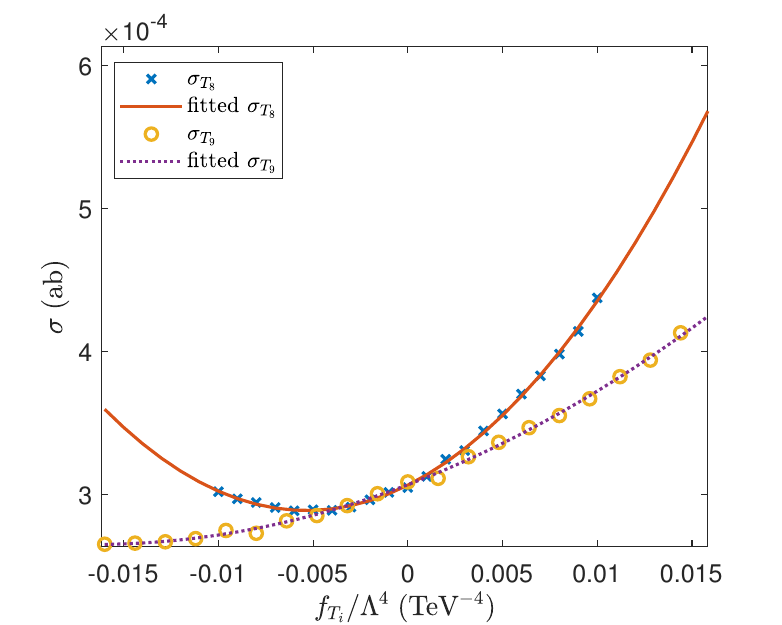}
\includegraphics[width=0.3\hsize]{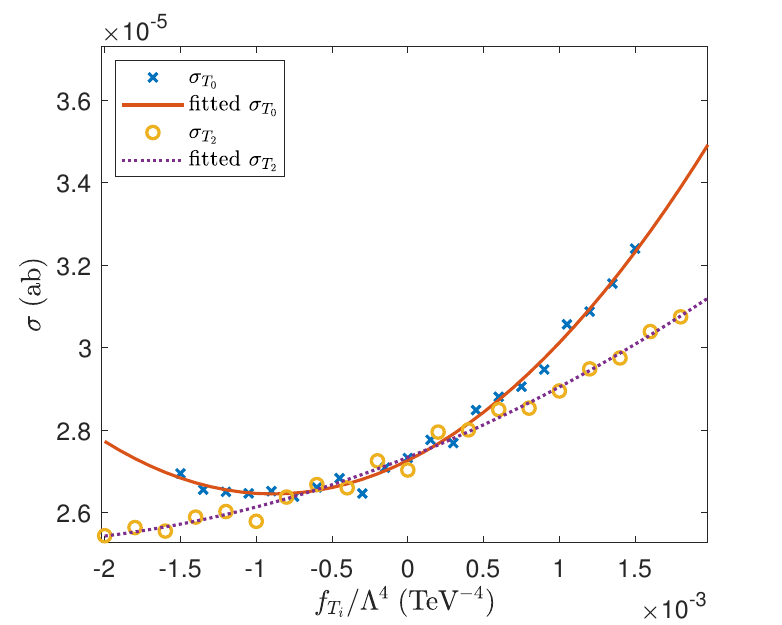}
\includegraphics[width=0.3\hsize]{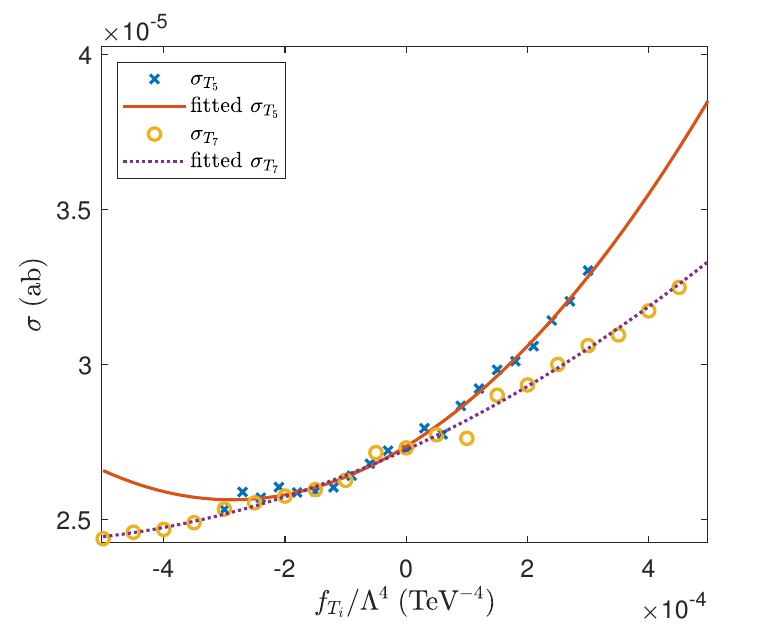}
\includegraphics[width=0.3\hsize]{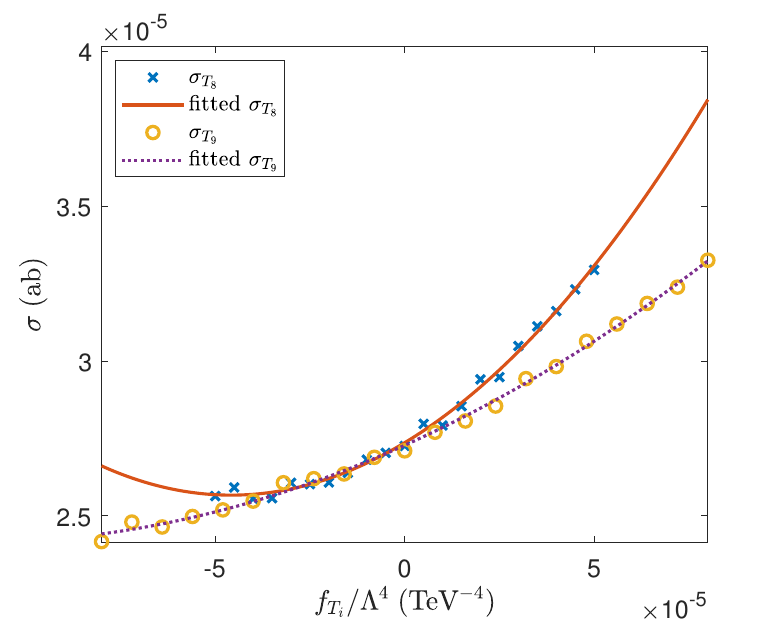}
\includegraphics[width=0.3\hsize]{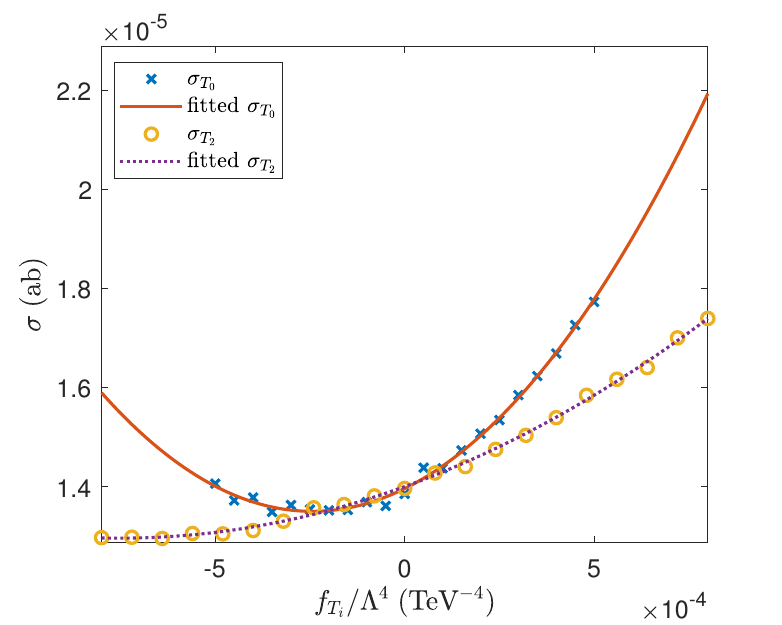}
\includegraphics[width=0.3\hsize]{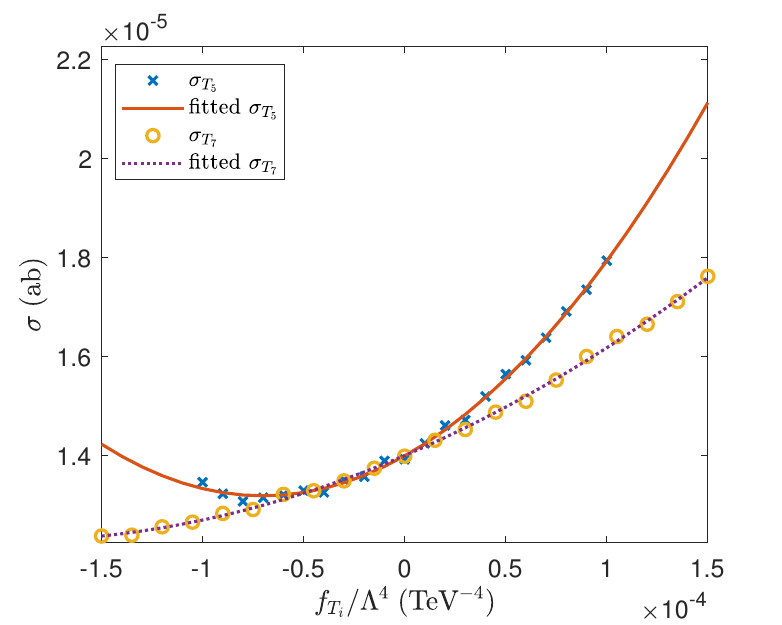}
\includegraphics[width=0.3\hsize]{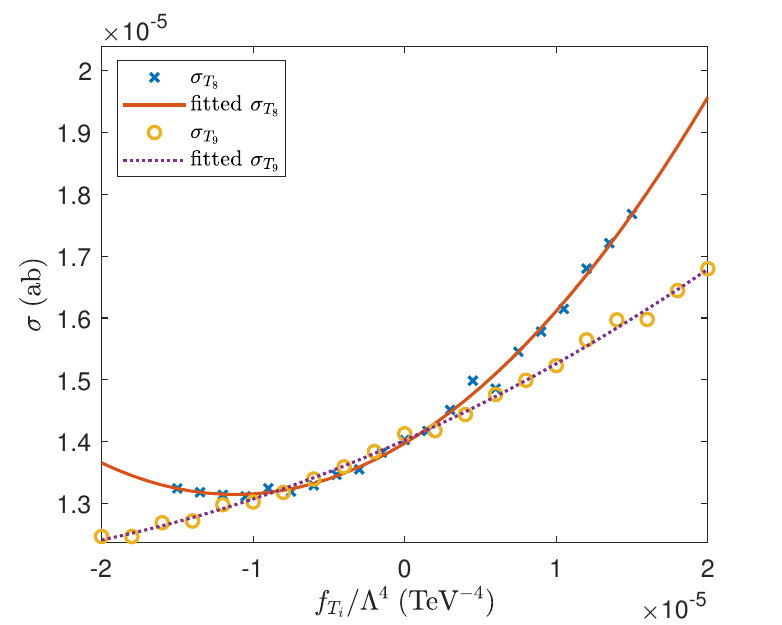}
\includegraphics[width=0.3\hsize]{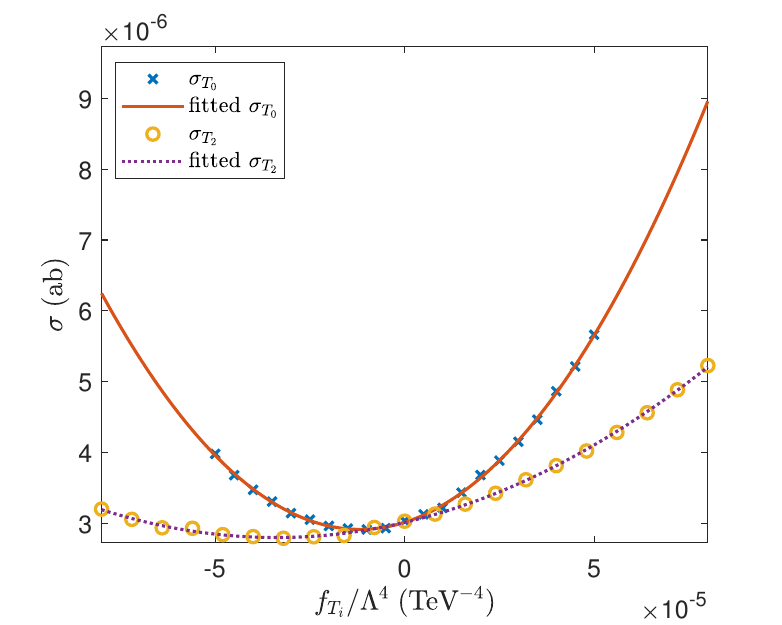}
\includegraphics[width=0.3\hsize]{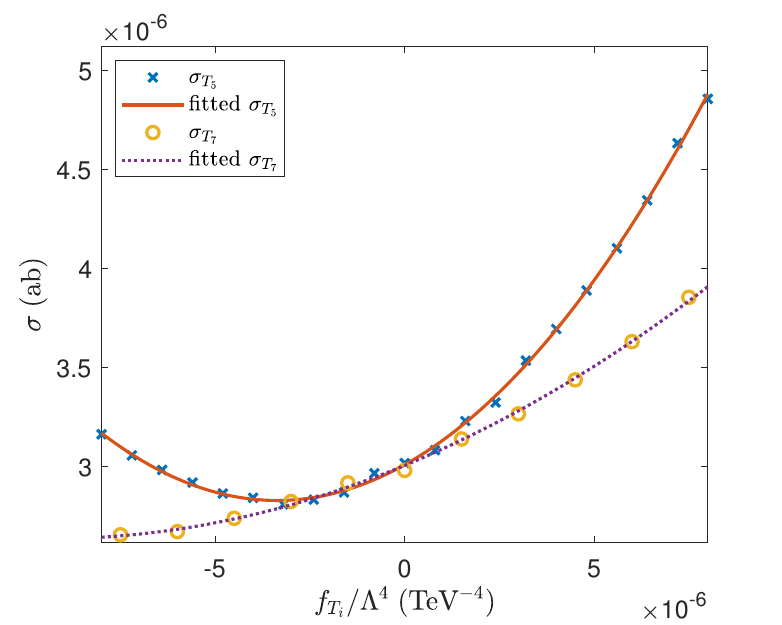}
\includegraphics[width=0.3\hsize]{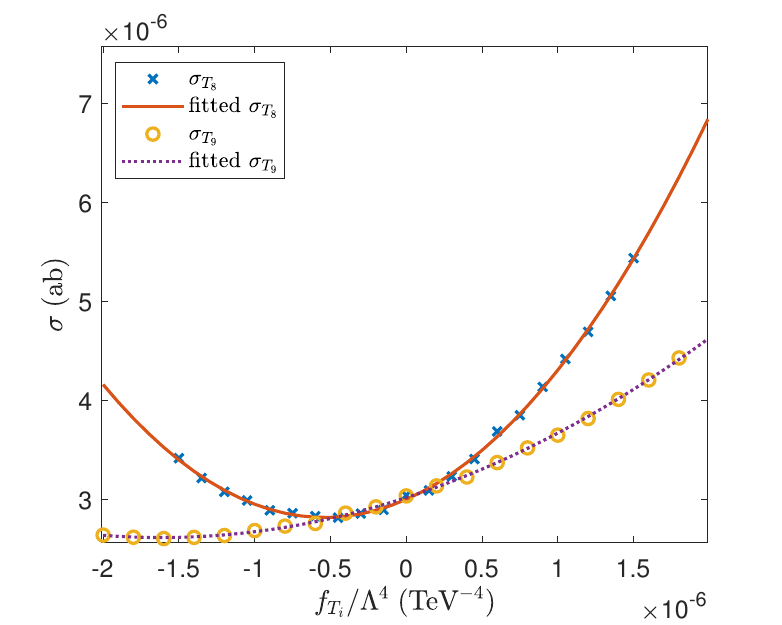}
\caption{\label{fig:crosssectionaqgc}
The cross-sections $\sigma$ after cuts compared with the fitted $\sigma$ according to Eq.~(\ref{eq.aqgccs}) of $f_{T_{i}}/\Lambda^4$ at different collision energies, $\sqrt{s} = 3\;{\rm TeV}$ (row $1$), $10\;{\rm TeV}$ (row $2$), $14\;{\rm TeV}$  (row $3$), and $30\;{\rm TeV}$  (row $4$).}
\end{center}
\end{figure*}

\begin{table}[htbp]
\centering
\begin{tabular}{c|c|c|c|c} 
\hline
 $\sqrt{s}$ & 3\;{\rm TeV} &10\;{\rm TeV} & $14\;{\rm TeV}$ & $30\;{\rm TeV}$ \\ 
\hline
$\left| f_{T_0}/\Lambda ^4\right|$ & $\leq 0.3$&$\leq 0.0015$ &$\leq 0.0005$& $\leq 0.00005$\\ 
\hline
$\left| f_{T_2}/\Lambda ^4\right|$ & $\leq 0.5$&$\leq 0.002$ & $\leq 0.0008$& $\leq 0.00008$\\
\hline
$\left| f_{T_5}/\Lambda ^4\right|$ & $\leq 0.06$ &$\leq 0.0003$ & $\leq 0.0001$& $\leq 0.000008$\\
\hline
$\left| f_{T_7}/\Lambda ^4\right|$ & $\leq 0.1$ & $\leq 0.0005$ &$\leq 0.00015$ &$\leq 0.000015$ \\
\hline
$\left| f_{T_8}\Lambda ^4\right|$ & $\leq 0.01$& $\leq 0.00005$ & $\leq 0.000015$ & $\leq 0.0000015$ \\ 
\hline
$\left| f_{T_9}/\Lambda ^4\right|$ & $\leq 0.016$& $\leq 0.00008$ & $\leq 0.00002$ & $\leq 0.000002$ \\  
\hline
\end{tabular}
\caption{The ranges of operator coefficients used in the scanning~(${\rm TeV}^{-4}$).}
\label{table:aqgccoefficientscan}
\end{table}
The test data are generated with the interference between the SM and NP considered and with one operator at a time.
We scan the parameters with the coefficients in the regions listed in Table~\ref{table:aqgccoefficientscan}.
After the events are generated, $10^5$ events are picked at each coefficient after the requirement of at least three photons in the final states.
The cross-section after cut can be written as,
\begin{equation}
\begin{split}
&\sigma(f_{T_i})=\sigma _{\rm SM} + f_{T_i}\sigma _{\rm int} + f^2_{T_i}\sigma _{\rm NP},
\end{split}
\label{eq.aqgccs}
\end{equation}
where $\sigma _{\rm SM}$, $\sigma _{\rm int}$ and $\sigma _{\rm NP}$ are parameters to be fitted representing the SM, interference and NP contributions.
In this subsection, we use a fixed $c'=200$, choosing $\hat{a}_{th}=0.35$ for the cases of $\sqrt{s}=3\;{\rm TeV}$, $10\;{\rm TeV}$ and $14\;{\rm TeV}$, and $\hat{a}_{th}=0.36$ for the case of $\sqrt{s}=30\;{\rm TeV}$, the cross-sections after cut and the fitted cross-sections are shown in Fig.~\ref{fig:crosssectionaqgc}.

\begin{table*}[htbp]
  \centering
  \begin{tabular}{c|c|c|c|c|c} 
  \hline
   &  &$3\;{\rm TeV} $&$10\;{\rm TeV}$ &$14\;{\rm TeV}$ &$30\;{\rm TeV} $\\
    &$\mathcal{S}_{stat}$& $1\;{\rm ab}^{-1}$&$10\;{\rm ab}^{-1}$&$10\;{\rm ab}^{-1}$ &$10\;{\rm ab}^{-1}$ \\
    &  &$(10^{-2}\;{\rm TeV^{-4}})$&$(10^{-4}\;{\rm TeV^{-4}})$&$(10^{-4}\;{\rm TeV^{-4}})$&$(10^{-5}\;{\rm TeV^{-4}})$ \\
  \hline
    & $2$ &$[-35.30, 13.12]$&$[-29.03, 11.28]$& $[-8.54, 3.67]$&$[-5.38, 3.01]$\\
  $f_{T_{0}}(f_{T_{1}})/\Lambda^4$& $3 $&$[-39.79, 17.62]$&$[-32.85, 15.10]$& $[-9.76, 4.89]$&$[-6.31, 3.95]$ \\
    &$ 5 $& $[-47.31, 25.14]$&$[-39.22, 21.47]$&$[-11.79, 6.92]$&$[-7.90, 5.53]$\\
  \hline
   & $2$ &$[-90.53, 21.33]$&$[-75.97, 17.78]$&$[-21.26, 6.21]$&$[-12.00, 5.22]$ \\
  $f_{T_{2}}/\Lambda^4$&$ 3$ &$[-98.78, 29.59]$&$[-82.88, 24.69]$&$[-23.57, 8.53]$&$[-13.78, 7.00]$ \\
    & $5$ &$[-113.06, 43.87]$&$[-94.83, 36.64]$&$[-27.52, 12.48]$&$[-16.83, 10.04]$ \\
  \hline
   & $2 $&$[-9.04, 2.44]$&$[-7.81, 2.07]$&$[-2.11, 0.70]$& $[-1.26, 0.58]$\\
  $f_{T_{5}}(f_{T_{6}})/\Lambda^4$&$ 3 $&$[-9.95, 3.35]$&$[-8.59,2.84]$&$[-2.36, 0.96]$&$[-1.45, 0.78]$ \\
    & $5$ &$[-11.51,4.92]$&$[-9.93, 4.18]$&$[-2.79, 1.38]$&$[-1.78, 1.11]$ \\
  \hline
  & $2$ &$[-23.68,3.76]$&$[-16.59, 3.08]$&$[-5.08,1.10]$& $[-2.84, 0.97]$\\
  $f_{T_{7}}/\Lambda^4$& $3$ &$[-25.26, 5.34]$&$[-17.85, 4.33]$&$[-5.52, 1.53]$&$[-3.20, 1.33]$ \\
    &$ 5$ &$[-28.07, 8.15]$&$[-20.06, 6.54]$&$[-6.28, 2.30]$& $[-3.82, 1.95]$ \\
  \hline
   &$2$ &$[-1.46, 0.39]$&$[-1.25, 0.33]$&$[-0.33,  0.11]$&$[-0.20, 0.09]$ \\
  $f_{T_{8}}/\Lambda^4$& $3$ &$[-1.61, 0.54]$&$[-1.38, 0.46]$&$[-0.37, 0.15]$&$[-0.23, 0.12]$ \\
    &$ 5$ &$[-1.86, 0.79]$&$[-1.59, 0.68]$&$[-0.44, 0.22]$& $[-0.28, 0.18]$\\
  \hline
   &$2$ &$[-3.92, 0.60]$&$[-2.78, 0.50]$&$[-0.94, 0.18]$&$[-0.48, 0.16]$ \\
  $f_{T_{9}}/\Lambda^4$& $3 $&$[-4.17, 0.85]$&$[-2.99, 0.71]$&$[-1.01, 0.25]$& $[-0.54, 0.22]$\\
    & $5$ &$[-4.63, 1.31]$&$[-3.35, 1.07]$&$[-1.14, 0.38]$&$[-0.64, 0.32]$\\
  \hline
  \end{tabular}
  \caption{In the ``conservative'' case, the projected sensitivities on the coefficients of the $O_{T_i}$ operators at the muon colliders for various c.m. energies and integrated luminosities.}
  \label{table:8}
  \end{table*}
    
 \begin{table}[htbp]
  \centering
  \begin{tabular}{c|c|c|c} 
  \hline
    &  &$14\;{\rm TeV}$ &$30\;{\rm TeV}$ \\
    &$\mathcal{S}_{stat}$& $20\;{\rm ab}^{-1}$&$90\;{\rm ab}^{-1}$ \\
    &  &$(10^{-5}\;{\rm TeV^{-4}})$&$(10^{-6}\;{\rm TeV^{-4}})$ \\
  \hline
    & $2$ &$[-77.26, 28.50]$&$[-37.51, 13.86]$ \\
  $f_{T_{0}}(f_{T_{1}})/\Lambda^4$& $3$ &$[-87.06, 38.30]$&$[-42.27, 18.63]$ \\
    & $5 $& $[-103.48, 54.72]$& $[-50.26, 26.61]$ \\
  \hline
   & $2$ &$[-197.39, 46.98]$&$[-90.17, 22.29]$ \\
  $f_{T_{2}}/\Lambda^4$&$ 3$ &$[-215.55, 65.15]$&$[-98.72, 30.84]$ \\
    &$ 5 $&$[-246.95, 96.54]$& $[-113.45, 45.58]$ \\
  \hline
   & $2 $&$[-19.41, 5.37]$&$[-9.28, 2.53]$ \\
  $f_{T_{5}}(f_{T_{6}})/\Lambda^4$&$ 3 $&$[-21.41, 7.37]$&$[-10.23,3.48]$ \\
    & $5 $& $[-24.83, 10.79]$&$[-11.85, 5.11]$ \\
  \hline
  & $2$ &$[-47.96, 8.15]$&$[-22.65, 3.91]$ \\
  $f_{T_{7}}/\Lambda^4$& $3$ &$[-51.34, 11.52]$&$[-24.27, 5.52]$ \\
    & $5 $&$[-57.33,17.52]$& $[-27.13, 8.39]$ \\
  \hline
   &$2 $&$[-3.08, 0.84]$&$[-1.48, 0.40]$ \\
  $f_{T_{8}}/\Lambda^4$&$ 3 $&$[-3.39, 1.15]$&$[-1.63, 0.55]$ \\
    &$5$ &$[-3.93, 1.69]$& $[-1.89, 0.81]$ \\
  \hline
   &$2 $&$[-8.92, 1.33]$&$[-3.86, 0.63]$ \\
  $f_{T_{9}}/\Lambda^4$&$ 3 $&$[-9.49, 1.89]$&$[-4.13, 0.89]$ \\
    & $5 $&$[-10.50, 2.90]$&$[-4.60, 1.36]$ \\
  \hline
  \end{tabular}
  \caption{Same as Table~\ref{table:8} but for the ``optimistic'' case.}
  \label{table:9}
  \end{table}

Using the designed luminosities for both the ``conservative'' and ``optimistic'' cases~\cite{Black:2022cth,Accettura:2023ked} and the fitted cross-sections, the expected constraints can be estimated using Eq.~(\ref{eq.ss}).
The results are listed in Tables~\ref{table:8} and \ref{table:9}.
It can be shown that the VQSN is effective in the search for aQGCs, the expected constraints generally outperform the ones obtained by a traditional event selection strategy~\cite{Yang:2020rjt}.

In the case of $l=10$, the number of $U$ gates is $64$ and the number of $CZ$ gates is $62$, the cc is estimated by average $r$ of the entire test dataset and $c'=200$ that $c'/r=564$, so the cc is $7.4\times 10^4$ for one test event.
If the (t-)KNN is used, the cc is $2^{17}d'=7.86\times 10^5$ for one test event, therefore the cc of VQSN is one order of magnitude smaller.

\subsection{Discussion on efficacy difference between (t-)KNN and (V)QSN}

In studies of several new physics scenarios, numerical results indicate that t-KNN achieves tighter constraints on coefficients compared to VQSN. 
Through testing, we identified that such difference comes from the fundamental distinction: t-KNN is intrinsically an algorithm considering only nearest neighbors in training dataset, whereas VQSN employs a weighted scoring over the entire training dataset.

While the VQSN method can technically be adapted to classify using only nearest neighbors~\cite{basheer2021quantumknearestneighborsalgorithm}, implementing this quantum variant demands additional gate operations. 
This not only fails to surpass the computational efficiency of classical t-KNN when processing classical data but also produces deeper circuits impractical for execution on current noisy quantum hardware. 
Consequently, we contend that a quantum algorithm achieving coefficient constraints on par with t-KNN~(within the same order of magnitude) while simultaneously reducing cc by an order of magnitude and remaining executable on current quantum hardware would hold significant value for high-luminosity collider experiments confronting exponentially growing datasets.

\begin{figure*}[htpb]
\begin{center}
\includegraphics[width=0.9\hsize]{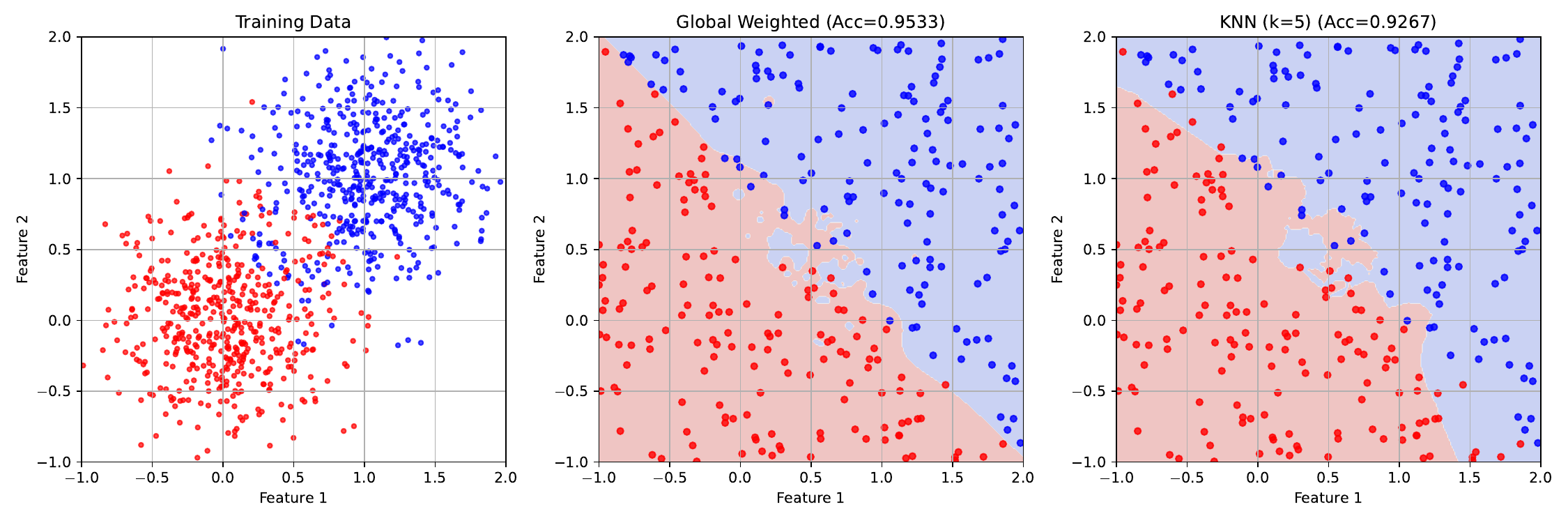}
\includegraphics[width=0.9\hsize]{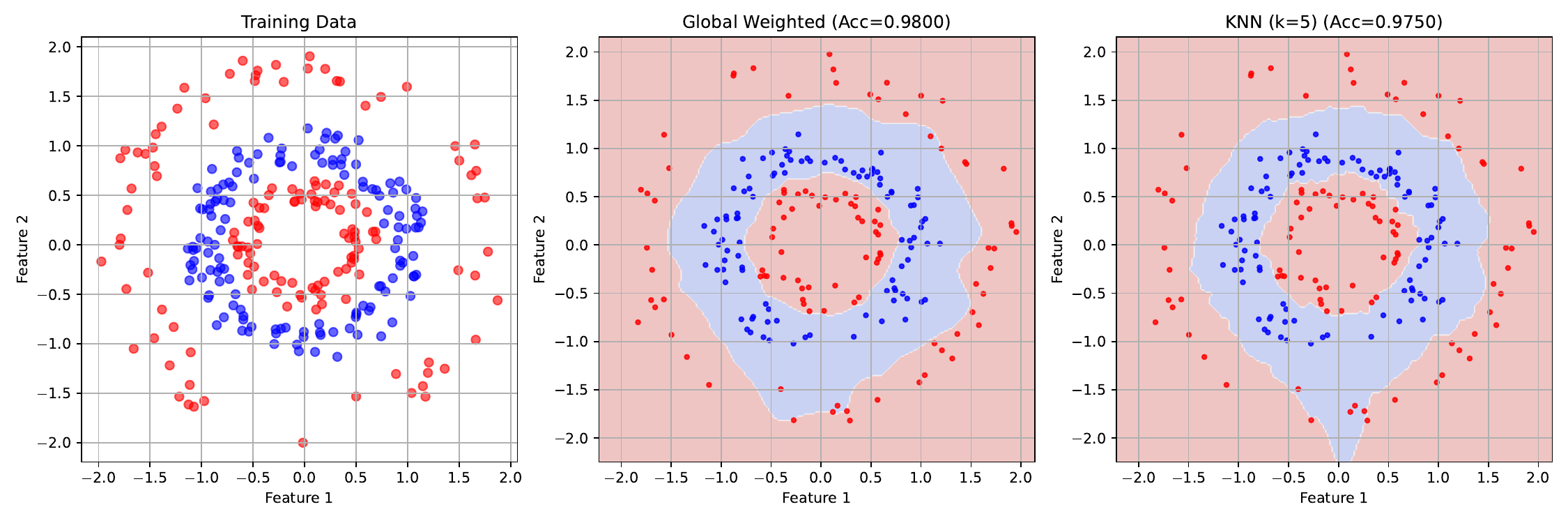}
\caption{\label{fig:qsnexample}
Two examples that global weighted can surpass the KNN.
The left panels are the points in training dataset with the labels $class_i=\pm 1$.
The middle panels are the points in test dataset and decision boundary calculated by the sign of $\sum _{i\in training} class_i/\left(v_{test}-v_i\right)^2$.
The right panels are the points in test dataset and decision boundary calculated by using KNN with $k=5$.
For the test set, blue points within red regions represent misclassifications, and vice versa. 
Accuracy~(denoted as Acc) is defined as the proportion of correctly classified data points in the test set.}
\end{center}
\end{figure*}
Furthermore, there is no inherent superiority between algorithms that consider only nearest neighbors and those utilizing weighted scoring across the entire training dataset. 
The relative efficacy depends fundamentally on the data distribution of the problem under study. 
To facilitate discussion, we restrict our analysis to classical algorithms: one employing distance inverse weighted scoring and another using nearest neighbor based unweighted voting~(KNN).
In Fig.~\ref{fig:qsnexample} we present two illustrative cases. 
For the upper row, points of both classes interpenetrate extensively. 
Within overlapping regions, some points are entirely surrounded by instances of the opposite class, causing KNN to perform worse than the global weighting approach. 
The bottom row in Fig.~\ref{fig:qsnexample} relates less to distribution characteristics and more to training dataset sampling. 
Consider the lowermost point in the training data, due to accidentally sparse sampling in its neighborhood, the locally focused KNN underperforms compared to the globally-aware weighting method.

Therefore, given VQSN's superior performance in cc and its potential to surpass t-KNN under specific conditions in efficacy, coupled with unexplored opportunities for feature space optimization, we conclude that VQSN represents a valuable algorithm for NP searches.

\section{Summary}

In this study, we present a (V)QSN algorithm for classification that is designed to operate on a quantum computer.
A numerical example running on the \textit{tianyan176-2} of the \textit{Tianyan quantum computing cloud platform} is established to show that the VQSN is suitable for a general proposed classification.
This algorithm is analogous to the KNN algorithm. 
We employ the (V)QSN and the (t-)KNN with the objective of identifying signals of NP in different processes.

The VQSN is used in the phenomenological study of the gQGCs in the process $\mu^+\mu^-\to jj\nu\bar{\nu}$ at a $\sqrt{s}=3$ TeV muon collider.
The VQSN event selection strategy is tested on the \textit{tianyan176-2}.
The expected coefficient constraints are obtained and found to be at the same order of magnitude as the traditional event selection strategy in Ref.~\cite{Yang:2023gos}, indicating that the VQSN is an effective method and possesses noise tolerance.

The gQGCs at the LHC are also studied, using the processes $pp\to jj \ell^+\ell^-\nu\bar{\nu}$, $pp\to j \ell^+\ell^-\nu\bar{\nu}$ and $pp\to \gamma j \ell^+\ell^-\nu\bar{\nu}$. 
These processes can be concluded to be sensitive to the gQGCs.
The constraints on the coefficient are of a similar magnitude to those obtained in the $pp\to \gamma\gamma$ process~\cite{Ellis:2018cos}. 
Consequently, they can serve as a valuable supplement to the constraints on the gQGCs. 
Moreover, the $pp\to jj \ell^+\ell^-\nu\bar{\nu}$ process contains a VBS process which is expected to be more significant at higher energies~\cite{Yang:2020rjt}, therefore is expected to be more important at future colliders such as the HE-LHC and FCC-hh~\cite{FCC:2018bvk,Benedikt:2022kan}. 

Another application is the contribution of the aQGCs in the tri-photon processes at muon colliders.
In this case, only the VQSN is studied.
The expected constraints are obtained, and found to be more strengthened than the ones obtained using a traditional event selection strategy~\cite{Yang:2020rjt}.

In terms of cc, the VQSN is capable of achieving the same order of magnitude of the discriminative power to that of the KNN, while requiring one order of magnitude less cc in the three applications.
It is important to note that the datasets used in this study are classical, i.e., even when working with classical data, the VQSN can demonstrate its superiority in efficiency in the sense of cc. 
When keeping the amount of test data much larger than the amount of training data, the larger training dataset and the larger feature space, the more efficient the VQSN is, the significance of VQSN is likely to increase as the processes under consideration become more intricate. 

This study has not addressed further improvements in encoding the datasets in optimistic ways~\cite{Havlicek:2018nqz,Lloyd:2020eeh} which are suitable for cooperation with (V)QSN.
Furthermore, it would be beneficial to investigate the simulation of the collision process on a quantum computer.
Reading results directly from a quantum computer is resource intensive, QSN can process data within the quantum computer, avoiding this consumption.
In particular, it is plausible that the training dataset originates from simulations on a quantum computer, while the test dataset is derived from classical data obtained from a collider. 
This scenario is well-suited to the QSN.

In conclusion, QSN and VQSN are suitable for the NP phenomenological study.

\acknowledgments
We sincerely thank the \textit{Tianyan quantum computing cloud platform} of China Telecom Quantum Information Technology Group Co., LTD. for providing quantum computer machine time and technical support.
This work was supported in part by the National Natural Science Foundation of China under Grants Nos. 12575106, 11875157 and 12147214, and the Natural Science Foundation of the Liaoning Scientific Committee No.~LJKZ0978.

\bibliography{qsvn}

@article{Farzan:2017xzy,
    author = "Farzan, Y. and Tortola, M.",
    title = "{Neutrino oscillations and Non-Standard Interactions}",
    eprint = "1710.09360",
    archivePrefix = "arXiv",
    primaryClass = "hep-ph",
    doi = "10.3389/fphy.2018.00010",
    journal = "Front. in Phys.",
    volume = "6",
    pages = "10",
    year = "2018"
}

@proceedings{Proceedings:2019qno,
    author = "Bhupal Dev, P. S. and others",
    title = "{Neutrino Non-Standard Interactions: A Status Report}",
    eprint = "1907.00991",
    archivePrefix = "arXiv",
    primaryClass = "hep-ph",
    reportNumber = "FERMILAB-CONF-19-299-T",
    doi = "10.21468/SciPostPhysProc.2.001",
    volume = "2",
    pages = "001",
    year = "2019"
}

@article{Arguelles:2022tki,
    author = {Arg\"uelles, C. A. and others},
    title = "{Snowmass white paper: beyond the standard model effects on neutrino flavor: Submitted to the proceedings of the US community study on the future of particle physics (Snowmass 2021)}",
    eprint = "2203.10811",
    archivePrefix = "arXiv",
    primaryClass = "hep-ph",
    doi = "10.1140/epjc/s10052-022-11049-7",
    journal = "Eur. Phys. J. C",
    volume = "83",
    number = "1",
    pages = "15",
    year = "2023"
}

@article{CDF:2022hxs,
    author = "Aaltonen, T. and others",
    collaboration = "CDF",
    title = "{High-precision measurement of the $W$ boson mass with the CDF II detector}",
    reportNumber = "FERMILAB-PUB-22-254-PPD",
    doi = "10.1126/science.abk1781",
    journal = "Science",
    volume = "376",
    number = "6589",
    pages = "170--176",
    year = "2022"
}

@article{deBlas:2022hdk,
    author = "de Blas, J. and Pierini, M. and Reina, L. and Silvestrini, L.",
    title = "{Impact of the Recent Measurements of the Top-Quark and W-Boson Masses on Electroweak Precision Fits}",
    eprint = "2204.04204",
    archivePrefix = "arXiv",
    primaryClass = "hep-ph",
    doi = "10.1103/PhysRevLett.129.271801",
    journal = "Phys. Rev. Lett.",
    volume = "129",
    number = "27",
    pages = "271801",
    year = "2022"
}

@article{Muong-2:2023cdq,
    author = "Aguillard, D. P. and others",
    collaboration = "Muon g-2",
    title = "{Measurement of the Positive Muon Anomalous Magnetic Moment to 0.20~ppm}",
    eprint = "2308.06230",
    archivePrefix = "arXiv",
    primaryClass = "hep-ex",
    reportNumber = "FERMILAB-PUB-23-385-AD-CSAID-PPD",
    doi = "10.1103/PhysRevLett.131.161802",
    journal = "Phys. Rev. Lett.",
    volume = "131",
    number = "16",
    pages = "161802",
    year = "2023"
}

@article{Muong-2:2021ojo,
    author = "Abi, B. and others",
    collaboration = "Muon g-2",
    title = "{Measurement of the Positive Muon Anomalous Magnetic Moment to 0.46 ppm}",
    eprint = "2104.03281",
    archivePrefix = "arXiv",
    primaryClass = "hep-ex",
    reportNumber = "FERMILAB-PUB-21-132-E",
    doi = "10.1103/PhysRevLett.126.141801",
    journal = "Phys. Rev. Lett.",
    volume = "126",
    number = "14",
    pages = "141801",
    year = "2021"
}

@article{Aoyama:2020ynm,
    author = "Aoyama, T. and others",
    title = "{The anomalous magnetic moment of the muon in the Standard Model}",
    eprint = "2006.04822",
    archivePrefix = "arXiv",
    primaryClass = "hep-ph",
    reportNumber = "FERMILAB-PUB-20-207-T, INT-PUB-20-021, KEK Preprint 2020-5,
  MITP/20-028, KEK Preprint 2020-5, MITP/20-028, CERN-TH-2020-075, IFT-UAM/CSIC-20-74, LMU-ASC 18/20, LTH 1234,
  LU TP 20-20, LTH 1234, LU TP 20-20, MAN/HEP/2020/003, PSI-PR-20-06, UWThPh 2020-14, ZU-TH 18/20",
    doi = "10.1016/j.physrep.2020.07.006",
    journal = "Phys. Rept.",
    volume = "887",
    pages = "1--166",
    year = "2020"
}

@inproceedings{Crivellin:2023zui,
    author = "Crivellin, Andreas and Mellado, Bruce",
    title = "{Anomalies in Particle Physics}",
    eprint = "2309.03870",
    archivePrefix = "arXiv",
    primaryClass = "hep-ph",
    reportNumber = "PSI-PR-23-34, ZU-TH 53/23, ICPP-73",
    month = "9",
    year = "2023"
}

@article{Ellis:2012zz,
    author = "Ellis, John",
    title = "{Outstanding questions: Physics beyond the Standard Model}",
    doi = "10.1098/rsta.2011.0452",
    journal = "Phil. Trans. Roy. Soc. Lond. A",
    volume = "370",
    pages = "818--830",
    year = "2012"
}

@article{Radovic:2018dip,
    author = "Radovic, Alexander and Williams, Mike and Rousseau, David and Kagan, Michael and Bonacorsi, Daniele and Himmel, Alexander and Aurisano, Adam and Terao, Kazuhiro and Wongjirad, Taritree",
    title = "{Machine learning at the energy and intensity frontiers of particle physics}",
    reportNumber = "FERMILAB-PUB-18-436-ND",
    doi = "10.1038/s41586-018-0361-2",
    journal = "Nature",
    volume = "560",
    number = "7716",
    pages = "41--48",
    year = "2018"
}

@article{Baldi:2014kfa,
    author = "Baldi, Pierre and Sadowski, Peter and Whiteson, Daniel",
    title = "{Searching for Exotic Particles in High-Energy Physics with Deep Learning}",
    eprint = "1402.4735",
    archivePrefix = "arXiv",
    primaryClass = "hep-ph",
    doi = "10.1038/ncomms5308",
    journal = "Nature Commun.",
    volume = "5",
    pages = "4308",
    year = "2014"
}

@article{Ren:2017ymm,
    author = "Ren, Jie and Wu, Lei and Yang, Jin Min and Zhao, Jun",
    title = "{Exploring supersymmetry with machine learning}",
    eprint = "1708.06615",
    archivePrefix = "arXiv",
    primaryClass = "hep-ph",
    doi = "10.1016/j.nuclphysb.2019.114613",
    journal = "Nucl. Phys. B",
    volume = "943",
    pages = "114613",
    year = "2019"
}

@article{Abdughani:2018wrw,
    author = "Abdughani, Murat and Ren, Jie and Wu, Lei and Yang, Jin Min",
    title = "{Probing stop pair production at the LHC with graph neural networks}",
    eprint = "1807.09088",
    archivePrefix = "arXiv",
    primaryClass = "hep-ph",
    doi = "10.1007/JHEP08(2019)055",
    journal = "JHEP",
    volume = "08",
    pages = "055",
    year = "2019"
}

@article{Ren:2019xhp,
    author = "Ren, Jie and Wu, Lei and Yang, Jin Min",
    title = "{Unveiling CP property of top-Higgs coupling with graph neural networks at the LHC}",
    eprint = "1901.05627",
    archivePrefix = "arXiv",
    primaryClass = "hep-ph",
    doi = "10.1016/j.physletb.2020.135198",
    journal = "Phys. Lett. B",
    volume = "802",
    pages = "135198",
    year = "2020"
}

@article{Letizia:2022xbe,
    author = "Letizia, Marco and Losapio, Gianvito and Rando, Marco and Grosso, Gaia and Wulzer, Andrea and Pierini, Maurizio and Zanetti, Marco and Rosasco, Lorenzo",
    title = "{Learning new physics efficiently with nonparametric methods}",
    eprint = "2204.02317",
    archivePrefix = "arXiv",
    primaryClass = "hep-ph",
    doi = "10.1140/epjc/s10052-022-10830-y",
    journal = "Eur. Phys. J. C",
    volume = "82",
    number = "10",
    pages = "879",
    year = "2022"
}

@article{DAgnolo:2019vbw,
    author = "D'Agnolo, Raffaele Tito and Grosso, Gaia and Pierini, Maurizio and Wulzer, Andrea and Zanetti, Marco",
    title = "{Learning multivariate new physics}",
    eprint = "1912.12155",
    archivePrefix = "arXiv",
    primaryClass = "hep-ph",
    reportNumber = "CERN-TH-2019-226",
    doi = "10.1140/epjc/s10052-021-08853-y",
    journal = "Eur. Phys. J. C",
    volume = "81",
    number = "1",
    pages = "89",
    year = "2021"
}

@article{DAgnolo:2018cun,
    author = "D'Agnolo, Raffaele Tito and Wulzer, Andrea",
    title = "{Learning New Physics from a Machine}",
    eprint = "1806.02350",
    archivePrefix = "arXiv",
    primaryClass = "hep-ph",
    doi = "10.1103/PhysRevD.99.015014",
    journal = "Phys. Rev. D",
    volume = "99",
    number = "1",
    pages = "015014",
    year = "2019"
}

@article{DeSimone:2018efk,
    author = "De Simone, Andrea and Jacques, Thomas",
    title = "{Guiding New Physics Searches with Unsupervised Learning}",
    eprint = "1807.06038",
    archivePrefix = "arXiv",
    primaryClass = "hep-ph",
    reportNumber = "SISSA 27/2018/FISI",
    doi = "10.1140/epjc/s10052-019-6787-3",
    journal = "Eur. Phys. J. C",
    volume = "79",
    number = "4",
    pages = "289",
    year = "2019"
}

@article{MdAli:2020yzb,
    author = "Md Ali, Mohd Adli and Badrud'din, Nu'man and Abdullah, Hafidzul and Kemi, Faiz",
    title = "{Alternate methods for anomaly detection in high-energy physics via semi-supervised learning}",
    doi = "10.1142/S0217751X20501316",
    journal = "Int. J. Mod. Phys. A",
    volume = "35",
    number = "23",
    pages = "2050131",
    year = "2020"
}

@article{Fol:2020tva,
    author = "Fol, E. and Tom\'as, R. and Coello de Portugal, J. and Franchetti, G.",
    title = "{Detection of faulty beam position monitors using unsupervised learning}",
    doi = "10.1103/PhysRevAccelBeams.23.102805",
    journal = "Phys. Rev. Accel. Beams",
    volume = "23",
    number = "10",
    pages = "102805",
    year = "2020"
}

@article{Kasieczka:2021xcg,
    author = "Kasieczka, Gregor and others",
    title = "{The LHC Olympics 2020 a community challenge for anomaly detection in high energy physics}",
    eprint = "2101.08320",
    archivePrefix = "arXiv",
    primaryClass = "hep-ph",
    doi = "10.1088/1361-6633/ac36b9",
    journal = "Rept. Prog. Phys.",
    volume = "84",
    number = "12",
    pages = "124201",
    year = "2021"
}

@article{Dong:2023nir,
    author = "Dong, Yi-Fei and Mao, Ying-Chen and Yang, Ji-Chong",
    title = "{Searching for anomalous quartic gauge couplings at muon colliders using principal component analysis}",
    eprint = "2304.01505",
    archivePrefix = "arXiv",
    primaryClass = "hep-ph",
    doi = "10.1140/epjc/s10052-023-11719-0",
    journal = "Eur. Phys. J. C",
    volume = "83",
    number = "7",
    pages = "555",
    year = "2023"
}

@article{CrispimRomao:2020ucc,
    author = "Crispim Rom\~ao, M. and Castro, N. F. and Pedro, R.",
    title = "{Finding New Physics without learning about it: Anomaly Detection as a tool for Searches at Colliders}",
    eprint = "2006.05432",
    archivePrefix = "arXiv",
    primaryClass = "hep-ph",
    doi = "10.1140/epjc/s10052-021-09813-2",
    journal = "Eur. Phys. J. C",
    volume = "81",
    number = "1",
    pages = "27",
    year = "2021",
    note = "[Erratum: Eur.Phys.J.C 81, 1020 (2021)]"
}

@article{vanBeekveld:2020txa,
    author = "van Beekveld, Melissa and Caron, Sascha and Hendriks, Luc and Jackson, Paul and Leinweber, Adam and Otten, Sydney and Patrick, Riley and Ruiz De Austri, Roberto and Santoni, Marco and White, Martin",
    title = "{Combining outlier analysis algorithms to identify new physics at the LHC}",
    eprint = "2010.07940",
    archivePrefix = "arXiv",
    primaryClass = "hep-ph",
    doi = "10.1007/JHEP09(2021)024",
    journal = "JHEP",
    volume = "09",
    pages = "024",
    year = "2021"
}

@article{Kuusela:2011aa,
    author = "Kuusela, Mikael and Vatanen, Tommi and Malmi, Eric and Raiko, Tapani and Aaltonen, Timo and Nagai, Yoshikazu",
    editor = "Teodorescu, Liliana and Britton, David and Glover, Nigel and Heinrich, Gudrun and Lauret, Jerome and Naumann, Axel and Speer, Thomas and Teixeira-Dias, Pedro",
    title = "{Semi-Supervised Anomaly Detection - Towards Model-Independent Searches of New Physics}",
    eprint = "1112.3329",
    archivePrefix = "arXiv",
    primaryClass = "physics.data-an",
    doi = "10.1088/1742-6596/368/1/012032",
    journal = "J. Phys. Conf. Ser.",
    volume = "368",
    pages = "012032",
    year = "2012"
}

@article{Weinberg:1979sa,
    author = "Weinberg, Steven",
    title = "{Baryon and Lepton Nonconserving Processes}",
    reportNumber = "HUTP-79-A050",
    doi = "10.1103/PhysRevLett.43.1566",
    journal = "Phys. Rev. Lett.",
    volume = "43",
    pages = "1566--1570",
    year = "1979"
}

@article{Grzadkowski:2010es,
    author = "Grzadkowski, B. and Iskrzynski, M. and Misiak, M. and Rosiek, J.",
    title = "{Dimension-Six Terms in the Standard Model Lagrangian}",
    eprint = "1008.4884",
    archivePrefix = "arXiv",
    primaryClass = "hep-ph",
    reportNumber = "IFT-9-2010, TTP10-35",
    doi = "10.1007/JHEP10(2010)085",
    journal = "JHEP",
    volume = "10",
    pages = "085",
    year = "2010"
}

@article{Willenbrock:2014bja,
    author = "Willenbrock, Scott and Zhang, Cen",
    title = "{Effective Field Theory Beyond the Standard Model}",
    eprint = "1401.0470",
    archivePrefix = "arXiv",
    primaryClass = "hep-ph",
    reportNumber = "CP3-14-02",
    doi = "10.1146/annurev-nucl-102313-025623",
    journal = "Ann. Rev. Nucl. Part. Sci.",
    volume = "64",
    pages = "83--100",
    year = "2014"
}

@article{Masso:2014xra,
    author = "Masso, Eduard",
    title = "{An Effective Guide to Beyond the Standard Model Physics}",
    eprint = "1406.6376",
    archivePrefix = "arXiv",
    primaryClass = "hep-ph",
    doi = "10.1007/JHEP10(2014)128",
    journal = "JHEP",
    volume = "10",
    pages = "128",
    year = "2014"
}

@article{Green:2016trm,
    author = "Green, Daniel R. and Meade, Patrick and Pleier, Marc-Andre",
    title = "{Multiboson interactions at the LHC}",
    eprint = "1610.07572",
    archivePrefix = "arXiv",
    primaryClass = "hep-ex",
    reportNumber = "FERMILAB-PUB-16-507-CMS, BNL-114542-2017-JA",
    doi = "10.1103/RevModPhys.89.035008",
    journal = "Rev. Mod. Phys.",
    volume = "89",
    number = "3",
    pages = "035008",
    year = "2017"
}

@article{Zhang:2020jyn,
    author = "Zhang, Cen and Zhou, Shuang-Yong",
    title = "{Convex Geometry Perspective on the (Standard Model) Effective Field Theory Space}",
    eprint = "2005.03047",
    archivePrefix = "arXiv",
    primaryClass = "hep-ph",
    reportNumber = "USTC-ICTS/PCFT-20-14",
    doi = "10.1103/PhysRevLett.125.201601",
    journal = "Phys. Rev. Lett.",
    volume = "125",
    number = "20",
    pages = "201601",
    year = "2020"
}

@article{Murphy:2020rsh,
    author = "Murphy, Christopher W.",
    title = "{Dimension-8 operators in the Standard Model Eective Field Theory}",
    eprint = "2005.00059",
    archivePrefix = "arXiv",
    primaryClass = "hep-ph",
    doi = "10.1007/JHEP10(2020)174",
    journal = "JHEP",
    volume = "10",
    pages = "174",
    year = "2020"
}

@article{Li:2020gnx,
    author = "Li, Hao-Lin and Ren, Zhe and Shu, Jing and Xiao, Ming-Lei and Yu, Jiang-Hao and Zheng, Yu-Hui",
    title = "{Complete set of dimension-eight operators in the standard model effective field theory}",
    eprint = "2005.00008",
    archivePrefix = "arXiv",
    primaryClass = "hep-ph",
    doi = "10.1103/PhysRevD.104.015026",
    journal = "Phys. Rev. D",
    volume = "104",
    number = "1",
    pages = "015026",
    year = "2021"
}

@article{Anders:2018oin,
    author = "Anders, C. F. and others",
    title = "{Vector boson scattering: Recent experimental and theory developments}",
    eprint = "1801.04203",
    archivePrefix = "arXiv",
    primaryClass = "hep-ph",
    reportNumber = "VBSCAN-PUB-01-17, FERMILAB-CONF-18-021-PPD, VBSCan-PUB-01-17",
    doi = "10.1016/j.revip.2018.11.001",
    journal = "Rev. Phys.",
    volume = "3",
    pages = "44--63",
    year = "2018"
}

@article{Henning:2015alf,
    author = "Henning, Brian and Lu, Xiaochuan and Melia, Tom and Murayama, Hitoshi",
    title = "{2, 84, 30, 993, 560, 15456, 11962, 261485, ...: Higher dimension operators in the SM EFT}",
    eprint = "1512.03433",
    archivePrefix = "arXiv",
    primaryClass = "hep-ph",
    reportNumber = "UCB-PTH-15-14, IPMU15-0207",
    doi = "10.1007/JHEP08(2017)016",
    journal = "JHEP",
    volume = "08",
    pages = "016",
    year = "2017",
    note = "[Erratum: JHEP 09, 019 (2019)]"
}

@article{qml2,
    author = {Maria Schuld and Ilya Sinayskiy and Francesco Petruccione},
    title = {An introduction to quantum machine learning},
    journal = {Contemporary Physics},
    volume = {56},
    number = {2},
    pages = {172-185},
    year  = {2015},
    publisher = {Taylor & Francis},
    doi = {10.1080/00107514.2014.964942},
}

@inproceedings{Garcia:2022cqq,
    author = "Garc\'\i{}a, David Peral and Cruz-Benito, Juan and Garc\'\i{}a-Pe\~nalvo, Francisco Jos\'e",
    title = "{Systematic Literature Review: Quantum Machine Learning and its applications}",
    eprint = "2201.04093",
    archivePrefix = "arXiv",
    primaryClass = "quant-ph",
    month = "1",
    year = "2022"
}

@article{Zhang:2018shp,
    author = "Zhang, Cen and Zhou, Shuang-Yong",
    title = "{Positivity bounds on vector boson scattering at the LHC}",
    eprint = "1808.00010",
    archivePrefix = "arXiv",
    primaryClass = "hep-ph",
    reportNumber = "USTC-ICTS-18-13",
    doi = "10.1103/PhysRevD.100.095003",
    journal = "Phys. Rev. D",
    volume = "100",
    number = "9",
    pages = "095003",
    year = "2019"
}

@article{Bi:2019phv,
    author = "Bi, Qi and Zhang, Cen and Zhou, Shuang-Yong",
    title = "{Positivity constraints on aQGC: carving out the physical parameter space}",
    eprint = "1902.08977",
    archivePrefix = "arXiv",
    primaryClass = "hep-ph",
    reportNumber = "USTC-ICTS-19-01",
    doi = "10.1007/JHEP06(2019)137",
    journal = "JHEP",
    volume = "06",
    pages = "137",
    year = "2019"
}

@article{ATLAS:2017vqm,
    author = "Aaboud, Morad and others",
    collaboration = "ATLAS",
    title = "{Studies of $Z\gamma$ production in association with a high-mass dijet system in $pp$ collisions at $\sqrt{s}=$ 8 TeV with the ATLAS detector}",
    eprint = "1705.01966",
    archivePrefix = "arXiv",
    primaryClass = "hep-ex",
    reportNumber = "CERN-EP-2017-046",
    doi = "10.1007/JHEP07(2017)107",
    journal = "JHEP",
    volume = "07",
    pages = "107",
    year = "2017"
}

@article{CMS:2017rin,
    author = "Khachatryan, Vardan and others",
    collaboration = "CMS",
    title = "{Measurement of the cross section for electroweak production of Z$\gamma$ in association with two jets and constraints on anomalous quartic gauge couplings in proton\textendash{}proton collisions at $\sqrt{s} = 8$ TeV}",
    eprint = "1702.03025",
    archivePrefix = "arXiv",
    primaryClass = "hep-ex",
    reportNumber = "CMS-SMP-14-018, CERN-EP-2016-308",
    doi = "10.1016/j.physletb.2017.04.071",
    journal = "Phys. Lett. B",
    volume = "770",
    pages = "380--402",
    year = "2017"
}

@article{CMS:2020ioi,
    author = "Sirunyan, Albert M and others",
    collaboration = "CMS",
    title = "{Measurement of the cross section for electroweak production of a Z boson, a photon and two jets in proton-proton collisions at $\sqrt{s} =$ 13 TeV and constraints on anomalous quartic couplings}",
    eprint = "2002.09902",
    archivePrefix = "arXiv",
    primaryClass = "hep-ex",
    reportNumber = "CMS-SMP-18-007, CERN-EP-2020-007",
    doi = "10.1007/JHEP06(2020)076",
    journal = "JHEP",
    volume = "06",
    pages = "076",
    year = "2020"
}

@article{CMS:2016gct,
    author = "Khachatryan, Vardan and others",
    collaboration = "CMS",
    title = "{Measurement of electroweak-induced production of W$\gamma$ with two jets in pp collisions at $ \sqrt{s}=8 $ TeV and constraints on anomalous quartic gauge couplings}",
    eprint = "1612.09256",
    archivePrefix = "arXiv",
    primaryClass = "hep-ex",
    reportNumber = "CMS-SMP-14-011, CERN-EP-2016-289",
    doi = "10.1007/JHEP06(2017)106",
    journal = "JHEP",
    volume = "06",
    pages = "106",
    year = "2017"
}

@article{CMS:2017zmo,
    author = "Sirunyan, Albert M and others",
    collaboration = "CMS",
    title = "{Measurement of vector boson scattering and constraints on anomalous quartic couplings from events with four leptons and two jets in proton\textendash{}proton collisions at $\sqrt{s}=$ 13 TeV}",
    eprint = "1708.02812",
    archivePrefix = "arXiv",
    primaryClass = "hep-ex",
    reportNumber = "CMS-SMP-17-006, CERN-EP-2017-177",
    doi = "10.1016/j.physletb.2017.10.020",
    journal = "Phys. Lett. B",
    volume = "774",
    pages = "682--705",
    year = "2017"
}

@article{CMS:2018ccg,
    author = "Sirunyan, Albert M. and others",
    collaboration = "CMS",
    title = "{Measurement of differential cross sections for Z boson pair production in association with jets at $\sqrt{s} =$ 8 and 13 TeV}",
    eprint = "1806.11073",
    archivePrefix = "arXiv",
    primaryClass = "hep-ex",
    reportNumber = "CMS-SMP-17-005, CERN-EP-2018-161",
    doi = "10.1016/j.physletb.2018.11.007",
    journal = "Phys. Lett. B",
    volume = "789",
    pages = "19--44",
    year = "2019"
}

@article{ATLAS:2018mxa,
    author = "Aaboud, Morad and others",
    collaboration = "ATLAS",
    title = "{Observation of electroweak $W^{\pm}Z$ boson pair production in association with two jets in $pp$ collisions at $\sqrt{s} =$ 13 TeV with the ATLAS detector}",
    eprint = "1812.09740",
    archivePrefix = "arXiv",
    primaryClass = "hep-ex",
    reportNumber = "CERN-EP-2018-286",
    doi = "10.1016/j.physletb.2019.05.012",
    journal = "Phys. Lett. B",
    volume = "793",
    pages = "469--492",
    year = "2019"
}

@article{CMS:2019uys,
    author = "Sirunyan, Albert M and others",
    collaboration = "CMS",
    title = "{Measurement of electroweak WZ boson production and search for new physics in WZ + two jets events in pp collisions at $\sqrt{s} =$ 13TeV}",
    eprint = "1901.04060",
    archivePrefix = "arXiv",
    primaryClass = "hep-ex",
    reportNumber = "CMS-SMP-18-001, CERN-EP-2018-333",
    doi = "10.1016/j.physletb.2019.05.042",
    journal = "Phys. Lett. B",
    volume = "795",
    pages = "281--307",
    year = "2019"
}

@article{CMS:2016rtz,
    author = "Khachatryan, Vardan and others",
    collaboration = "CMS",
    title = "{Evidence for exclusive $\gamma\gamma \to W^+ W^-$ production and constraints on anomalous quartic gauge couplings in $pp$ collisions at $ \sqrt{s}=7 $ and 8 TeV}",
    eprint = "1604.04464",
    archivePrefix = "arXiv",
    primaryClass = "hep-ex",
    reportNumber = "CMS-FSQ-13-008, CERN-EP-2016-073",
    doi = "10.1007/JHEP08(2016)119",
    journal = "JHEP",
    volume = "08",
    pages = "119",
    year = "2016"
}

@article{CMS:2017fhs,
    author = "Sirunyan, Albert M and others",
    collaboration = "CMS",
    title = "{Observation of electroweak production of same-sign W boson pairs in the two jet and two same-sign lepton final state in proton-proton collisions at $\sqrt{s} = $ 13 TeV}",
    eprint = "1709.05822",
    archivePrefix = "arXiv",
    primaryClass = "hep-ex",
    reportNumber = "CMS-SMP-17-004, CERN-EP-2017-232",
    doi = "10.1103/PhysRevLett.120.081801",
    journal = "Phys. Rev. Lett.",
    volume = "120",
    number = "8",
    pages = "081801",
    year = "2018"
}

@inproceedings{Ellis:2023ucy,
    author = "Ellis, John and He, Hong-Jian and Xiao, Rui-Qing",
    title = "{Probing Neutral Triple Gauge Couplings with $Z^* \gamma\, (\nu \bar \nu \gamma)$ Production at Hadron Colliders}",
    eprint = "2308.16887",
    archivePrefix = "arXiv",
    primaryClass = "hep-ph",
    reportNumber = "KCL-PH-TH/2023-39, CERN-TH-2023-129",
    month = "8",
    year = "2023"
}

@article{Spor:2022hhn,
    author = "Spor, S.",
    title = "{Probe of the anomalous neutral triple gauge couplings in photon-induced collision at future muon colliders}",
    eprint = "2207.11585",
    archivePrefix = "arXiv",
    primaryClass = "hep-ph",
    doi = "10.1016/j.nuclphysb.2023.116198",
    journal = "Nucl. Phys. B",
    volume = "991",
    pages = "116198",
    year = "2023"
}

@article{Spor:2022zob,
    author = {Spor, S. and Gurkanli, E. and K\"oksal, M.},
    title = "{Search for the anomalous ZZ\ensuremath{\gamma} and Z\ensuremath{\gamma}\ensuremath{\gamma} couplings via \ensuremath{\nu}\ensuremath{\nu}\ensuremath{\gamma} production at the CLIC}",
    eprint = "2203.02352",
    archivePrefix = "arXiv",
    primaryClass = "hep-ph",
    doi = "10.1016/j.nuclphysb.2022.115785",
    journal = "Nucl. Phys. B",
    volume = "979",
    pages = "115785",
    year = "2022"
}

@article{Yilmaz:2021ule,
    author = "Yilmaz, Ali",
    title = "{Search for the limits on anomalous neutral triple gauge couplings via ZZ production in the $\ell\ell\nu\nu$ channel at FCC-hh}",
    eprint = "2102.01989",
    archivePrefix = "arXiv",
    primaryClass = "hep-ph",
    doi = "10.1016/j.nuclphysb.2021.115471",
    journal = "Nucl. Phys. B",
    volume = "969",
    pages = "115471",
    year = "2021"
}

@article{Ellis:2020ljj,
    author = "Ellis, John and He, Hong-Jian and Xiao, Rui-Qing",
    title = "{Probing new physics in dimension-8 neutral gauge couplings at e$^{+}$e$^{-}$ colliders}",
    eprint = "2008.04298",
    archivePrefix = "arXiv",
    primaryClass = "hep-ph",
    reportNumber = "KCL-PH-TH/2020-28, CERN-TH-2020-076",
    doi = "10.1007/s11433-020-1617-3",
    journal = "Sci. China Phys. Mech. Astron.",
    volume = "64",
    number = "2",
    pages = "221062",
    year = "2021"
}

@inproceedings{Senol:2019swu,
    author = "Senol, A. and Denizli, H. and Yilmaz, A. and Turk Cakir, I. and Cakir, O.",
    title = "{Study on Anomalous Neutral Triple Gauge Boson Couplings from Dimension-eight Operators at the HL-LHC}",
    eprint = "1906.04589",
    archivePrefix = "arXiv",
    primaryClass = "hep-ph",
    doi = "10.5506/APhysPolB.50.1597",
    month = "6",
    year = "2019"
}

@article{Yilmaz:2019cue,
    author = "Yilmaz, A. and Senol, A. and Denizli, H. and Turk Cakir, I. and Cakir, O.",
    title = "{Sensitivity on Anomalous Neutral Triple Gauge Couplings via $ZZ$ Production at FCC-hh}",
    eprint = "1906.03911",
    archivePrefix = "arXiv",
    primaryClass = "hep-ph",
    doi = "10.1140/epjc/s10052-020-7731-2",
    journal = "Eur. Phys. J. C",
    volume = "80",
    number = "2",
    pages = "173",
    year = "2020"
}

@article{Ellis:2019zex,
    author = "Ellis, John and Ge, Shao-Feng and He, Hong-Jian and Xiao, Rui-Qing",
    title = "{Probing the scale of new physics in the  $ZZ\gamma$ coupling at $e^+e^-$  colliders}",
    eprint = "1902.06631",
    archivePrefix = "arXiv",
    primaryClass = "hep-ph",
    reportNumber = "KCL-PH-TH/2019-11, CERN-TH/2019-008, CERN-TH-2019-008, IPMU19-0021",
    doi = "10.1088/1674-1137/44/6/063106",
    journal = "Chin. Phys. C",
    volume = "44",
    number = "6",
    pages = "063106",
    year = "2020"
}

@article{Guo:2020lim,
    author = "Guo, Yu-Chen and Wang, Ying-Ying and Yang, Ji-Chong and Yue, Chong-Xing",
    title = "{Constraints on anomalous quartic gauge couplings via $W\gamma jj$ production at the LHC}",
    eprint = "2002.03326",
    archivePrefix = "arXiv",
    primaryClass = "hep-ph",
    doi = "10.1088/1674-1137/abb4d2",
    journal = "Chin. Phys. C",
    volume = "44",
    number = "12",
    pages = "123105",
    year = "2020"
}

@article{Guo:2019agy,
    author = "Guo, Yu-Chen and Wang, Ying-Ying and Yang, Ji-Chong",
    title = "{Constraints on anomalous quartic gauge couplings by $\gamma\gamma \to W^+W^-$ scattering}",
    eprint = "1912.10686",
    archivePrefix = "arXiv",
    primaryClass = "hep-ph",
    doi = "10.1016/j.nuclphysb.2020.115222",
    journal = "Nucl. Phys. B",
    volume = "961",
    pages = "115222",
    year = "2020"
}

@article{Yang:2021pcf,
    author = "Yang, Ji-Chong and Guo, Yu-Chen and Yue, Chong-Xing and Fu, Qing",
    title = "{Constraints on anomalous quartic gauge couplings via Z\ensuremath{\gamma}jj production at the LHC}",
    eprint = "2107.01123",
    archivePrefix = "arXiv",
    primaryClass = "hep-ph",
    doi = "10.1103/PhysRevD.104.035015",
    journal = "Phys. Rev. D",
    volume = "104",
    number = "3",
    pages = "035015",
    year = "2021"
}

@article{Fu:2021mub,
    author = "Fu, Qing and Yang, Ji-Chong and Yue, Chong-Xing and Guo, Yu-Chen",
    title = "{The study of neutral triple gauge couplings in the process e+e\ensuremath{-}\textrightarrow{}Z\ensuremath{\gamma} including unitarity bounds}",
    eprint = "2102.03623",
    archivePrefix = "arXiv",
    primaryClass = "hep-ph",
    doi = "10.1016/j.nuclphysb.2021.115543",
    journal = "Nucl. Phys. B",
    volume = "972",
    pages = "115543",
    year = "2021"
}

@article{Yang:2020rjt,
    author = "Yang, Ji-Chong and Qing, Zhi-Bin and Han, Xue-Ying and Guo, Yu-Chen and Li, Tong",
    title = "{Tri-photon at muon collider: a new process to probe the anomalous quartic gauge couplings}",
    eprint = "2204.08195",
    archivePrefix = "arXiv",
    primaryClass = "hep-ph",
    doi = "10.1007/JHEP07(2022)053",
    journal = "JHEP",
    volume = "22",
    pages = "053",
    year = "2020"
}

@article{Ellis:2018cos,
    author = "Ellis, John and Ge, Shao-Feng",
    title = "{Constraining Gluonic Quartic Gauge Coupling Operators with gg\textrightarrow{}\ensuremath{\gamma}\ensuremath{\gamma}}",
    eprint = "1802.02416",
    archivePrefix = "arXiv",
    primaryClass = "hep-ph",
    reportNumber = "KCL-PH-TH-2018-02, CERN-TH-2018-014, IPMU-18-0022",
    doi = "10.1103/PhysRevLett.121.041801",
    journal = "Phys. Rev. Lett.",
    volume = "121",
    number = "4",
    pages = "041801",
    year = "2018"
}

@article{Ellis:2021dfa,
    author = "Ellis, John and Ge, Shao-Feng and Ma, Kai",
    title = "{Hadron collider probes of the quartic couplings of gluons to the photon and Z boson}",
    eprint = "2112.06729",
    archivePrefix = "arXiv",
    primaryClass = "hep-ph",
    reportNumber = "KCL-PH-TH/2021-95, CERN-TH-2021-215",
    doi = "10.1007/JHEP04(2022)123",
    journal = "JHEP",
    volume = "04",
    pages = "123",
    year = "2022"
}

@inproceedings{Yang:2023gos,
    author = "Yang, Ji-Chong and Guo, Yu-Chen and Dong, Yi-Fei",
    title = "{Study of the gluonic quartic gauge couplings at muon colliders}",
    eprint = "2307.04207",
    archivePrefix = "arXiv",
    primaryClass = "hep-ph",
    doi = "10.1088/1572-9494/acfd14",
    journal = "Commun. Theor. Phys.",
    volume = "75",
    number = "11",
    pages = "115201",
    year = "2023"
}

@article{Born:1934gh,
    author = "Born, M. and Infeld, L.",
    title = "{Foundations of the new field theory}",
    doi = "10.1098/rspa.1934.0059",
    journal = "Proc. Roy. Soc. Lond. A",
    volume = "144",
    number = "852",
    pages = "425--451",
    year = "1934"
}

@article{Fradkin:1985qd,
    author = "Fradkin, E. S. and Tseytlin, Arkady A.",
    title = "{Nonlinear Electrodynamics from Quantized Strings}",
    reportNumber = "LEBEDEV-85-193",
    doi = "10.1016/0370-2693(85)90205-9",
    journal = "Phys. Lett. B",
    volume = "163",
    pages = "123--130",
    year = "1985"
}

@inproceedings{Tseytlin:1999dj,
    author = "Tseytlin, Arkady A.",
    editor = "Shifman, Mikhail A.",
    title = "{Born-Infeld action, supersymmetry and string theory}",
    eprint = "hep-th/9908105",
    archivePrefix = "arXiv",
    reportNumber = "IMPERIAL-TP-98-99-67",
    doi = "10.1142/9789812793850_0025",
    pages = "417--452",
    month = "8",
    year = "1999"
}

@article{Cheung:2018oki,
    author = "Cheung, Clifford and Kampf, Karol and Novotny, Jiri and Shen, Chia-Hsien and Trnka, Jaroslav and Wen, Congkao",
    title = "{Vector Effective Field Theories from Soft Limits}",
    eprint = "1801.01496",
    archivePrefix = "arXiv",
    primaryClass = "hep-th",
    reportNumber = "CALT-TH-2017-074",
    doi = "10.1103/PhysRevLett.120.261602",
    journal = "Phys. Rev. Lett.",
    volume = "120",
    number = "26",
    pages = "261602",
    year = "2018"
}

@article{Alwall:2014hca,
    author = "Alwall, J. and Frederix, R. and Frixione, S. and Hirschi, V. and Maltoni, F. and Mattelaer, O. and Shao, H. -S. and Stelzer, T. and Torrielli, P. and Zaro, M.",
    title = "{The automated computation of tree-level and next-to-leading order differential cross sections, and their matching to parton shower simulations}",
    eprint = "1405.0301",
    archivePrefix = "arXiv",
    primaryClass = "hep-ph",
    reportNumber = "CERN-PH-TH-2014-064, CP3-14-18, LPN14-066, MCNET-14-09, ZU-TH-14-14",
    doi = "10.1007/JHEP07(2014)079",
    journal = "JHEP",
    volume = "07",
    pages = "079",
    year = "2014"
}

@article{Christensen:2008py,
    author = "Christensen, Neil D. and Duhr, Claude",
    title = "{FeynRules - Feynman rules made easy}",
    eprint = "0806.4194",
    archivePrefix = "arXiv",
    primaryClass = "hep-ph",
    reportNumber = "MSUHEP-080616, CP3-08-20",
    doi = "10.1016/j.cpc.2009.02.018",
    journal = "Comput. Phys. Commun.",
    volume = "180",
    pages = "1614--1641",
    year = "2009"
}

@article{Ball:2013hta,
    author = "Ball, Richard D. and Bertone, Valerio and Carrazza, Stefano and Del Debbio, Luigi and Forte, Stefano and Guffanti, Alberto and Hartland, Nathan P. and Rojo, Juan",
    collaboration = "NNPDF",
    title = "{Parton distributions with QED corrections}",
    eprint = "1308.0598",
    archivePrefix = "arXiv",
    primaryClass = "hep-ph",
    reportNumber = "EDINBURGH-2013-20, FR-PHENO-2013-008, CERN-PH-TH-2013-075, Edinburgh 2013/20, IFUM-1014-FT, FR-PHENO-2013-008,
  CERN-PH-TH/2013-075",
    doi = "10.1016/j.nuclphysb.2013.10.010",
    journal = "Nucl. Phys. B",
    volume = "877",
    pages = "290--320",
    year = "2013"
}

@article{Sjostrand:2014zea,
    author = {Sj\"ostrand, Torbj\"orn and Ask, Stefan and Christiansen, Jesper R. and Corke, Richard and Desai, Nishita and Ilten, Philip and Mrenna, Stephen and Prestel, Stefan and Rasmussen, Christine O. and Skands, Peter Z.},
    title = "{An introduction to PYTHIA 8.2}",
    eprint = "1410.3012",
    archivePrefix = "arXiv",
    primaryClass = "hep-ph",
    reportNumber = "LU-TP-14-36, MCNET-14-22, CERN-PH-TH-2014-190, FERMILAB-PUB-14-316-CD, DESY-14-178, SLAC-PUB-16122",
    doi = "10.1016/j.cpc.2015.01.024",
    journal = "Comput. Phys. Commun.",
    volume = "191",
    pages = "159--177",
    year = "2015"
}

@article{deFavereau:2013fsa,
    author = "de Favereau, J. and Delaere, C. and Demin, P. and Giammanco, A. and Lema\^\i{}tre, V. and Mertens, A. and Selvaggi, M.",
    collaboration = "DELPHES 3",
    title = "{DELPHES 3, A modular framework for fast simulation of a generic collider experiment}",
    eprint = "1307.6346",
    archivePrefix = "arXiv",
    primaryClass = "hep-ex",
    doi = "10.1007/JHEP02(2014)057",
    journal = "JHEP",
    volume = "02",
    pages = "057",
    year = "2014"
}

@article{Guo:2023nfu,
    author = "Guo, Yu-Chen and Feng, Fan and Di, An and Lu, Shi-Qi and Yang, Ji-Chong",
    title = "{MLAnalysis: An open-source program for high energy physics analyses}",
    eprint = "2305.00964",
    archivePrefix = "arXiv",
    primaryClass = "hep-ph",
    doi = "10.1016/j.cpc.2023.108957",
    journal = "Comput. Phys. Commun.",
    volume = "294",
    pages = "108957",
    year = "2024"
}

@article{Degrande:2011ua,
    author = "Degrande, Celine and Duhr, Claude and Fuks, Benjamin and Grellscheid, David and Mattelaer, Olivier and Reiter, Thomas",
    title = "{UFO - The Universal FeynRules Output}",
    eprint = "1108.2040",
    archivePrefix = "arXiv",
    primaryClass = "hep-ph",
    reportNumber = "CP3-11-25, IPHC-PHENO-11-04, IPPP-11-39, DCPT-11-78, MPP-2011-68",
    doi = "10.1016/j.cpc.2012.01.022",
    journal = "Comput. Phys. Commun.",
    volume = "183",
    pages = "1201--1214",
    year = "2012"
}

@article{Cowan:2010js,
    author = "Cowan, Glen and Cranmer, Kyle and Gross, Eilam and Vitells, Ofer",
    title = "{Asymptotic formulae for likelihood-based tests of new physics}",
    eprint = "1007.1727",
    archivePrefix = "arXiv",
    primaryClass = "physics.data-an",
    doi = "10.1140/epjc/s10052-011-1554-0",
    journal = "Eur. Phys. J. C",
    volume = "71",
    pages = "1554",
    year = "2011",
    note = "[Erratum: Eur.Phys.J.C 73, 2501 (2013)]"
}

@article{ParticleDataGroup:2020ssz,
    author = "Zyla, P. A. and others",
    collaboration = "Particle Data Group",
    title = "{Review of Particle Physics}",
    doi = "10.1093/ptep/ptaa104",
    journal = "PTEP",
    volume = "2020",
    number = "8",
    pages = "083C01",
    year = "2020"
}

@article{Layssac:1993vfp,
    author = "Layssac, J. and Renard, F. M. and Gounaris, G. J.",
    title = "{Unitarity constraints for transverse gauge bosons at LEP and supercolliders}",
    eprint = "hep-ph/9311370",
    archivePrefix = "arXiv",
    reportNumber = "PM-93-37, THES-TP-93-11",
    doi = "10.1016/0370-2693(94)90872-9",
    journal = "Phys. Lett. B",
    volume = "332",
    pages = "146--152",
    year = "1994"
}

@article{Corbett:2017qgl,
    author = "Corbett, Tyler and \'Eboli, O. J. P. and Gonzalez-Garcia, M. C.",
    title = "{Unitarity Constraints on Dimension-six Operators II: Including Fermionic Operators}",
    eprint = "1705.09294",
    archivePrefix = "arXiv",
    primaryClass = "hep-ph",
    doi = "10.1103/PhysRevD.96.035006",
    journal = "Phys. Rev. D",
    volume = "96",
    number = "3",
    pages = "035006",
    year = "2017"
}

@article{Perez:2018kav,
    author = "Perez, Genessis and Sekulla, Marco and Zeppenfeld, Dieter",
    title = "{Anomalous quartic gauge couplings and unitarization for the vector boson scattering process $pp\rightarrow W^+W^+jjX\rightarrow \ell ^+\nu _\ell \ell ^+\nu _\ell jjX$}",
    eprint = "1807.02707",
    archivePrefix = "arXiv",
    primaryClass = "hep-ph",
    reportNumber = "KA-TP-15-2018",
    doi = "10.1140/epjc/s10052-018-6230-1",
    journal = "Eur. Phys. J. C",
    volume = "78",
    number = "9",
    pages = "759",
    year = "2018"
}

@article{Almeida:2020ylr,
    author = "Almeida, Eduardo da Silva and \'Eboli, O. J. P. and Gonzalez\textendash{}Garcia, M. C.",
    title = "{Unitarity constraints on anomalous quartic couplings}",
    eprint = "2004.05174",
    archivePrefix = "arXiv",
    primaryClass = "hep-ph",
    reportNumber = "YITP-SB-2020-8",
    doi = "10.1103/PhysRevD.101.113003",
    journal = "Phys. Rev. D",
    volume = "101",
    number = "11",
    pages = "113003",
    year = "2020"
}

@article{Kilian:2018bhs,
    author = "Kilian, Wolfgang and Sun, Sichun and Yan, Qi-Shu and Zhao, Xiaoran and Zhao, Zhijie",
    title = "{Multi-Higgs boson production and unitarity in vector-boson fusion at future hadron colliders}",
    eprint = "1808.05534",
    archivePrefix = "arXiv",
    primaryClass = "hep-ph",
    reportNumber = "SI-HEP-2018-27, CP3-18-55, MCnet-18-21",
    doi = "10.1103/PhysRevD.101.076012",
    journal = "Phys. Rev. D",
    volume = "101",
    number = "7",
    pages = "076012",
    year = "2020"
}

@article{Donoho_2004,
	doi = {10.1214/009053604000000265},
  
	url = {https://doi.org/10.1214%2F009053604000000265},
  
	year = 2004,
	month = {jun},
  
	publisher = {Institute of Mathematical Statistics},
  
	volume = {32},
  
	number = {3},
  
	author = {David Donoho and Jiashun Jin},
  
	title = {Higher criticism for detecting sparse heterogeneous mixtures},
  
	journal = {The Annals of Statistics}
}

@article{Jahedi:2022duc,
    author = "Jahedi, Sahabub and Lahiri, Jayita",
    title = "{Probing anomalous ZZ\ensuremath{\gamma} and Z\ensuremath{\gamma}\ensuremath{\gamma} couplings at the e$^{+}$e$^{-}$ colliders using optimal observable technique}",
    eprint = "2212.05121",
    archivePrefix = "arXiv",
    primaryClass = "hep-ph",
    doi = "10.1007/JHEP04(2023)085",
    journal = "JHEP",
    volume = "04",
    pages = "085",
    year = "2023"
}

@inproceedings{Jahedi:2023myu,
    author = "Jahedi, Sahabub",
    title = "{Optimal estimation of Dimension-8 Neutral Triple Gauge Couplings at $e^+ e^-$ Colliders}",
    eprint = "2305.11266",
    archivePrefix = "arXiv",
    primaryClass = "hep-ph",
    month = "5",
    year = "2023"
}

@article{Bauer:2022hpo,
    author = "Bauer, Christian W. and others",
    title = "{Quantum Simulation for High-Energy Physics}",
    eprint = "2204.03381",
    archivePrefix = "arXiv",
    primaryClass = "quant-ph",
    reportNumber = "UMD-PP-022-04, LA-UR-22-22100, RIKEN-iTHEMS-Report-22, RIKEN-iTHEMS-Report-22,
  FERMILAB-PUB-22-249-SQMS-T, IQuS@UW-21-027, MITRE-21-03848-2, FERMILAB-PUB-22-249-SQMS-T",
    doi = "10.1103/PRXQuantum.4.027001",
    journal = "PRX Quantum",
    volume = "4",
    number = "2",
    pages = "027001",
    year = "2023"
}

@article{Chang:2013aya,
    author = "Chang, Jung and Cheung, Kingman and Lu, Chih-Ting and Yuan, Tzu-Chiang",
    title = "{WW scattering in the era of post-Higgs-boson discovery}",
    eprint = "1303.6335",
    archivePrefix = "arXiv",
    primaryClass = "hep-ph",
    doi = "10.1103/PhysRevD.87.093005",
    journal = "Phys. Rev. D",
    volume = "87",
    pages = "093005",
    year = "2013"
}

@article{CMS:2019qfk,
    author = "Sirunyan, Albert M and others",
    collaboration = "CMS",
    title = "{Search for anomalous electroweak production of vector boson pairs in association with two jets in proton-proton collisions at 13 TeV}",
    eprint = "1905.07445",
    archivePrefix = "arXiv",
    primaryClass = "hep-ex",
    reportNumber = "CMS-SMP-18-006, CERN-EP-2019-089",
    doi = "10.1016/j.physletb.2019.134985",
    journal = "Phys. Lett. B",
    volume = "798",
    pages = "134985",
    year = "2019"
}

@article{Eboli:2006wa,
    author = "Eboli, O. J. P. and Gonzalez-Garcia, M. C. and Mizukoshi, J. K.",
    title = "{p p ---\ensuremath{>} j j e+- mu+- nu nu and j j e+- mu-+ nu nu at O( alpha(em)**6) and O(alpha(em)**4 alpha(s)**2) for the study of the quartic electroweak gauge boson vertex at CERN LHC}",
    eprint = "hep-ph/0606118",
    archivePrefix = "arXiv",
    reportNumber = "YITP-SB-06-10, IFUSP-1620-2006",
    doi = "10.1103/PhysRevD.74.073005",
    journal = "Phys. Rev. D",
    volume = "74",
    pages = "073005",
    year = "2006"
}

@article{Eboli:2016kko,
    author = "\'Eboli, O. J. P. and Gonzalez-Garcia, M. C.",
    title = "{Classifying the bosonic quartic couplings}",
    eprint = "1604.03555",
    archivePrefix = "arXiv",
    primaryClass = "hep-ph",
    reportNumber = "YITP-SB-16-09",
    doi = "10.1103/PhysRevD.93.093013",
    journal = "Phys. Rev. D",
    volume = "93",
    number = "9",
    pages = "093013",
    year = "2016"
}

@article{Mottonen:2004vly,
    author = {M\"ott\"onen, Mikko and Vartiainen, Juha J. and Bergholm, Ville and Salomaa, Martti M.},
    title = "{Quantum Circuits for General Multiqubit Gates}",
    doi = "10.1103/PhysRevLett.93.130502",
    journal = "Phys. Rev. Lett.",
    volume = "93",
    number = "13",
    pages = "130502",
    year = "2004"
}

@article{Jones:2019knd,
    author = "Jones, Tyson and Brown, Anna and Bush, Ian and Benjamin, Simon C.",
    title = "{QuEST and High Performance Simulation of Quantum Computers}",
    doi = "10.1038/s41598-019-47174-9",
    journal = "Sci. Rep.",
    volume = "9",
    number = "1",
    pages = "10736",
    year = "2019"
}

@article{Black:2022cth,
    author = "Black, K. M. and others",
    title = "{Muon Collider Forum report}",
    eprint = "2209.01318",
    archivePrefix = "arXiv",
    primaryClass = "hep-ex",
    reportNumber = "FERMILAB-FN-1194",
    doi = "10.1088/1748-0221/19/02/T02015",
    journal = "JINST",
    volume = "19",
    number = "02",
    pages = "T02015",
    year = "2024"
}

@article{Accettura:2023ked,
    author = "Accettura, Carlotta and others",
    title = "{Towards a muon collider}",
    eprint = "2303.08533",
    archivePrefix = "arXiv",
    primaryClass = "physics.acc-ph",
    reportNumber = "FERMILAB-PUB-23-123-AD-PPD-T",
    doi = "10.1140/epjc/s10052-023-11889-x",
    journal = "Eur. Phys. J. C",
    volume = "83",
    number = "9",
    pages = "864",
    year = "2023",
    note = "[Erratum: Eur.Phys.J.C 84, 36 (2024)]"
}

@article{Guan:2020bdl,
    author = "Guan, Wen and Perdue, Gabriel and Pesah, Arthur and Schuld, Maria and Terashi, Koji and Vallecorsa, Sofia and Vlimant, Jean-Roch",
    title = "{Quantum Machine Learning in High Energy Physics}",
    eprint = "2005.08582",
    archivePrefix = "arXiv",
    primaryClass = "quant-ph",
    reportNumber = "FERMILAB-PUB-20-184-QIS",
    doi = "10.1088/2632-2153/abc17d",
    journal = "Mach. Learn. Sci. Tech.",
    volume = "2",
    pages = "011003",
    year = "2021"
}

@article{Terashi:2020wfi,
    author = "Terashi, Koji and Kaneda, Michiru and Kishimoto, Tomoe and Saito, Masahiko and Sawada, Ryu and Tanaka, Junichi",
    title = "{Event Classification with Quantum Machine Learning in High-Energy Physics}",
    eprint = "2002.09935",
    archivePrefix = "arXiv",
    primaryClass = "physics.comp-ph",
    doi = "10.1007/s41781-020-00047-7",
    journal = "Comput. Softw. Big Sci.",
    volume = "5",
    number = "1",
    pages = "2",
    year = "2021"
}

@article{Havlicek:2018nqz,
    author = "Havlicek, Vojtech and C\'orcoles, Antonio D. and Temme, Kristan and Harrow, Aram W. and Kandala, Abhinav and Chow, Jerry M. and Gambetta, Jay M.",
    title = "{Supervised learning with quantum-enhanced feature spaces}",
    eprint = "1804.11326",
    archivePrefix = "arXiv",
    primaryClass = "quant-ph",
    doi = "10.1038/s41586-019-0980-2",
    journal = "Nature",
    volume = "567",
    pages = "209--212",
    year = "2019"
}

@article{Chen:2021num,
    title={Exponential suppression of bit or phase errors with cyclic error correction},
    volume={595},
    ISSN={1476-4687},
    url={http://dx.doi.org/10.1038/s41586-021-03588-y},
    DOI={10.1038/s41586-021-03588-y},
    number={7867},
    journal={Nature},
    publisher={Springer Science and Business Media LLC},
    author={Chen, Zijun and Satzinger, Kevin J. and Atalaya, Juan and Korotkov, Alexander N. and Dunsworth, Andrew and Sank, Daniel and Quintana, Chris and McEwen, Matt and Barends, Rami and Klimov, Paul V. and Hong, Sabrina and Jones, Cody and Petukhov, Andre and Kafri, Dvir and Demura, Sean and Burkett, Brian and Gidney, Craig and Fowler, Austin G. and Paler, Alexandru and Putterman, Harald and Aleiner, Igor and Arute, Frank and Arya, Kunal and Babbush, Ryan and Bardin, Joseph C. and Bengtsson, Andreas and Bourassa, Alexandre and Broughton, Michael and Buckley, Bob B. and Buell, David A. and Bushnell, Nicholas and Chiaro, Benjamin and Collins, Roberto and Courtney, William and Derk, Alan R. and Eppens, Daniel and Erickson, Catherine and Farhi, Edward and Foxen, Brooks and Giustina, Marissa and Greene, Ami and Gross, Jonathan A. and Harrigan, Matthew P. and Harrington, Sean D. and Hilton, Jeremy and Ho, Alan and Huang, Trent and Huggins, William J. and Ioffe, L. B. and Isakov, Sergei V. and Jeffrey, Evan and Jiang, Zhang and Kechedzhi, Kostyantyn and Kim, Seon and Kitaev, Alexei and Kostritsa, Fedor and Landhuis, David and Laptev, Pavel and Lucero, Erik and Martin, Orion and McClean, Jarrod R. and McCourt, Trevor and Mi, Xiao and Miao, Kevin C. and Mohseni, Masoud and Montazeri, Shirin and Mruczkiewicz, Wojciech and Mutus, Josh and Naaman, Ofer and Neeley, Matthew and Neill, Charles and Newman, Michael and Niu, Murphy Yuezhen and O’Brien, Thomas E. and Opremcak, Alex and Ostby, Eric and Pató, Bálint and Redd, Nicholas and Roushan, Pedram and Rubin, Nicholas C. and Shvarts, Vladimir and Strain, Doug and Szalay, Marco and Trevithick, Matthew D. and Villalonga, Benjamin and White, Theodore and Yao, Z. Jamie and Yeh, Ping and Yoo, Juhwan and Zalcman, Adam and Neven, Hartmut and Boixo, Sergio and Smelyanskiy, Vadim and Chen, Yu and Megrant, Anthony and Kelly, Julian},
    year={2021},
    month=jul, 
    pages={383–387},
    eprint = "2102.06132",
    archivePrefix = "arXiv"
}

@article{Arute:2019zxq,
    author = "Arute, Frank and others",
    title = "{Quantum supremacy using a programmable superconducting processor}",
    eprint = "1910.11333",
    archivePrefix = "arXiv",
    primaryClass = "quant-ph",
    doi = "10.1038/s41586-019-1666-5",
    journal = "Nature",
    volume = "574",
    number = "7779",
    pages = "505--510",
    year = "2019"
}

@article{Preskill:2018jim,
    author = "Preskill, John",
    title = "{Quantum Computing in the NISQ era and beyond}",
    eprint = "1801.00862",
    archivePrefix = "arXiv",
    primaryClass = "quant-ph",
    doi = "10.22331/q-2018-08-06-79",
    journal = "Quantum",
    volume = "2",
    pages = "79",
    year = "2018"
}

@article{Zhang:2023ykh,
    author = "Zhang, Shuai and Guo, Yu-Chen and Yang, Ji-Chong",
    title = "{Optimize the event selection strategy to study the anomalous quartic gauge couplings at muon colliders using the support vector machine and quantum support vector machine}",
    eprint = "2311.15280",
    archivePrefix = "arXiv",
    primaryClass = "hep-ph",
    doi = "10.1140/epjc/s10052-024-13208-4",
    journal = "Eur. Phys. J. C",
    volume = "84",
    number = "8",
    pages = "833",
    year = "2024"
}

@article{Kilian:2021whd,
    author = "Kilian, Wolfgang and Sun, Sichun and Yan, Qi-Shu and Zhao, Xiaoran and Zhao, Zhijie",
    title = "{Highly Boosted Higgs Bosons and Unitarity in Vector-Boson Fusion at Future Hadron Colliders}",
    eprint = "2101.12537",
    archivePrefix = "arXiv",
    primaryClass = "hep-ph",
    reportNumber = "SI-HEP-2021-05, CP3-21-02",
    doi = "10.1007/JHEP05(2021)198",
    journal = "JHEP",
    volume = "05",
    pages = "198",
    year = "2021"
}

@article{Yue:2021snv,
    author = "Yue, Chong-Xing and Cheng, Xue-Jia and Yang, Ji-Chong",
    title = "{Charged-current non-standard neutrino interactions at the LHC and HL-LHC*}",
    eprint = "2110.01204",
    archivePrefix = "arXiv",
    primaryClass = "hep-ph",
    doi = "10.1088/1674-1137/acb993",
    journal = "Chin. Phys. C",
    volume = "47",
    number = "4",
    pages = "043111",
    year = "2023"
}

@article{Bravo-Prieto:2019kld,
    author = "Bravo-Prieto, Carlos and LaRose, Ryan and Cerezo, M. and Subasi, Yigit and Cincio, Lukasz and Coles, Patrick J.",
    title = "{Variational Quantum Linear Solver}",
    eprint = "1909.05820",
    archivePrefix = "arXiv",
    primaryClass = "quant-ph",
    reportNumber = "LA-UR-19-29101",
    doi = "10.22331/q-2023-11-22-1188",
    journal = "Quantum",
    volume = "7",
    pages = "1188",
    year = "2023"
}

@article{Pedregosa:2011ork,
    author = "Pedregosa, Fabian and others",
    title = "{Scikit-learn: Machine Learning in Python}",
    eprint = "1201.0490",
    archivePrefix = "arXiv",
    primaryClass = "cs.LG",
    journal = "J. Machine Learning Res.",
    volume = "12",
    pages = "2825--2830",
    year = "2011"
}

@article{Yang:2022ilt,
    author = "Yang, Ji-Chong and Guo, Yu-Chen and Liu, Bing and Li, Tong",
    title = "{Shining light on magnetic monopoles through high-energy muon colliders}",
    eprint = "2208.02188",
    archivePrefix = "arXiv",
    primaryClass = "hep-ph",
    doi = "10.1016/j.nuclphysb.2023.116097",
    journal = "Nucl. Phys. B",
    volume = "987",
    pages = "116097",
    year = "2023"
}

@article{ATLAS:2014jzl,
    author = "Aad, Georges and others",
    collaboration = "ATLAS",
    title = "{Evidence for Electroweak Production of $W^{\pm}W^{\pm}jj$ in $pp$ Collisions at $\sqrt{s}=8$ TeV with the ATLAS Detector}",
    eprint = "1405.6241",
    archivePrefix = "arXiv",
    primaryClass = "hep-ex",
    reportNumber = "CERN-PH-EP-2014-079",
    doi = "10.1103/PhysRevLett.113.141803",
    journal = "Phys. Rev. Lett.",
    volume = "113",
    number = "14",
    pages = "141803",
    year = "2014"
}

@article{CMS:2020gfh,
    author = "Sirunyan, Albert M and others",
    collaboration = "CMS",
    title = "{Measurements of production cross sections of WZ and same-sign WW boson pairs in association with two jets in proton-proton collisions at $\sqrt{s} =$ 13 TeV}",
    eprint = "2005.01173",
    archivePrefix = "arXiv",
    primaryClass = "hep-ex",
    reportNumber = "CMS-SMP-19-012, CERN-EP-2020-064",
    doi = "10.1016/j.physletb.2020.135710",
    journal = "Phys. Lett. B",
    volume = "809",
    pages = "135710",
    year = "2020"
}

@article{CMS:2020ypo,
    author = "Sirunyan, Albert M and others",
    collaboration = "CMS",
    title = "{Observation of electroweak production of W$\gamma$ with two jets in proton-proton collisions at $\sqrt {s}$ = 13 TeV}",
    eprint = "2008.10521",
    archivePrefix = "arXiv",
    primaryClass = "hep-ex",
    reportNumber = "CMS-SMP-19-008, CERN-EP-2020-143",
    doi = "10.1016/j.physletb.2020.135988",
    journal = "Phys. Lett. B",
    volume = "811",
    pages = "135988",
    year = "2020"
}

@article{CMS:2020fqz,
    author = "Sirunyan, Albert M and others",
    collaboration = "CMS",
    title = "{Evidence for electroweak production of four charged leptons and two jets in proton-proton collisions at $\sqrt {s}$ = 13 TeV}",
    eprint = "2008.07013",
    archivePrefix = "arXiv",
    primaryClass = "hep-ex",
    reportNumber = "CMS-SMP-20-001, CERN-EP-2020-127",
    doi = "10.1016/j.physletb.2020.135992",
    journal = "Phys. Lett. B",
    volume = "812",
    pages = "135992",
    year = "2021"
}

@article{Wu:2020cye,
    author = "Wu, Sau Lan and others",
    title = "{Application of quantum machine learning using the quantum variational classifier method to high energy physics analysis at the LHC on IBM quantum computer simulator and hardware with 10 qubits}",
    eprint = "2012.11560",
    archivePrefix = "arXiv",
    primaryClass = "quant-ph",
    reportNumber = "FERMILAB-PUB-20-675-DI-QIS",
    doi = "10.1088/1361-6471/ac1391",
    journal = "J. Phys. G",
    volume = "48",
    number = "12",
    pages = "125003",
    year = "2021"
}

@article{Wu:2021xsj,
    author = "Wu, Sau Lan and others",
    title = "{Application of quantum machine learning using the quantum kernel algorithm on high energy physics analysis at the LHC}",
    eprint = "2104.05059",
    archivePrefix = "arXiv",
    primaryClass = "quant-ph",
    reportNumber = "FERMILAB-PUB-21-552-DI-QIS",
    doi = "10.1103/PhysRevResearch.3.033221",
    journal = "Phys. Rev. Res.",
    volume = "3",
    number = "3",
    pages = "033221",
    year = "2021"
}

@article{Fadol:2022umw,
    author = "Fadol, Abdualazem and Sha, Qiyu and Fang, Yaquan and Li, Zhan and Qian, Sitian and Xiao, Yuyang and Zhang, Yu and Zhou, Chen",
    title = "{Application of quantum machine learning in a Higgs physics study at the CEPC}",
    eprint = "2209.12788",
    archivePrefix = "arXiv",
    primaryClass = "hep-ex",
    doi = "10.1142/S0217751X24500076",
    journal = "Int. J. Mod. Phys. A",
    volume = "39",
    number = "01",
    pages = "2450007",
    year = "2024"
}

@article{Kandala:2017vok,
    author = "Kandala, Abhinav and Mezzacapo, Antonio and Temme, Kristan and Takita, Maika and Brink, Markus and Chow, Jerry M. and Gambetta, Jay M.",
    title = "{Hardware-efficient variational quantum eigensolver for small molecules and quantum magnets}",
    eprint = "1704.05018",
    archivePrefix = "arXiv",
    primaryClass = "quant-ph",
    doi = "10.1038/nature23879",
    journal = "Nature",
    volume = "549",
    number = "7671",
    pages = "242--246",
    year = "2017"
}

@inproceedings{Park:2024rim,
    author = "Park, Chae-Yeun and Kang, Minhyeok and Huh, Joonsuk",
    title = "{Hardware-efficient ansatz without barren plateaus in any depth}",
    eprint = "2403.04844",
    archivePrefix = "arXiv",
    primaryClass = "quant-ph",
    month = "3",
    year = "2024"
}

@article{ATLAS:2019bwq,
    author = "Aad, Georges and others",
    collaboration = "ATLAS",
    title = "{ATLAS b-jet identification performance and efficiency measurement with $t{\bar{t}}$ events in pp collisions at $\sqrt{s}=13$ TeV}",
    eprint = "1907.05120",
    archivePrefix = "arXiv",
    primaryClass = "hep-ex",
    reportNumber = "CERN-EP-2019-132",
    doi = "10.1140/epjc/s10052-019-7450-8",
    journal = "Eur. Phys. J. C",
    volume = "79",
    number = "11",
    pages = "970",
    year = "2019"
}

@article{Czakon:2011xx,
    author = "Czakon, Michal and Mitov, Alexander",
    title = "{Top++: A Program for the Calculation of the Top-Pair Cross-Section at Hadron Colliders}",
    eprint = "1112.5675",
    archivePrefix = "arXiv",
    primaryClass = "hep-ph",
    reportNumber = "CERN-PH-TH-2011-303, TTK-11-58",
    doi = "10.1016/j.cpc.2014.06.021",
    journal = "Comput. Phys. Commun.",
    volume = "185",
    pages = "2930",
    year = "2014"
}

@article{CMS:2020grm,
    author = "Sirunyan, Albert M and others",
    collaboration = "CMS",
    title = "{Measurement of the cross section for $\text{t}\bar{\text{t}}$ production with additional jets and b jets in pp collisions at $\sqrt{s}=$ 13 TeV}",
    eprint = "2003.06467",
    archivePrefix = "arXiv",
    primaryClass = "hep-ex",
    reportNumber = "CMS-TOP-18-002, CERN-EP-2020-011",
    doi = "10.1007/JHEP07(2020)125",
    journal = "JHEP",
    volume = "07",
    pages = "125",
    year = "2020"
}

@inproceedings{ATLAS:2024aht,
    author = "Aad, Georges and others",
    collaboration = "ATLAS",
    title = "{Measurement of $t\bar{t}$ production in association with additional $b$-jets in the $e\mu$ final state in proton-proton collisions at $\sqrt{s}$=13 TeV with the ATLAS detector}",
    eprint = "2407.13473",
    archivePrefix = "arXiv",
    primaryClass = "hep-ex",
    reportNumber = "CERN-EP-2024-191",
    month = "7",
    year = "2024"
}

@article{Carena:2022kpg,
    author = "Carena, Marcela and Lamm, Henry and Li, Ying-Ying and Liu, Wanqiang",
    title = "{Improved Hamiltonians for Quantum Simulations of Gauge Theories}",
    eprint = "2203.02823",
    archivePrefix = "arXiv",
    primaryClass = "hep-lat",
    reportNumber = "FERMILAB-PUB-21-674-T",
    doi = "10.1103/PhysRevLett.129.051601",
    journal = "Phys. Rev. Lett.",
    volume = "129",
    number = "5",
    pages = "051601",
    year = "2022"
}

@article{Gustafson:2022xdt,
    author = "Gustafson, Erik J. and Lamm, Henry and Lovelace, Felicity and Musk, Damian",
    title = "{Primitive quantum gates for an SU(2) discrete subgroup: Binary tetrahedral}",
    eprint = "2208.12309",
    archivePrefix = "arXiv",
    primaryClass = "quant-ph",
    reportNumber = "FERMILAB-PUB-22-583-SQMS-T",
    doi = "10.1103/PhysRevD.106.114501",
    journal = "Phys. Rev. D",
    volume = "106",
    number = "11",
    pages = "114501",
    year = "2022"
}

@article{Lamm:2024jnl,
    author = "Lamm, Henry and Li, Ying-Ying and Shu, Jing and Wang, Yi-Lin and Xu, Bin",
    title = "{Block encodings of discrete subgroups on a quantum computer}",
    eprint = "2405.12890",
    archivePrefix = "arXiv",
    primaryClass = "hep-lat",
    reportNumber = "USTC-ICTS/PCFT-24-15, FERMILAB-PUB-24-0242-T",
    doi = "10.1103/PhysRevD.110.054505",
    journal = "Phys. Rev. D",
    volume = "110",
    number = "5",
    pages = "054505",
    year = "2024"
}

@article{Carena:2024dzu,
    author = "Carena, Marcela and Lamm, Henry and Li, Ying-Ying and Liu, Wanqiang",
    title = "{Quantum error thresholds for gauge-redundant digitizations of lattice field theories}",
    eprint = "2402.16780",
    archivePrefix = "arXiv",
    primaryClass = "hep-lat",
    reportNumber = "USTC-ICTS/PCFT-24-06, FERMILAB-PUB-23-570-T",
    doi = "10.1103/PhysRevD.110.054516",
    journal = "Phys. Rev. D",
    volume = "110",
    number = "5",
    pages = "054516",
    year = "2024"
}

@article{Atas:2021ext,
    author = "Atas, Yasar Y. and Zhang, Jinglei and Lewis, Randy and Jahanpour, Amin and Haase, Jan F. and Muschik, Christine A.",
    title = "{SU(2) hadrons on a quantum computer via a variational approach}",
    eprint = "2102.08920",
    archivePrefix = "arXiv",
    primaryClass = "quant-ph",
    doi = "10.1038/s41467-021-26825-4",
    journal = "Nature Commun.",
    volume = "12",
    number = "1",
    pages = "6499",
    year = "2021"
}

@article{Li:2023vwx,
    author = "Li, Ying-Ying and Sajid, Muhammad Omer and Unmuth-Yockey, Judah",
    title = "{Lattice holography on a quantum computer}",
    eprint = "2312.10544",
    archivePrefix = "arXiv",
    primaryClass = "hep-lat",
    reportNumber = "USTC-ICTS/PCFT-23-25, FERMILAB-PUB-23-0817-T",
    doi = "10.1103/PhysRevD.110.034507",
    journal = "Phys. Rev. D",
    volume = "110",
    number = "3",
    pages = "034507",
    year = "2024"
}

@article{Cui:2019sfz,
    author = "Cui, Xiaopeng and Shi, Yu and Yang, Ji-Chong",
    title = "{Circuit-based digital adiabatic quantum simulation and pseudoquantum simulation as new approaches to lattice gauge theory}",
    eprint = "1910.08020",
    archivePrefix = "arXiv",
    primaryClass = "quant-ph",
    doi = "10.1007/JHEP08(2020)160",
    journal = "JHEP",
    volume = "08",
    pages = "160",
    year = "2020"
}

@article{Zou:2021pvl,
    author = "Zou, Yi-Tong and Bo, Yu-Jiao and Yang, Ji-Chong",
    title = "{Optimize quantum simulation using a force-gradient integrator}",
    eprint = "2103.05876",
    archivePrefix = "arXiv",
    primaryClass = "quant-ph",
    doi = "10.1209/0295-5075/135/10004",
    journal = "EPL",
    volume = "135",
    pages = "10004",
    year = "2021"
}

@article{Georgescu:2013oza,
    author = "Georgescu, I. M. and Ashhab, S. and Nori, Franco",
    title = "{Quantum Simulation}",
    eprint = "1308.6253",
    archivePrefix = "arXiv",
    primaryClass = "quant-ph",
    doi = "10.1103/RevModPhys.86.153",
    journal = "Rev. Mod. Phys.",
    volume = "86",
    pages = "153",
    year = "2014"
}

@article{Lamm:2019uyc,
    author = "Lamm, Henry and Lawrence, Scott and Yamauchi, Yukari",
    collaboration = "NuQS",
    title = "{Parton physics on a quantum computer}",
    eprint = "1908.10439",
    archivePrefix = "arXiv",
    primaryClass = "hep-lat",
    doi = "10.1103/PhysRevResearch.2.013272",
    journal = "Phys. Rev. Res.",
    volume = "2",
    number = "1",
    pages = "013272",
    year = "2020"
}

@article{Li:2021kcs,
    author = "Li, Tianyin and Guo, Xingyu and Lai, Wai Kin and Liu, Xiaohui and Wang, Enke and Xing, Hongxi and Zhang, Dan-Bo and Zhu, Shi-Liang",
    collaboration = "QuNu",
    title = "{Partonic collinear structure by quantum computing}",
    eprint = "2106.03865",
    archivePrefix = "arXiv",
    primaryClass = "hep-ph",
    doi = "10.1103/PhysRevD.105.L111502",
    journal = "Phys. Rev. D",
    volume = "105",
    number = "11",
    pages = "L111502",
    year = "2022"
}

@article{Echevarria:2020wct,
    author = "Echevarria, M. G. and Egusquiza, I. L. and Rico, E. and Schnell, G.",
    title = "{Quantum simulation of light-front parton correlators}",
    eprint = "2011.01275",
    archivePrefix = "arXiv",
    primaryClass = "quant-ph",
    doi = "10.1103/PhysRevD.104.014512",
    journal = "Phys. Rev. D",
    volume = "104",
    number = "1",
    pages = "014512",
    year = "2021"
}

@article{Perez-Salinas:2020nem,
    author = "P\'erez-Salinas, Adri\'an and Cruz-Martinez, Juan and Alhajri, Abdulla A. and Carrazza, Stefano",
    title = "{Determining the proton content with a quantum computer}",
    eprint = "2011.13934",
    archivePrefix = "arXiv",
    primaryClass = "hep-ph",
    reportNumber = "TIF-UNIMI-2020-30",
    doi = "10.1103/PhysRevD.103.034027",
    journal = "Phys. Rev. D",
    volume = "103",
    number = "3",
    pages = "034027",
    year = "2021"
}

@article{Jordan:2011ci,
    author = "Jordan, Stephen P. and Lee, Keith S. M. and Preskill, John",
    title = "{Quantum Computation of Scattering in Scalar Quantum Field Theories}",
    eprint = "1112.4833",
    archivePrefix = "arXiv",
    primaryClass = "hep-th",
    journal = "Quant. Inf. Comput.",
    volume = "14",
    pages = "1014--1080",
    year = "2014"
}

@article{Mueller:2019qqj,
    author = "Mueller, Niklas and Tarasov, Andrey and Venugopalan, Raju",
    title = "{Deeply inelastic scattering structure functions on a hybrid quantum computer}",
    eprint = "1908.07051",
    archivePrefix = "arXiv",
    primaryClass = "hep-th",
    doi = "10.1103/PhysRevD.102.016007",
    journal = "Phys. Rev. D",
    volume = "102",
    number = "1",
    pages = "016007",
    year = "2020"
}

@article{Biamonte:2016ugo,
    author = "Biamonte, Jacob and Wittek, Peter and Pancotti, Nicola and Rebentrost, Patrick and Wiebe, Nathan and Lloyd, Seth",
    title = "{Quantum machine learning}",
    eprint = "1611.09347",
    archivePrefix = "arXiv",
    primaryClass = "quant-ph",
    doi = "10.1038/nature23474",
    journal = "Nature",
    volume = "549",
    number = "7671",
    pages = "195--202",
    year = "2017"
}

@article{Innocente:1992gq,
    author = "Innocente, V. and Wang, Y. F. and Zhang, Z. P.",
    title = "{Identification of tau decays using a neural network}",
    reportNumber = "CERN-PPE-92-098, CERN-PPE-92-98",
    doi = "10.1016/0168-9002(92)90011-R",
    journal = "Nucl. Instrum. Meth. A",
    volume = "323",
    pages = "647--656",
    year = "1992"
}

@article{Lee:2018xtt,
    author = "Lee, Junho and Chanon, Nicolas and Levin, Andrew and Li, Jing and Lu, Meng and Li, Qiang and Mao, Yajun",
    title = "{Polarization fraction measurement in same-sign WW scattering using deep learning}",
    eprint = "1812.07591",
    archivePrefix = "arXiv",
    primaryClass = "hep-ph",
    reportNumber = "VBSCAN-PUB-09-18",
    doi = "10.1103/PhysRevD.99.033004",
    journal = "Phys. Rev. D",
    volume = "99",
    number = "3",
    pages = "033004",
    year = "2019"
}

@article{Albertsson:2018maf,
    author = "Albertsson, Kim and others",
    title = "{Machine Learning in High Energy Physics Community White Paper}",
    eprint = "1807.02876",
    archivePrefix = "arXiv",
    primaryClass = "physics.comp-ph",
    reportNumber = "FERMILAB-PUB-18-318-CD-DI-PPD",
    doi = "10.1088/1742-6596/1085/2/022008",
    journal = "J. Phys. Conf. Ser.",
    volume = "1085",
    number = "2",
    pages = "022008",
    year = "2018"
}

@article{Guest:2018yhq,
    author = "Guest, Dan and Cranmer, Kyle and Whiteson, Daniel",
    title = "{Deep Learning and its Application to LHC Physics}",
    eprint = "1806.11484",
    archivePrefix = "arXiv",
    primaryClass = "hep-ex",
    doi = "10.1146/annurev-nucl-101917-021019",
    journal = "Ann. Rev. Nucl. Part. Sci.",
    volume = "68",
    pages = "161--181",
    year = "2018"
}

@article{Feynman:1981tf,
    author = "Feynman, Richard P.",
    editor = "Brown, L. M.",
    title = "{Simulating physics with computers}",
    doi = "10.1007/BF02650179",
    journal = "Int. J. Theor. Phys.",
    volume = "21",
    pages = "467--488",
    year = "1982"
}

@article{Roggero:2018hrn,
    author = "Roggero, Alessandro and Carlson, Joseph",
    title = "{Dynamic linear response quantum algorithm}",
    eprint = "1804.01505",
    archivePrefix = "arXiv",
    primaryClass = "quant-ph",
    reportNumber = "LA-UR-18-22120",
    doi = "10.1103/PhysRevC.100.034610",
    journal = "Phys. Rev. C",
    volume = "100",
    number = "3",
    pages = "034610",
    year = "2019"
}

@article{Roggero:2019myu,
    author = "Roggero, Alessandro and Li, Andy C. Y. and Carlson, Joseph and Gupta, Rajan and Perdue, Gabriel N.",
    title = "{Quantum Computing for Neutrino-Nucleus Scattering}",
    eprint = "1911.06368",
    archivePrefix = "arXiv",
    primaryClass = "quant-ph",
    reportNumber = "LA-UR-19-31323, INT-PUB-19-052, FERMILAB-PUB-19-547-QIS",
    doi = "10.1103/PhysRevD.101.074038",
    journal = "Phys. Rev. D",
    volume = "101",
    number = "7",
    pages = "074038",
    year = "2020"
}

@article{Grant:2018oml,
    author = "Grant, Edward and Benedetti, Marcello and Cao, Shuxiang and Hallam, Andrew and Lockhart, Joshua and Stojevic, Vid and Green, Andrew G. and Severini, Simone",
    title = "{Hierarchical quantum classifiers}",
    doi = "10.1038/s41534-018-0116-9",
    journal = "npj Quantum Inf.",
    volume = "4",
    pages = "65",
    year = "2018"
}

@inproceedings{Lloyd:2020eeh,
    author = "Lloyd, Seth and Schuld, Maria and Ijaz, Aroosa and Izaac, Josh and Killoran, Nathan",
    title = "{Quantum embeddings for machine learning}",
    eprint = "2001.03622",
    archivePrefix = "arXiv",
    primaryClass = "quant-ph",
    month = "1",
    year = "2020"
}

@inproceedings{Zhang:2024ebl,
    author = "Zhang, Shuai and Chen, Ke-Xin and Yang, Ji-Chong",
    title = "{Detect anomalous quartic gauge couplings at muon colliders with quantum kernel k-means}",
    eprint = "2409.07010",
    archivePrefix = "arXiv",
    primaryClass = "hep-ph",
    month = "9",
    year = "2024"
}

@inproceedings{Zhu:2024own,
    author = "Zhu, Yongfeng and Zhuang, Weifeng and Qian, Chen and Ma, Yunheng and Liu, Dong E. and Ruan, Manqi and Zhou, Chen",
    title = "{A Novel Quantum Realization of Jet Clustering in High-Energy Physics Experiments}",
    eprint = "2407.09056",
    archivePrefix = "arXiv",
    primaryClass = "quant-ph",
    month = "7",
    year = "2024"
}

@article{Komiske:2019fks,
    author = "Komiske, Patrick T. and Metodiev, Eric M. and Thaler, Jesse",
    title = "{Metric Space of Collider Events}",
    eprint = "1902.02346",
    archivePrefix = "arXiv",
    primaryClass = "hep-ph",
    reportNumber = "MIT-CTP 5102",
    doi = "10.1103/PhysRevLett.123.041801",
    journal = "Phys. Rev. Lett.",
    volume = "123",
    number = "4",
    pages = "041801",
    year = "2019"
}

@inproceedings{Brassard:2000xvp,
    author = "Brassard, Gilles and Hoyer, Peter and Mosca, Michele and Tapp, Alain",
    title = "{Quantum amplitude amplification and estimation}",
    eprint = "quant-ph/0005055",
    archivePrefix = "arXiv",
    doi = "10.1090/conm/305/05215",
    month = "5",
    year = "2000"
}

@article{Sim:2019yyv,
    author = "Sim, Sukin and Johnson, Peter D. and Aspuru-Guzik, Al\'an",
    title = "{Expressibility and Entangling Capability of Parameterized Quantum Circuits for Hybrid Quantum-Classical Algorithms}",
    eprint = "1905.10876",
    archivePrefix = "arXiv",
    primaryClass = "quant-ph",
    doi = "10.1002/qute.201900070",
    journal = "Adv. Quantum Technol.",
    volume = "2",
    number = "12",
    pages = "1900070",
    year = "2019"
}

@inproceedings{Kingma:2014vow,
    author = "Kingma, Diederik P. and Ba, Jimmy",
    title = "{Adam: A Method for Stochastic Optimization}",
    eprint = "1412.6980",
    archivePrefix = "arXiv",
    primaryClass = "cs.LG",
    month = "12",
    year = "2014"
}

@article{FCC:2018bvk,
    author = "Abada, A. and others",
    collaboration = "FCC",
    title = "{HE-LHC: The High-Energy Large Hadron Collider}: {Future Circular Collider Conceptual Design Report Volume 4}",
    reportNumber = "CERN-ACC-2018-0059",
    doi = "10.1140/epjst/e2019-900088-6",
    journal = "Eur. Phys. J. ST",
    volume = "228",
    number = "5",
    pages = "1109--1382",
    year = "2019"
}

@inproceedings{Benedikt:2022kan,
    author = "Benedikt, M. and others",
    title = "{Future Circular Hadron Collider FCC-hh: Overview and Status}",
    eprint = "2203.07804",
    archivePrefix = "arXiv",
    primaryClass = "physics.acc-ph",
    reportNumber = "FERMILAB-CONF-22-182-AD",
    month = "3",
    year = "2022"
}

@article{Zhu:2021gkn,
    author = "Zhu, Qingling and others",
    title = "{Quantum computational advantage via 60-qubit 24-cycle random circuit sampling}",
    eprint = "2109.03494",
    archivePrefix = "arXiv",
    primaryClass = "quant-ph",
    doi = "10.1016/j.scib.2021.10.017",
    journal = "Sci. Bull.",
    volume = "67",
    pages = "240--245",
    year = "2022"
}

@inproceedings{Lloyd:2013lby,
    author = "Lloyd, Seth and Mohseni, Masoud and Rebentrost, Patrick",
    title = "{Quantum algorithms for supervised and unsupervised machine learning}",
    eprint = "1307.0411",
    archivePrefix = "arXiv",
    primaryClass = "quant-ph",
    month = "7",
    year = "2013"
}

@article{Wiebe:2014imz,
    author = "Wiebe, Nathan and Kapoor, Ashish and Svore, Krysta M.",
    title = "{Quantum algorithms for nearest-neighbor methods for supervised and unsupervised learning}",
    eprint = "1401.2142",
    archivePrefix = "arXiv",
    primaryClass = "quant-ph",
    doi = "10.26421/qic15.3-4-7",
    journal = "Quant. Inf. Comput.",
    volume = "15",
    number = "3&4",
    pages = "316--356",
    year = "2015"
}

@misc{basheer2021quantumknearestneighborsalgorithm,
      title={Quantum $k$-nearest neighbors algorithm}, 
      author={Afrad Basheer and A. Afham and Sandeep K. Goyal},
      year={2021},
      eprint={2003.09187},
      archivePrefix={arXiv},
      primaryClass={quant-ph},
      url={https://arxiv.org/abs/2003.09187}, 
}
\bibliographystyle{JHEP}

\end{document}